\pdfoutput=1

\documentclass[hdvipdfm]{ptephy_v1}

\preprintnumber{XXXX-XXXX} 

\newcommand{\pslash}{p \hspace{-5pt}/ }

\usepackage{amsmath,amssymb}
\usepackage{bm}
\usepackage{graphics}
\usepackage{subfigure}
\usepackage{verbatim}
\usepackage{wrapfig}
\usepackage{ascmac}
\usepackage{makeidx}
\usepackage{braket}
\usepackage{caption}
\usepackage{here}
\usepackage{dcolumn}
\usepackage{color}
\newcommand\sout{\bgroup \color{red} \ULdepth=-.5ex \ULset}

\begin{document}

\title{$KN$ scattering amplitude revisited in a chiral unitary approach and 
a possible broad resonance in $S=+1$ channel}


\author[1,2]{Kenji Aoki}
\author[2,1]{Daisuke Jido}
\affil[1]{Department of Physics, Tokyo Metropolitan University, 
1-1 Minami-Osawa, Hachioji, Tokyo 192-0397, Japan\email{aoki-kenji1@ed.tmu.ac.jp}}
\affil[2]{Department of Physics, Tokyo Institute of Technology, 
2-12-1 Ookayama, Megro, Tokyo 152-8551, Japan}

%
%


\begin{abstract}%
We revisit the $KN$ scattering amplitude in order to investigate the possibility 
for the existence of  a broad resonance in the $I=0$ $KN$ channel around 
the energy of 1617 MeV with 305 MeV width. We use the chiral unitary model
to describe the $KN$ scattering amplitudes and 
determine the model
parameters so as to reproduce the differential cross sections of the $K^{+}N$ scatterings 
and the $I=0$ and 1 total cross sections up to $p_{\rm lab} = 800$ MeV/c, from which inelastic 
contributions start to be significant. 
Performing analytic continuation 
of the determined amplitude to the complex energy plane, we find 
a pole for a broad resonance state.
We point out 
that the rapid increase appearing in the $I=0$ total cross section around $p_{\rm lab}=500$
MeV/c is a hint of the possible broad resonance of strangeness $S=+1$.
\end{abstract}

\subjectindex{D32}

\maketitle

\section{Introduction}
The study of meson-baryon scattering is very important to understand 
the properties of hadron resonances. 
Because hadron resonances are short living and decay immediately by the strong interaction,
they appear only in the scattering processes and 
one can deduce the properties of the hadron resonances only from the investigation of 
the scattering process. 
Description of scattering amplitude is one of the first steps to 
investigate the hadronic resonances.
Once one obtains realistic scattering amplitudes reproducing the scattering cross sections
in terms of analytic functions, one can perform analytic continuation 
to complex energy plane 
and obtain properties of the resonances, such as 
their masses, widths and coupling strengths. 
For the purpose of description of the scattering amplitude in an analytic way, 
one of the theoretical tools is the chiral effective theory in which 
the low energy theorems by chiral symmetry constrain the hadronic interactions.
Chiral perturbation theory describes scattering amplitudes for the lowest channels, while
some unitarization procedure is necessary when one encounters resonances and open channels,
where hadronic dynamics plays an important role. For instance, chiral perturbation theory works
well for the $\pi N$ scattering at low energies, while, for the $\bar KN$ channel, since the 
$\Lambda(1405)$ resonance is located in the $I=0$ channel below the threshold and the 
$\pi\Sigma$ and $\pi\Lambda$ channels are open, one needs unitarization of the amplitude and
takes into account of coupled channels. 

In this article, we reexamine the elastic scattering amplitude of $KN$ in low energies, 
$p_{\rm lab}<800$ MeV/c, based on the chiral unitary approach
and study the possibility of an $S=+1$ exotic resonance in $I=0$ channel.
Baryons with strangeness $S=+1$ are so-called exotic hadrons, because
their quantum numbers cannot be described by three constituent quarks.
For the $S=+1$ baryon, one needs at least one anti-strange quark and more than three 
quarks to compensate the negative baryon number of the anti-strange quark
to have baryon number $+1$ in total. 
Thus, the minimal quark contents are $uudd\bar{s}$ for charge $Q=+1$.
Although there are no reasons to forbid the existence of such states
in quantum chromodynamics, 
the experimental evidence for existence of the $S=+1$ baryons is not well confirmed.

The scattering amplitudes of the $K^{+}N$ scattering in low energies have been studied for a long time. 
A comprehensive review can be found in Ref. \cite{dover1982}.
There are three $K^{+}N$ amplitudes, $K^{+}p \to K^{+}p$, $K^{+}n \to K^{+}n$ and $K^{+}n \to K^{0}p$
and isospin symmetry reduces two independent amplitudes for $I=0$ and $I=1$.
The $K^{+}p$ amplitude can be observed directly from the $K^{+}p \to K^{+}p$ scattering experiment
and provides the $I=1$ amplitude,
while for the $K^{+}n$ amplitudes one needs nuclear targets, such as deuterium, 
and obtains the $I=0$ amplitude using the $I=1$ amplitude.  
It is known that 
the $K^{+}N$ scatterings are almost elastic for $p_{\rm lab} < 800$ MeV/c and inelastic contributions 
are not significant~\cite{Bland:1969cb}.  For low energies, the $K^{+}p$ scattering is described 
by $S$-wave~\cite{Goldhaber:1962zz}. In addition, the differential cross section of the $K^{+}p$ scattering
in low energies shows constructive interference between the Coulomb and strong interactions
at very forward angles. This implies that the low energy $K^{+}p$ scattering is to be repulsive~\cite{Goldhaber:1962zz,cameron1974}. 
In contrast, the $I=0$ amplitude is more ambiguous. 
In Refs.~\cite{Slater:1961zz,PhysRev.134.B1111}, 
it was shown that the scattering amplitude for $I=0$ has $P$-wave contribution 
to reproduce the $K^{+}d$ scattering up to $p_{\rm lab} < 500$ MeV/c. 
The phase shift analysis up to $1.5$ GeV/c by Ref.~\cite{Giacomelli:1974az} found several solutions,
in which one solution implies that the low energy scattering is described dominantly by $S$-wave,
while another solution reproduces the amplitude mainly by $P$-wave. 
The phase shift analyses with new data performed by Refs.~\cite{Sakitt:1975hu,Sakitt:1976ny,Glasser:1977xs}
supported the latter $P$-wave solution. 
The analysis carried out by Ref.~\cite{martin1975} treated 
both $I=0$ and $I=1$ amplitudes at the same time,
and found that the $P$-wave contribution was significant for the low energy $I=0$ amplitude.

The search for $S=+1$ resonance has been carried out in the past.
In the earlier studies, 
a possible $S=+1$ broad resonance $Z^{\ast}$ in the $KN$ scattering with $I=0$ 
were discussed~\cite{Cool:1966zz,Tyson:1967zz,bugg1968,abrams1969,Giacomelli:1972uj,wilson1972,Giacomelli:1973ed,carroll1973,Giacomelli:1974az}.
In Ref.~\cite{abrams1969}, it was pointed out that there the $K^{+}N$ total cross 
sections \cite{Cool:1966zz} and $K^{-}$ photoproduction \cite{Tyson:1967zz} showed 
two bump structures which would have risen possible $KN$ resonances with $I=0$. 
Although the phase shift analysis by Martin~\cite{martin1975} found that there were 
no significant resonances in the partial wave amplitude, the Argand diagram suggested 
that there could be some broad resonances appearing in $P_{01}$ and $D_{03}$ \cite{Lea:1968,Watts:1980qs,Robertson:1980ma}.
These resonances were reported as broad resonances above 
the energies where the inelastic contributions start to be significant.
The studies of the bump structure in cross sections and the behavior of the Argand diagrams 
are not sufficient to fix the existence of the resonance states. One of the promising ways 
is to analyze the scattering amplitude as a analytic function and to carry out analytic continuation
of the amplitude into the complex energy plane. Resonance states are expressed as poles of the
scattering amplitude. 
Another kind of the resonance with $S=+1$ was suggested by LEPS collaboration
in photoproduction experiments~\cite{nakano}. They claimed a narrow resonance, $\Theta^{+}$, 
with $S=+1$ and 1.5 GeV/${\rm c}^{2}$ mass~\cite{nakano,Nakano:2008ee}. 
This experiment was motivated by a theoretical work \cite{diakonov} 
predicting a resonance with $S=+1$ around the mass 1540 MeV/${\rm c}^{2}$ and narrow width $\Gamma < 15$ MeV.
Further studies in the chiral soliton model were performed in Refs. \cite{jaffe, weigel}.
The $\Theta^{+}$ resonance is obviously different from the previous $Z^{\ast}$ resonance.
Here we revisit the possibility of the existence of a $Z^{*}$-type
broad resonance with $S=+1$ and $I=0$
at lower energies than where the inelastic contributions start to be significant. 

As mentioned in the above, it is important to understand the resonance properties by 
studying the scattering amplitude, especially in terms of an analytic function.
There are several approaches to describe baryon resonance.
We use chiral unitary model, in which scattering problem is solved in a simplified manner
by considering elastic unitarity of the two-body scattering and the elementary interaction 
is given based on chiral perturbation theory, firstly suggested in Ref. \cite{kaiser1995} and 
developed in \cite{oset1998}, for the $S=-1$ channel,  
and we can find recent progress in a review article~\cite{hyodo2012}.
Chiral unitary model impose unitary condition by infinite summation of the specific diagram
and describes the scattering amplitude as an analytic function. 
Thus, it is easy to perform analytic continuation of the scattering amplitude
and to pin down the position of the resonance state in the complex energy plane. 
One of the most successful examples of this approach is the finding of the
double pole structure of the $\Lambda(1405)$ and investigation of 
its physical significance~\cite{Oller:2000fj,Jido:2003cb}.
It was reported in Refs.~\cite{Hyodo:2006yk,Hyodo:2006kg} that the Tomozawa-Weinberg interactions 
for the exotic channels do not provide enough attraction to make two-body bound states 
for a Nambu-Goldstone boson and a hadron. Actually the Tomozawa-Weinberg term vanishes 
for $I=0$ and $S=+1$. Here the next-to-leading contributions are responsible for the attraction
to provide a broad resonance. 

This paper is organized as follows.
In Sec. \ref{sec:sec2}, we construct $KN$ scattering amplitude using chiral perturbation theory
and chiral unitary model.
In Sec. \ref{sec:sec3}, we determine the parameters which reproduce $KN$ scattering data.
The total cross section and differential cross section data are compared with our results.
Using the constructed amplitude, we discuss the possibility of resonance with large width.
In Sec. \ref{sec:sec4}, we summarize the results of this paper.

\section{Formulation}
\label{sec:sec2}
For our theoretical investigation, we would like to represent the $KN$ scattering amplitude 
in an analytic function of the center of mass energy $W$. Once we parametrize 
the scattering amplitude in an analytic function,  analytic continuation allows us
to extend the amplitude to the complex energy plane, where resonances 
are represented as poles, and extract the properties of resonances,
such as mass, decay width and coupling strength. 
For this purpose, we describe the $KN$ elastic scattering amplitude based on the chiral unitary approach by solving 
Lippmann-Schwinger equation 
\begin{equation}
    T = V + VGT
\end{equation}
in a simplified way. In the chiral unitary approach,
the interaction kernel $V$ is given by chiral perturbation theory and 
we restrict the intermediate state to the elastic channel.
The model parameters are determined so as to reproduce the observed $KN$ cross section. 

\subsection{Scattering amplitude}
\label{sec:amp}

Let us call the momenta of the kaon and nucleon in the initial (final) state 
by $p_{1}$ and $p_{2}$ ($p_{3}$ and $p_{4}$), respectively. According to Lorentz invariance, 
the $T$-matrix of the $KN$ scattering can be written in terms of two Lorentz invariant functions,
$A(s,t)$ and $B(s,t)$, in general, as
\begin{equation}
  T(s,t) = \bar u (\vec p_{4},s_{4}) \left[A(s,t) + \frac12(\pslash_{1}+\pslash_{3}) B(s,t) \right] u(\vec p_{2},s_{2})
\end{equation}
with the Mandelstam variables $s=(p_{1}+p_{2})^{2}$ and $t=(p_{1}-p_{3})^{2}$ and 
the on-shell Dirac spinor $u(\vec p, s)$ for nucleon with momentum $p$ and spin $s$. The Dirac spinor $u(\vec p, s)$ is normalized by $\bar u(\vec p,s) u(\vec p,s^{\prime}) = 2M_{N} \delta_{ss^{\prime}}$ with the nucleon mass $M_{N}$.
For partial wave decomposition, we write the $T$-matrix in the center of mass system
with two functions $f$ and $g$ in terms of the spin-nonflip and spin-flip parts as
\begin{equation}
  T(s,t) 
   = \chi^{\dagger}(\lambda_{4})\left[ f(W,\theta) - i (\vec \sigma \cdot \hat n) g(W,\theta)\right]
   \chi(\lambda_{2})
\end{equation}
where $W$ and $\theta$ are the total energy and scattering angle (angle between $\vec p_{1}$ and $\vec p_{3}$)
in the center of mass system, respectively, $\hat n$ is the normal vector of the scattering plane
defined by $\hat n = (\vec p_{3} \times \vec p_{1})/ | \vec p_{3} \times \vec p_{1}|$, and
$\chi(\lambda)$ is Pauli spinor of nucleon with helicity $\lambda$. 
The relation between $A, B$ and $f, g$ in the $KN$ elastic scattering is given by
\begin{eqnarray}
   f(W,\theta) &=&  (E_{N}+M_{N}) (A + \omega B ) 
   + k^{2} B
   +  \frac{(E_{N}+M_{N}+\omega)B-A}{E_{N}+M_{N}} k^{2} \cos\theta, \\
   g(W, \theta) &=& \frac{A- (E_{N} + M_{N} + \omega)B}{(E_{N}+M_{N})} k^{2} \sin\theta 
\end{eqnarray}
with the kaon energy $\omega$, the 3-momentum in the center of mass system $k$ and
the nucleon energy $E_{N}$. 

The amplitudes $f(W,\theta)$ and $g(W,\theta)$ can be decomposed into partial waves with
Legendre polynomials $P_{\ell}(x)$ as
\begin{eqnarray}
   f(W,\theta) &=& \sum_{\ell=0}^{\infty} f_{\ell}(W) P_{\ell}(\cos\theta) ,\\
   g(W,\theta) &=& \sum_{\ell=1}^{\infty} g_{\ell}(W) \sin\theta \frac{dP_{\ell}(\cos\theta)}{d\cos\theta}.
\end{eqnarray}
It is convenient to introduce the amplitude $T_{\ell \pm}$ having definite total angular momentum 
$j=\ell \pm \frac12$ by
\begin{eqnarray}
   f_{\ell}(W) &=& (\ell+1) T_{\ell+}(W) + \ell T_{\ell-}(W), \\
   g_{\ell}(W) &=& T_{\ell+} (W) - T_{\ell-}(W),
\end{eqnarray}
or equivalently 
\begin{eqnarray}
    T_{\ell+}(W) &=&  \frac{1}{2\ell+1} (f_{\ell} (W) + \ell g_{\ell}(W)), \\
    T_{\ell-}(W) &=& \frac{1}{2\ell +1} (f_{\ell}(W) - (\ell+1) g_{\ell}(W)).
\end{eqnarray}
We also introduce the partial-wave decomposed interaction kernels $V_{\ell+}(W)$ and $V_{\ell-}(W)$ 
in the same way.  

Here we show the $KN$ scattering amplitudes in the isospin channels, $T^{I=0}$ and $T^{I=1}$. 
The amplitudes in the particle basis can be obtained as
\begin{eqnarray}
    T_{K^{+}p \to K^{+}p} &=& T^{I=1}, \\
    T_{K^{+}n \to K^{+}n} &=& \frac{1}{2} (T^{I=1} + T^{I=0}), \label{eq:Kn} \\
    T_{K^{+}n \to K^{0}p} &=& \frac{1}{2} (T^{I=1} - T^{I=0}). \label{eq:CE}
\end{eqnarray}

Taking spin average in the initial state and spin summation in the final state
for nucleon, 
we calculate the differential cross section in the center of mass frame as
\begin{equation}
   \frac{d \sigma}{d \Omega} = \frac{1}{64 \pi^{2} s} \left( |f(W,\theta)|^{2} + |g(W,\theta)|^{2} \right)
\end{equation}
and the total cross section by integrating the differential cross section in terms of the scattering angle as
\begin{equation}
  \sigma = \frac{1}{32 \pi s} \int_{-1}^{1} d\cos\theta \left( |f(W,\theta)|^{2} + |g(W,\theta)|^{2} \right).
\end{equation}

\subsection{Chiral Lagrangian}

The leading order chiral Lagrangian for the baryon field $B$ reads 
\begin{eqnarray}
{\cal L}_{MB}^{(1)}=
{\rm Tr} \left[\bar{B}(i D \hspace{-7pt} / \, - M_{0}) B\right]
- \frac{D}{2} {\rm Tr} \left(\bar{B} \gamma_{\mu} \gamma_{5} \{u^{\mu},B\} \right)
-\frac{F}{2} {\rm Tr} \left (\bar{B} \gamma_{\mu} \gamma_{5} [u^{\mu},B] \right)
\label{eq:MB_1}, 
\end{eqnarray}
where $M_{0}$ is the baryon mass at the chiral limit, 
the baryon and meson fields, $B$ and $\Phi$, are written in the SU(3) matrix form
\begin{eqnarray}
B&=&\left(
\begin{array}{ccc}
\frac{\Sigma^{0}}{\sqrt{2}}+\frac{\Lambda}{\sqrt{6}} &\Sigma^{+}& p \\
\Sigma^{-} & -\frac{\Sigma^{0}}{\sqrt{2}}+\frac{\Lambda}{\sqrt{6}} & n \\
\Xi^{-}& \Xi^{0} &  -\frac{2\Lambda}{\sqrt{6}}
\end{array}
\right), \\
\Phi&=&\left(
\begin{array}{ccc}
\frac{\pi^{0}}{\sqrt{2}}+\frac{\eta}{\sqrt{6}} & \pi^{+} &K^{+} \\
\pi^{-} & -\frac{\pi^{0}}{\sqrt{2}}+\frac{\eta}{\sqrt{6}} & K^{0} \\
K^{-}& \bar{K}^{0} &  -\frac{2\eta}{\sqrt{6}}
\end{array}
\right).
\end{eqnarray}
We parametrize the chiral field $U$ in the CCWZ form as
\begin{equation}
   U = \xi^{2} = \exp \left( i \frac{\sqrt 2}{f} \Phi\right) 
\end{equation}
with a scale parameter $f$, which is turned to be identified as the meson decay constant in the leading order 
calculation of chiral perturbation theory, 
the covariant derivative for the baryon field is introduced as
\begin{equation}
   D_{\mu} B= \partial_{\mu} B + [ \Gamma_{\mu}, B ], 
\end{equation}
with the mesonic vector current 
\begin{equation}
   \Gamma_{\mu} = \frac{1}{2} ( \xi^{\dagger} \partial_{\mu} \xi + \xi \partial_{\mu} \xi^{\dagger} ),
\end{equation}
and the meson-baryon coupling is given through the mesonic axial vector current
\begin{equation}
   u_{\mu} = i \left(\xi^{\dagger} \partial_{\mu} \xi - \xi \partial_{\mu} \xi^{\dagger} \right)
\end{equation}
with low energy constants $D$ and $F$. The parameters $D$ and $F$ are to be determined 
by the axial couplings of the baryons at tree level. 

The next-leading order of the chiral Lagrangian
is composed of several terms 
\begin{align}
{\cal L}_{MB}^{(2)}
=&b_{D} {\rm Tr} \left(\bar{B} \{ \chi_{+}, B \} \right)
+b_{F} {\rm Tr} \left(\bar{B} [\chi_{+}, B] \right)
+b_{0} {\rm Tr} (\bar{B}B) {\rm Tr} (\chi_{+})
+d_{1} {\rm Tr} \left(\bar{B} \{ u_{\mu}, [u^{\mu}, B] \} \right)  
\nonumber \\ &
+d_{2} {\rm Tr} \left(\bar{B} [u_{\mu}, [u^{\mu}, B]] \right)
+d_{3} {\rm Tr} \left(\bar{B} u_{\mu} ) {\rm Tr} (u^{\mu} B \right) 
+d_{4} {\rm Tr} \left(\bar{B} B) {\rm Tr} (u^{\mu} u_{\mu} \right)
\nonumber \\ 
&-\frac{g_{1}}{8M_{N}^{2}} {\rm Tr}  \left( \bar B \{ u_{\mu}, [ u_{\nu}, \{D^{\mu},D^{\nu}\}B] \} \right) 
-\frac{g_{2}}{8M_{N}^{2}} {\rm Tr}  \left( \bar B [ u_{\mu}, [ u_{\nu}, \{D^{\mu},D^{\nu}\}B] ] \right) 
\nonumber \\ 
&-\frac{g_{3}}{8M_{N}^{2}} {\rm Tr}  (\bar B  u_{\mu} ) {\rm Tr}  ( u_{\nu}, \{D^{\mu},D^{\nu}\}B) 
-\frac{g_{4}}{8M_{N}^{2}} {\rm Tr}  (\bar B\{D^{\mu},D^{\nu}\}B) {\rm Tr} (u_{\mu} u_{\nu}) 
\nonumber \\ 
&-\frac{h_{1}}{4} {\rm Tr}  \left( \bar B [\gamma^{\mu},\gamma^{\nu}] B u_{\mu} u_{\nu} \right) 
-\frac{h_{2}}{4} {\rm Tr}  \left( \bar B [\gamma^{\mu},\gamma^{\nu}] u_{\mu} [u_{\nu}, B] \right) 
\nonumber \\ 
&-\frac{h_{3}}{4} {\rm Tr}  \left(\bar B [\gamma^{\mu},\gamma^{\nu}] u_{\mu} \{u_{\nu}, B\} \right) 
-\frac{h_{4}}{4} {\rm Tr}  (\bar B [\gamma^{\mu},\gamma^{\nu}] u_{\mu}) {\rm Tr} (u_{\nu} B) + {\rm h.c.}
\label{eq:MB_2}
\end{align}
where $b_{i}$, $d_{i}$, $g_{i}$ and $h_{i}$ are low energy constants. 
The terms with $b_{i}$ and $d_{i}$ appear in the typical SU(3) chiral Lagrangians,
while the terms with $g_{i}$ and $h_{i}$ are introduced as an extension of the SU(2) chiral Lagrangian \cite{Bernard:1995dp,Fettes:2000gb}. (See also Ref.~\cite{Lutz:2001yb}.)
The scalar field $\chi_{+}$ is given by
\begin{equation}
   \chi_{+} = 2 B_{0} \left (\xi {\cal M} \xi + \xi^{\dagger} {\cal M} \xi^{\dagger} \right),
\end{equation}
with the quark mass matrix
\begin{equation}
    {\cal M}={\rm diag} \left(\hat{m}, \hat{m}, m_{s} \right),
\end{equation}
where $\hat m$ stands for the mass of the $u$ and $d$ quarks by assuming isospin symmetry,
while $m_{s}$ means the strange quark mass.
Parameter $B_{0}$ is a positive constant related to the meson mass and always appears
together with the quark mass.
The parameter is fixed by $B_{0}=M_{K}^{2}/(\hat{m}+m_{s})$ where $M_{K}$ is the kaon mass.
The low energy constants some combinations of 
$b_{i}$, $d_{i}$, $g_{i}$ and $h_{i}$ are to be fixed
by the $KN$ scattering cross sections. 

\subsection{Interaction kernel}

The tree level amplitude of the $KN$ scattering up to the next-to-leading order
is composed by three parts, the leading order contact term, the hyperon crossed Born term,
and the next-to-leading contact term. 
The leading order contact term is called Tomozawa-Weinberg term, 
and is determined by the SU(3) group structure of hadrons 
without the low energy constants. It is known to be absent for the $KN$ channel with $I=0$:
\begin{eqnarray}
   V^{I=0}_{\rm TW} = 0, \qquad
   V^{I=1}_{\rm TW} = \frac{1}{2f_{K}^{2}} 
   \bar{u}(\vec{p}_4,s_4)(\pslash_1+\pslash_{3}) u(\vec{p}_2,s_2).
\end{eqnarray}
The corresponding invariant amplitudes read
\begin{eqnarray}
  A^{I=0}_{\rm TW} = B^{I=0}_{\rm TW} = A^{I=1}_{\rm TW} = 0, \qquad
  B^{I=1}_{\rm TW} = \frac{1}{f_{K}^{2}}.
\end{eqnarray}
%
For the Born term, we do not consider explicit baryonic states with strangeness $S=+1$. 
The pentaquark $\Theta^{+}$ is a candidate for such a state, but it is known to have a 
narrow width and a very weak coupling to $KN$. The hyperons with $S=-1$, $\Sigma$ and $\Lambda$,
contribute to the $KN$ amplitude as crossed Born terms.  
With the chiral Lagrangian (\ref{eq:MB_1}), we obtain the crossed Born terms as
\begin{eqnarray}
V^{I=0}_{{\rm Born}}&=& -\frac{3}{4}\frac{(D-F)^{2}}{f^{2}_{K}}
\bar{u}(\vec{p}_4,s_4)\pslash_1\gamma_5
\frac{M_{\Sigma} + (\pslash_2 - \pslash_3) }{M_{\Sigma}^{2} - (p_{2} - p_{3} )^{2} - i\epsilon}
\pslash_3\gamma_5u(\vec{p}_2,s_2)
\nonumber\\
&&+\frac{1}{12} \frac{(3F+D)^{2}}{f^{2}_{K}}\bar{u}(\vec{p}_4 ,s_4)\pslash_1\gamma_5
\frac{M_{\Lambda} + (\pslash_2 - \pslash_{3}) }{M_{\Lambda}^{2} - (p_{2} - p_{3} )^{2} - i\epsilon}
\pslash_3\gamma_5u(\vec{p}_2,s_2), \label{eq:I02} \\
V^{I=1}_{\rm Born} &=&  -\frac{1}{4} \frac{(D-F)^{2}}{f^{2}_{K}}
\bar{u}(\vec{p}_4,s_4)\pslash_1\gamma_5
\frac{M_{\Sigma} + (\pslash_2 - \pslash_{3}) }{M_{\Sigma}^{2} - (p_{2} - p_{3} )^{2} - i\epsilon}
\pslash_3\gamma_5 u(\vec{p}_2,s_2)
\nonumber\\
&&-\frac{1}{12}\frac{(3F+D)^{2}}{f^{2}_{K}}\bar{u}(\vec{p}_4 ,s_4)\pslash_1\gamma_5
\frac{M_{\Lambda} + (\pslash_2 - \pslash_{3} ) }{M_{\Lambda}^{2} - (p_{2} - p_{3} )^{2} - i\epsilon}
\pslash_3\gamma_5u(\vec{p}_2,s_2) \label{eq:I12}
\end{eqnarray}
with the $\Sigma$ mass $M_{\Sigma}$ and $\Lambda$ mass $M_{\Lambda}$. 
The invariant amplitudes are written as
\begin{eqnarray}
   A^{I=0}_{\rm Born} &=& \frac34 \frac{(D-F)^{2}}{f^{2}_{K}}
     \frac{(M_{N}+M_{\Sigma})(M_{N}^{2}-u)}{u - M_{\Sigma}^{2}} 
    -\frac{1}{12} \frac{(3F+D)^{2}}{f^{2}_{K}}
    \frac{(M_{N}+M_{\Lambda})(M_{N}^{2} - u)}{u - M_{\Lambda}^{2}},
     \\
   B^{I=0}_{\rm Born} &=& -\frac34 \frac{(D-F)^{2}}{f^{2}_{K}}
    \frac{u+M_{N}^{2} + 2 M_{\Sigma} M_{N} }{u - M_{\Sigma}^{2}} 
    +\frac{1}{12} \frac{(3F+D)^{2}}{f^{2}_{K}}
    \frac{u+M_{N}^{2} +2 M_{\Lambda} M_{N}}{u - M_{\Lambda}^{2}},\\
   A^{I=1}_{\rm Born} &=& \frac14 \frac{(D-F)^{2}}{f^{2}_{K}}
     \frac{(M_{N}+M_{\Sigma})(M_{N}^{2}-u)}{u - M_{\Sigma}^{2}} 
    +\frac{1}{12} \frac{(3F+D)^{2}}{f^{2}_{K}}
    \frac{(M_{N}+M_{\Lambda})(M_{N}^{2} - u)}{u - M_{\Lambda}^{2}},
     \quad \\
   B^{I=1}_{\rm Born} &=& -\frac14 \frac{(D-F)^{2}}{f^{2}_{K}}
    \frac{u+M_{N}^{2} + 2 M_{\Sigma} M_{N} }{u - M_{\Sigma}^{2}} 
    -\frac{1}{12} \frac{(3F+D)^{2}}{f^{2}_{K}}
    \frac{u+M_{N}^{2} +2 M_{\Lambda} M_{N}}{u - M_{\Lambda}^{2}}
\end{eqnarray}
with Mandelstam variable $u = (p_{1}-p_{4})^{2} = 2M_{N}^{2}+2M_{K}^{2} - s -t $,
the nucleon mass $M_{N}$ and the kaon mass $M_{K}$.
The $KN$ invariant amplitudes at the next-to-leading order chiral perturbation theory 
are calculated for each isospin channel as
\begin{eqnarray}
V^{I}_{\rm NLO}&=&
\left[
\frac{4B_0}{f^{2}_{K}}(\hat{m}+m_s)b^{I}
+\frac{2}{f^{2}_{K}}  (p_1\cdot p_3) d^{I} 
 \right .\nonumber \\ && \quad \left.
+\frac{(p_{2}\cdot p_{1})(p_{2}\cdot p_{3}) + (p_{4}\cdot p_{1})(p_{4}\cdot p_{3})}{2M_{N}^{2} f_{K}^{2}} g^{I}
\right]
\bar{u}(\vec{p}_4,s_4)u(\vec{p}_2, s_2) 
\nonumber \\ && 
 - \frac{ h^{I}}{2f_{K}^{2}}  p_{1}^{\mu} p_{3}^{\nu}\,  \bar{u}(\vec{p}_4,s_4) [\gamma_{\mu}, \gamma_{\nu}] 
u(\vec{p}_2, s_2),
 \label{eq:I03}
\end{eqnarray}
and the corresponding invariant amplitudes $A$ and $B$ read
\begin{eqnarray}
   A_{\rm NLO}^{I} &=& \frac{4B_0}{f^{2}_{K}}(\hat{m}+m_s)b^{I}
+\frac{2}{f^{2}_{K}}  (p_1\cdot p_3) d^{I}
  \nonumber \\ && 
+\frac{(p_{2}\cdot p_{1})(p_{2}\cdot p_{3}) + (p_{4}\cdot p_{1})(p_{4}\cdot p_{3})}{2M_{N}^{2} f_{K}^{2}} g^{I}
+ \frac{p_{1}\cdot(p_{2}+p_{4})}{f_{K}^{2}} h^{I}\\
   B_{\rm NLO}^{I} &=& -\frac{2 M_{N}}{f_{K}^{2}} h^{I}.
\end{eqnarray}
In these equations, the parameters $b^{I}$, $d^{I}$, $g^{I}$ and $h^{I}$ are defined by
\begin{align}
  b^{I=0} &= b_{0} - b_{F},                    & b^{I=1} &= b_{0} + b_{D}, \\ 
  d^{I=0} &= 2d_{1} + d_{3} - 2d_{4},    & d^{I=1} &= -2d_{2} - d_{3} - 2d_{4}, \\
  g^{I=0} &=  2g_{1} + g_{3} - 2 g_{4},   & g^{I=1} &=  -2g_{2} - g_{3} - 2 g_{4}, \\
  h^{I=0} &= h_{1}+h_{2}+h_{3}+h_{4},    & h^{I=1} &= h_{1}-h_{2}-h_{3}-h_{4},
\end{align}
in terms of the low energy constants appearing in Lagrangian (\ref{eq:MB_2}). 
We treat these combinations of the low energy constants as free parameters
to be adjusted to reproduce observed $KN$ cross sections.  

\subsection{Unitarization}

Unitarization is performed in each partial wave amplitude~\cite{Jido:2002zk}. 
Because the total angular momentum is a good quantum number, Lippmann-Schwinger equation is also
decomposed into partial waves as
\begin{equation}
   T_{\ell \pm}^{I} = V_{\ell \pm}^{I} + V_{\ell \pm}^{I} G T_{\ell \pm}^{I}
\end{equation}
where we have assume that we use a non-relativistic Green's function and do not consider
so-called zig-zag diagrams which mix the large and small components of the Dirac spinor. 
Supposing that we take only the on-shell contribution of the interaction kernel 
in the loop integral, we can solve Lippmann-Schwinger equation algebraically 
\begin{equation}
\label{eq:less_amp}
   T_{\ell \pm}^{I} = (1 - V_{\ell \pm}^{I}G)^{-1} V_{\ell \pm}^{I}
\end{equation}
where the loop contribution $G$ for the $KN$ channel is given 
as a function of the center of mass energy, $W$, by 
\begin{equation}
   G(W) = i\int \frac{d^{4} q}{(2\pi)^{4}} \frac{1}{(P-q)^{2} - M_{N}^{2} + i \epsilon}
   \frac{1}{q^{2} - M_{K}^{2} + i \epsilon} .
\end{equation}
This integral can be
performed by the dimensional regularization as
\begin{eqnarray}
G(W)&=&\frac{1}{(4 \pi )^{2}}
\biggl \{ a(\mu) 
+\ln \frac{M_{N}^{2}}{\mu^{2}} 
+\frac{M_{K}^{2}-M_{N}^{2} +s}{2 s} \ln \frac{M_{K}^{2}}{M_{N}^{2}}  \nonumber \\
&&+\frac{k}{\sqrt{s}} \Bigl [ \ln(s - (M_{N}^{2} - M_{K}^{2}) 
+ 2\sqrt{s} k) + \ln \left(s + \left(M_{N}^{2} - M_{K}^{2} \right) +2\sqrt{s} k \right) \nonumber \\
&&-\ln \left(-s + \left(M_{N}^{2} - M_{K}^{2} \right) 
+2\sqrt{s} k \right) 
-\ln\left(-s-\left(M_{N}^{2}-M_{K}^{2}\right)
+2\sqrt{s} k \right) \Bigl]  \biggl \},
\label{eq:loop}
\end{eqnarray}
where $\mu$ is the scale parameter of the dimensional regularization and
$a(\mu)$ is the subtraction constant depending on $\mu$.
We take away the infinite part as renormalization procedure, and 
the subtraction constant is determined so as to reproduce experiments. 
For the interaction kernel, here we take the chiral perturbation amplitudes 
calculated in the previous section up to the next-to-leading order as
\begin{equation}
    V^{I} = V_{\rm WT}^{I} + V_{\rm Born}^{I} + V_{\rm NLO}^{I}
\end{equation}
and perform partial wave decomposition in the way explained in Sec.~\ref{sec:amp}.

\subsection{Coulomb correction}
For the $K^{+}p$ amplitude, we introduce the Coulomb correction 
as done in Ref.~\cite{hashimoto}. 
To the strong interaction part of the $K^{+}p$ scattering amplitude
calculated in the center of mass frame,
we add
the Coulomb amplitude 
\begin{equation}
  f_{C} = -\frac{\alpha}{2 k v \sin^{2}(\theta/2)} \exp\left[ - i \frac{\alpha}{v} 
  \ln \left(\sin^{2} \frac{\theta}{2} \right)\right]
\end{equation}
with the scattering angle $\theta$, the fine structure constant $\alpha$
and the $KN$ relative velocity $v$ defined by
\begin{equation}
  v = \frac{ k (E_{K} + E_{p})}{E_{K}E_{p}},
\end{equation}
and multiply the Coulomb phase shift factor $e^{2i \Phi_{\ell}}$
with
\begin{equation}
   \Phi_{\ell} = \sum_{n=1}^{\ell} \tan^{-1} \frac{\alpha}{n v}
\end{equation}
for $\ell >0$ ($\Phi_{0} = 0$) as
\begin{eqnarray}
   f^{K^{+}p} &=& \sum_{\ell=0}^{\infty} \left[ (\ell+1) T_{\ell+}^{I=1} 
   + \ell T_{\ell-}^{I=1} \right] e^{2i\Phi_{\ell}} P_{\ell}(\cos\theta) 
   - 8 \pi \sqrt s f_{C}, \\
   g^{K^{+}p} &=& \sum_{\ell=1}^{\infty} 
   \left[T_{\ell+}^{I=1} - T_{\ell-}^{I=1}\right]    e^{2i\Phi_{\ell}} 
   \sin\theta \frac{dP_{\ell}(\cos\theta)}{d\cos\theta} .
\end{eqnarray}

\section{Results}
\label{sec:sec3}
%

In the previous section, we have constructed the unitarized $KN$ amplitudes using the chiral unitary model.
In this section, we carry out the $\chi^{2}$ fitting of the unitarized amplitude to the experimental data 
and determine the low-energy constants appearing in the next-to-leading order.
We assume to fix the subtraction constants for $I=0$ and $1$ at $a^{I=0,1} = -1.150$
with the regularization scale $\mu=1$ GeV 
to suppress the number of the parameters in the situation that the experimental data have somewhat large error
and disagreement in different experiments.
Moderate change of the subtraction constants can be absorbed into the low-energy constants.
We assume isospin symmetry and use the isospin averaged masses for the hadrons.
We take the kaon decay constant as $f_{K} = (1.19 \pm 0.01)f_{\pi} = 110.0$ MeV where $f_{\pi}$ = 92.4 MeV.
The low-energy constants at the leading-order chiral Lagrangian $D$ and $F$
are already determined by the semileptonic hyperon beta decay reported in \cite{DandF} as
\begin{eqnarray}
D=0.80, \enspace F= 0.46.
\end{eqnarray}
The values of these parameters are summarized in Table~\ref{tab:masses}.

\begin{table}[t]
\begin{center}
\caption{The values of the fixed parameters. We take the isospin averaged masses. }
\label{tab:masses}
\begin{tabular}{ccccccc}
\hline
$M_{N}$ & $M_{K}$ & $M_{\Lambda}$ & $M_{\Sigma}$ & $f_{K}$ & $D$ & $F$ \\
\hline\hline
$938.9$ MeV & $495.6$ MeV & $1115.7$ MeV & $1193.2$ MeV & $110.0$ MeV & 0.80 & 0.46 \\
\hline
\end{tabular}
\end{center}
\end{table}


\subsection{Determining amplitude} 
We determine the $KN$ amplitudes for $I=1$ and $I=0$ 
reproducing the experimental data using the $\chi^{2}$ fitting of the low-energy constants to minimize 
the $\chi^{2}$ function
\begin{eqnarray}
\chi^{2} = \sum_{i}^{N} \left(\frac{y_{i} - f(x_{i})}{\sigma_{i}} \right)^{2},
\end{eqnarray}
where $y_{i}, f(x_{i}), \sigma_{i}$ and $N$ are the experimental data,
the theoretical calculations with the parameters, the errors of the data and the number of the data, respectively.
In our analysis, we consider the partial waves up to the $D$-wave ($l=2$). 
We restrict the energy region up to $p_{\rm lab} = 800$ MeV/c, where the inelastic contribution 
such as pion production starts to be significant.


\begin{table}[t]
\begin{center}
\caption{The determined parameters for the $I=1$ and $I=0$ amplitudes. There are two 
parameter sets.
Solutions 1 and 2 are characteristic in reproduction of the experimental data. (See text in details.) 
The values of the parameters $b^{I}, d^{I}, g^{I}$ and $h^{I}$ are shown in unit of $10^{-3}\ {\rm MeV}^{-1}$.
The subtraction constants for $I=0$ and $I=1$ are fixed as $a^{I=0, 1}=-1.150$.}
\begin{tabular}{c c| D{.}{.}{4}D{.}{.}{4} } 
\hline
  & & \multicolumn{1}{c}{Solution 1}  & \multicolumn{1}{c}{Solution 2}  \\ 
 \hline  \hline
          &$b^{I=1}$         &0.54        &0.30        \\
          &$d^{I=1}$         &-0.29       &-0.24         \\
$I=1$  &$g^{I=1}$         &0.05       &0.72        \\
           &$h^{I=1}$         &0.03       &1.05       \\
           & $\chi^{2}/N$    &2.96       &2.97     \\
\hline
          &$b^{I=0}$         &0.11        &-0.46          \\
          &$d^{I=0}$         &0.33       &0.73       \\
$I=0$  &$g^{I=0}$         &-0.42       &0.56        \\
           &$h^{I=0}$         &1.14        &-3.54     \\
           & $\chi^{2}/N$    &4.54       &4.06     \\
\hline
\end{tabular} 
\label{tab:parameter}
\end{center}
\end{table}

\begin{figure}[]
    \begin{tabular}{c}
      \begin{minipage}[t]{1.0\hsize}
        \centering
        \includegraphics[keepaspectratio, scale=0.8,bb=0 0 360 252]{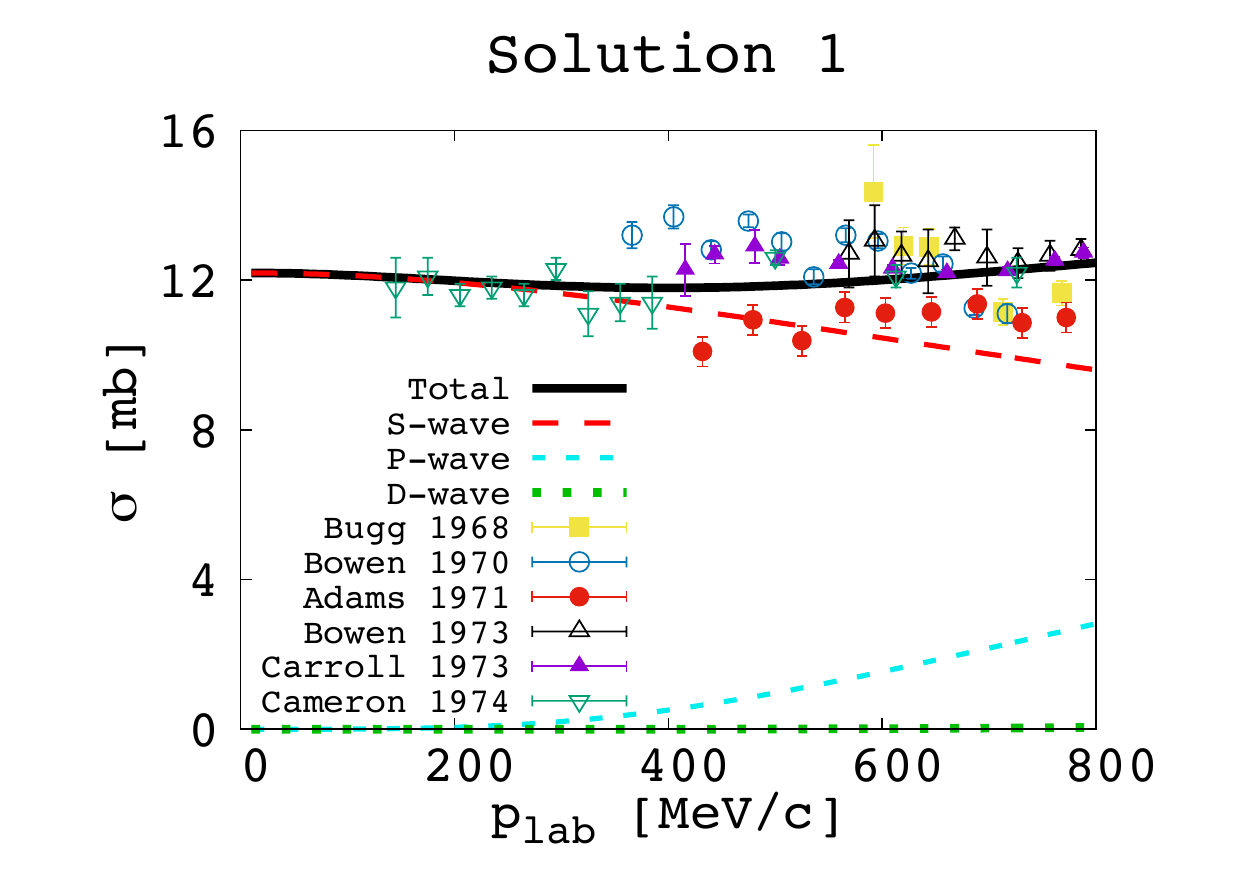}
      \end{minipage} \\
      \begin{minipage}[t]{1.0\hsize}
        \centering
        \includegraphics[keepaspectratio, scale=0.8,bb=0 0 360 252]{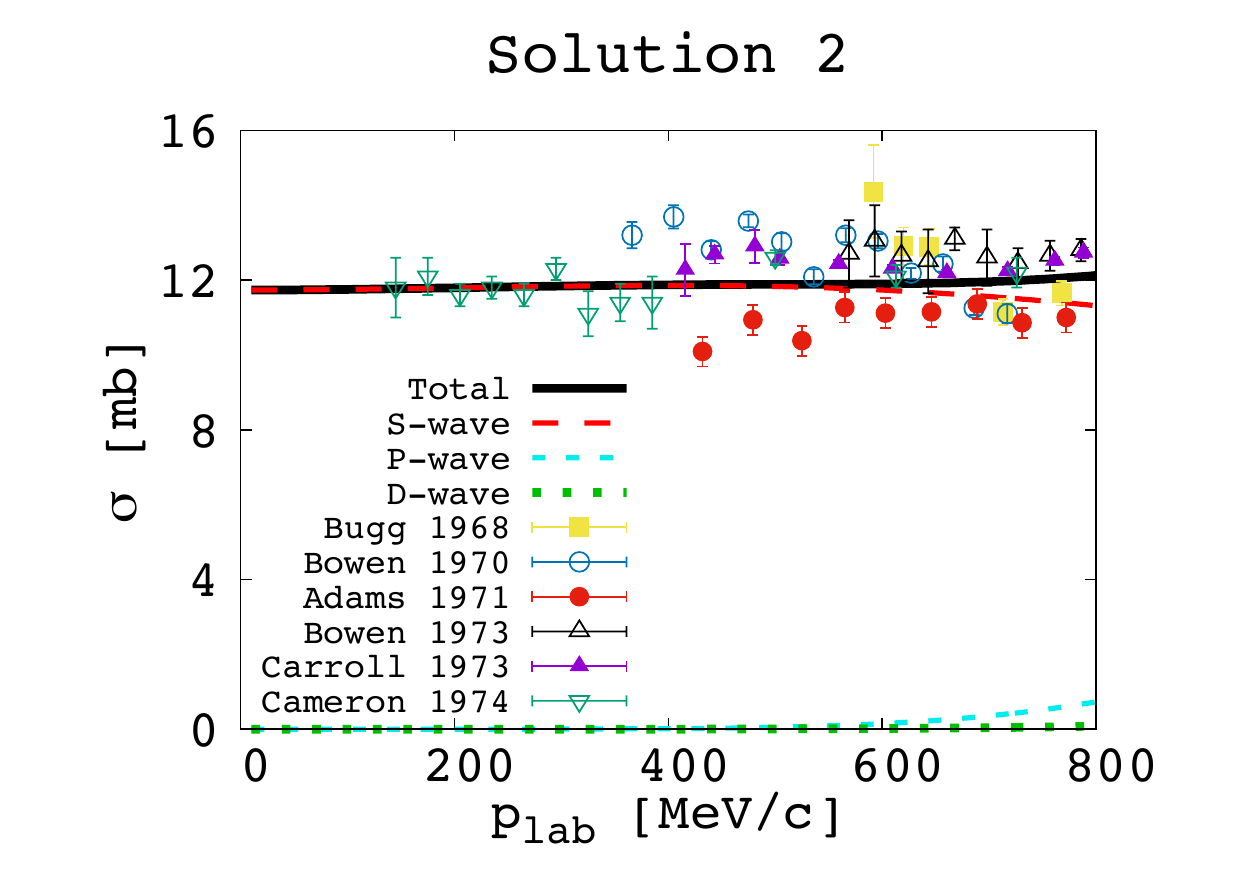}
      \end{minipage} \\
    \end{tabular}
     \caption{The calculated $I=1$ total cross sections using Solutions 1 and 2 in comparison 
with the experimental data \cite{bugg1968, bowen1970, adams1971,
bowen1973, carroll1973, cameron1974}.
  The partial wave components are also described by the dashed line.
  The horizontal axis means the $K^{+}$ meson incident momentum 
  in the lab frame $p_{{\rm lab}}$ in the unit of MeV/c
  and the vertical axis is the total cross section $\sigma$ in the unit of mb.
}
\label{fig:i1_tot}
  \end{figure}

\begin{figure}[]
\begin{minipage}{0.5\hsize}
  \begin{center}
   \includegraphics[width=80mm,bb=0 0 360 252]{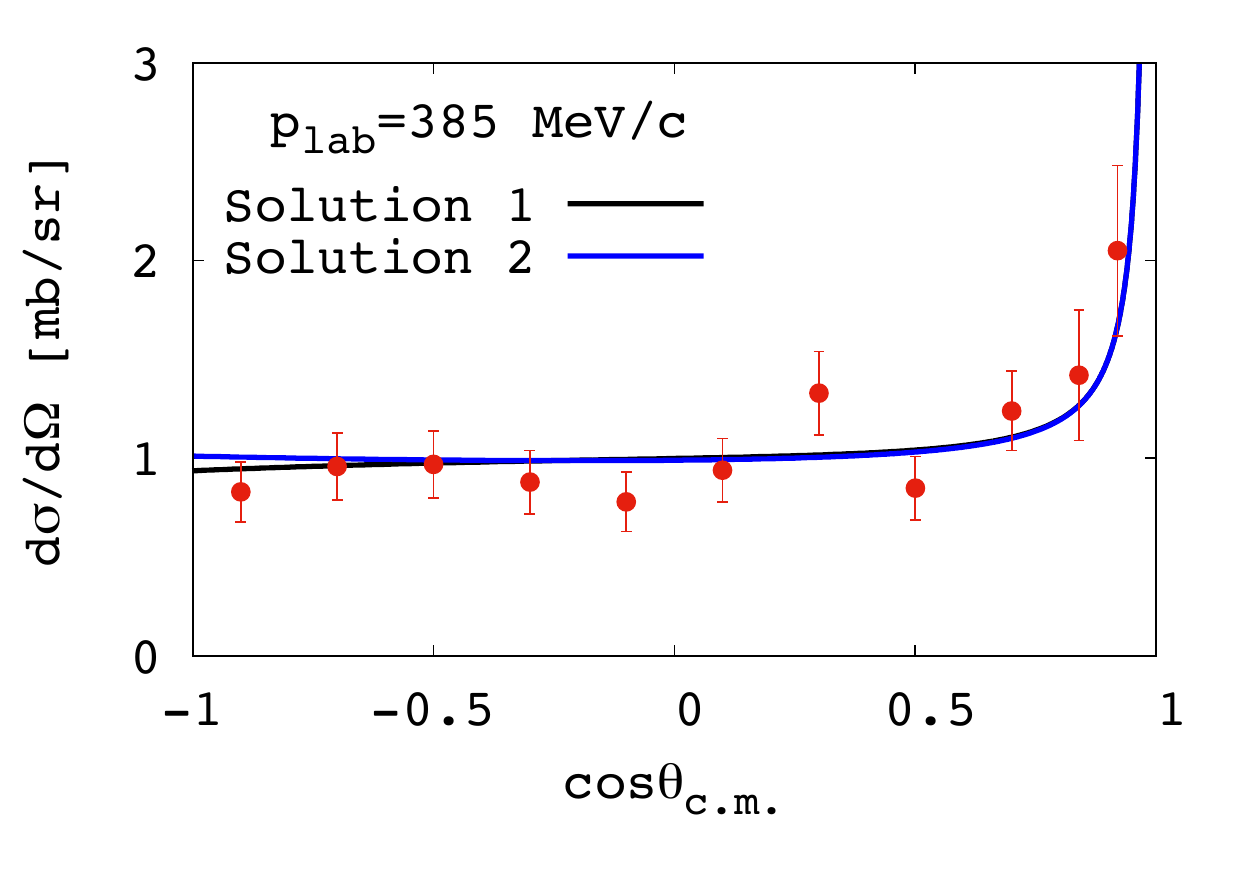}
  \end{center}
 \end{minipage}
\begin{minipage}{0.5\hsize}
  \begin{center}
   \includegraphics[width=80mm,bb=0 0 360 252]{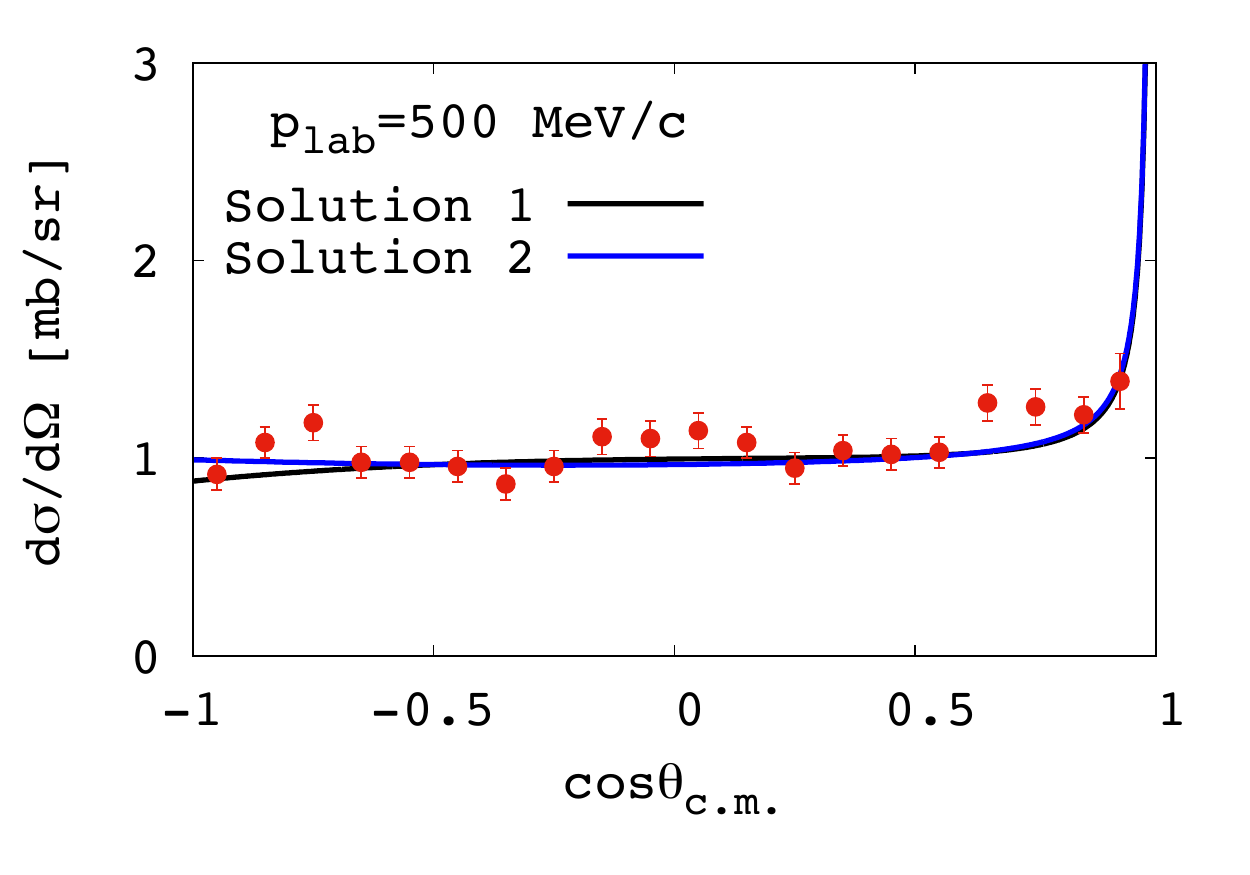}
  \end{center}
 \end{minipage}
\begin{minipage}{0.5\hsize}
  \begin{center}
   \includegraphics[width=80mm,bb=0 0 360 252]{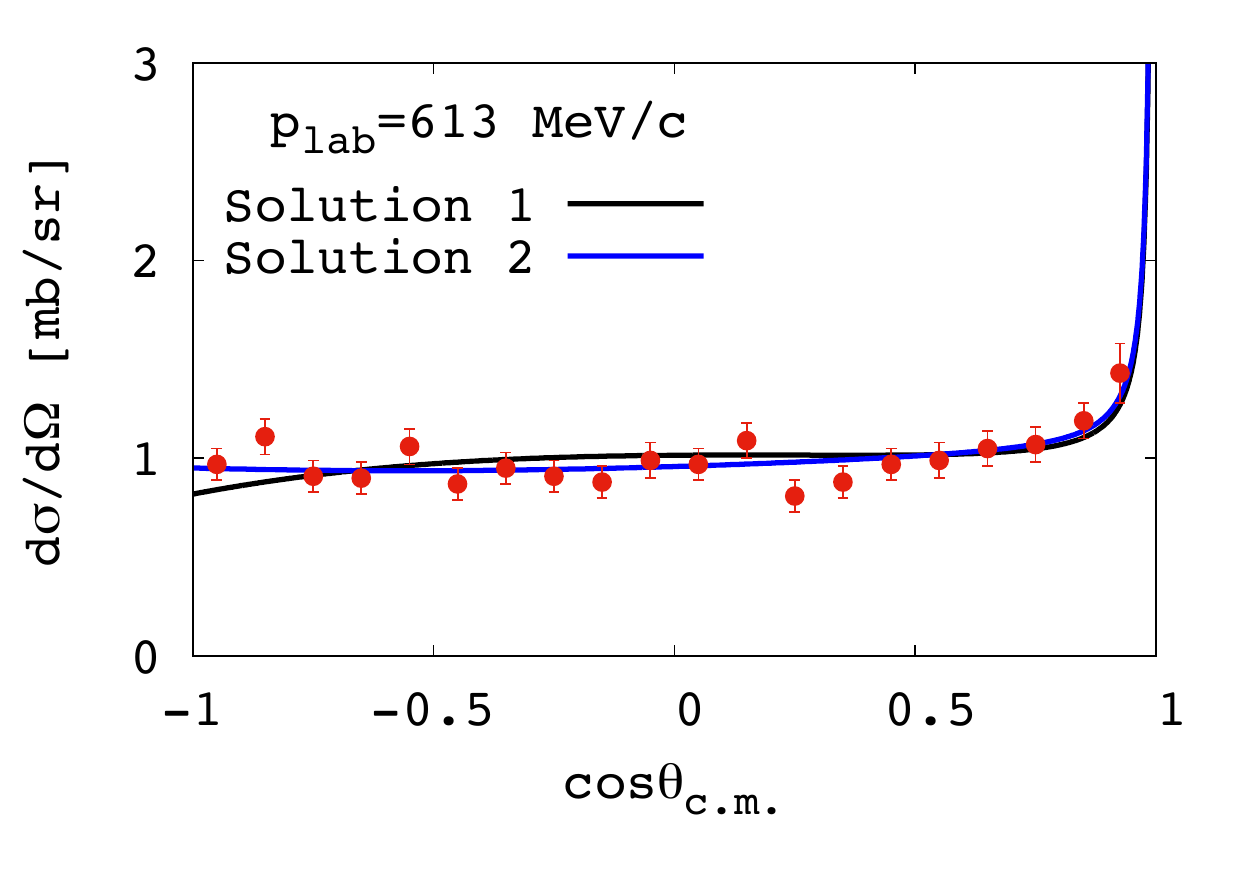}
  \end{center}
 \end{minipage}
\begin{minipage}{0.5\hsize}
  \begin{center}
   \includegraphics[width=80mm,bb=0 0 360 252]{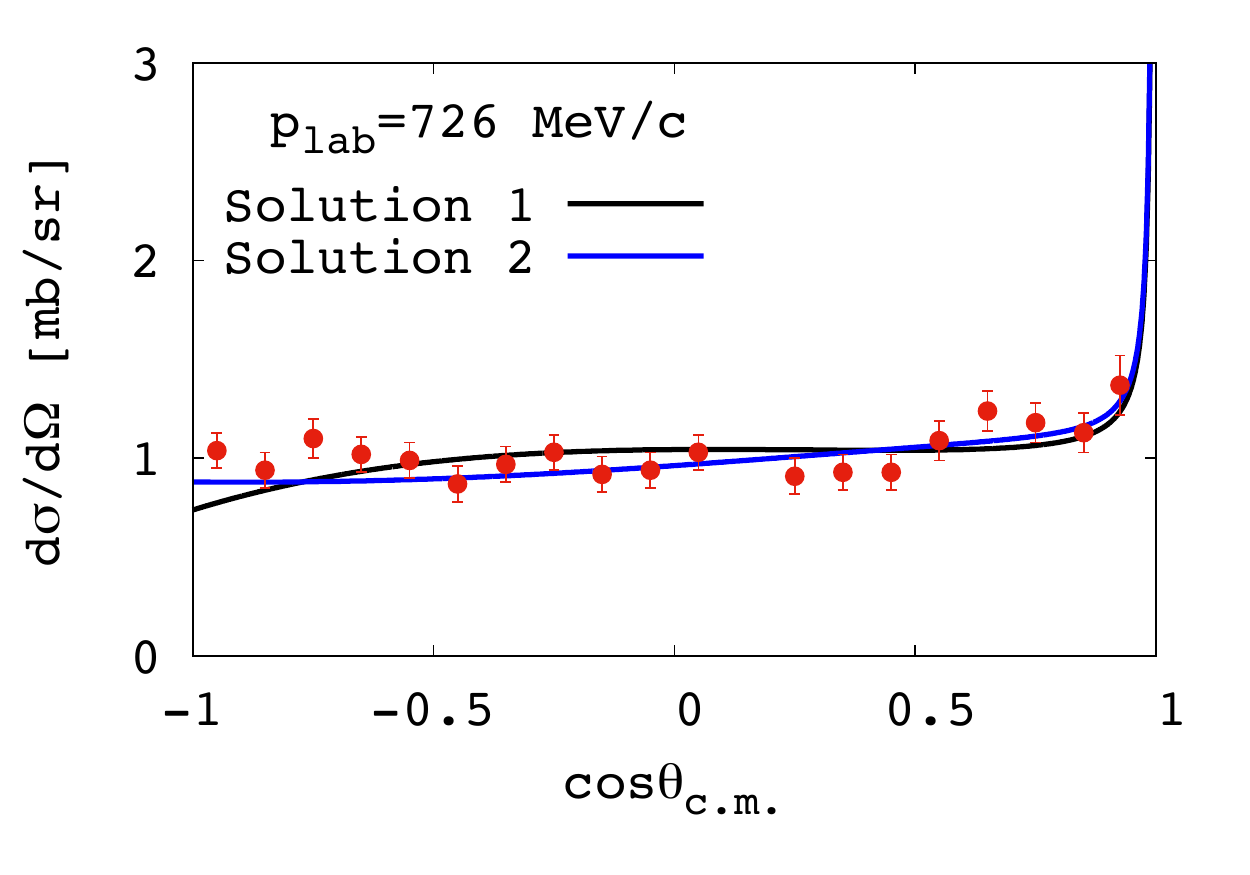}
  \end{center}
 \end{minipage}
 \vspace{-0.5cm}
\caption{
The calculated differential cross sections of the $K^{+}p$ elastic scattering 
using Solutions 1 and 2 
in comparison with the experimental data of Ref. \cite{cameron1974}.
}
\label{fig:kp_diff}
\end{figure}

First of all, we determine the $I=1$ scattering amplitude from the $K^{+}p$ 
elastic scattering data, which were observed well with small 
variation and certainly constrain the $I=1$ parameters.  
To fix the $I=1$ low-energy constants $b^{I=1}, d^{I=1}, g^{I=1}$ and $h^{I=1}$,
we use the $K^{+}p$ differential cross section between 
$p_{{\rm lab}} = 145$ to 726 MeV/c \cite{cameron1974}
and the total cross section between $p_{{\rm lab}} = 145$ to 788 MeV/c
\cite{bugg1968, bowen1970, adams1971,bowen1973, carroll1973, cameron1974}.
The fitted values of the parameters are summarized in the Table \ref{tab:parameter}.
We find two solutions for $I=1$, which equivalently reproduce the cross sections. 
In Fig.~\ref{fig:i1_tot}, we present the total cross sections of the $I=1$ $KN$ elastic scattering
obtained with the fitted parameters and compare with the experimental 
data~\cite{bugg1968, bowen1970, adams1971,bowen1973, carroll1973, cameron1974}.
The calculated amplitude gives a good reproduction of the data up to $p_{{\rm lab}}=$800~MeV/c.
It shows that the $S$-wave contribution is dominated 
and the contributions from the partial waves higher than the $P$-wave are negligibly small. 
This is consistent with the old observation.
In Fig.~\ref{fig:kp_diff}, we show the calculated differential cross sections  
and make a comparison with the experimental data. 
The figure shows that the obtained amplitudes reproduce the experimental data well
for all the energies which we consider here.

\begin{figure}[]
 \begin{minipage}{0.5\hsize}
  \begin{center}
   \includegraphics[width=80mm,bb=0 0 360 252]{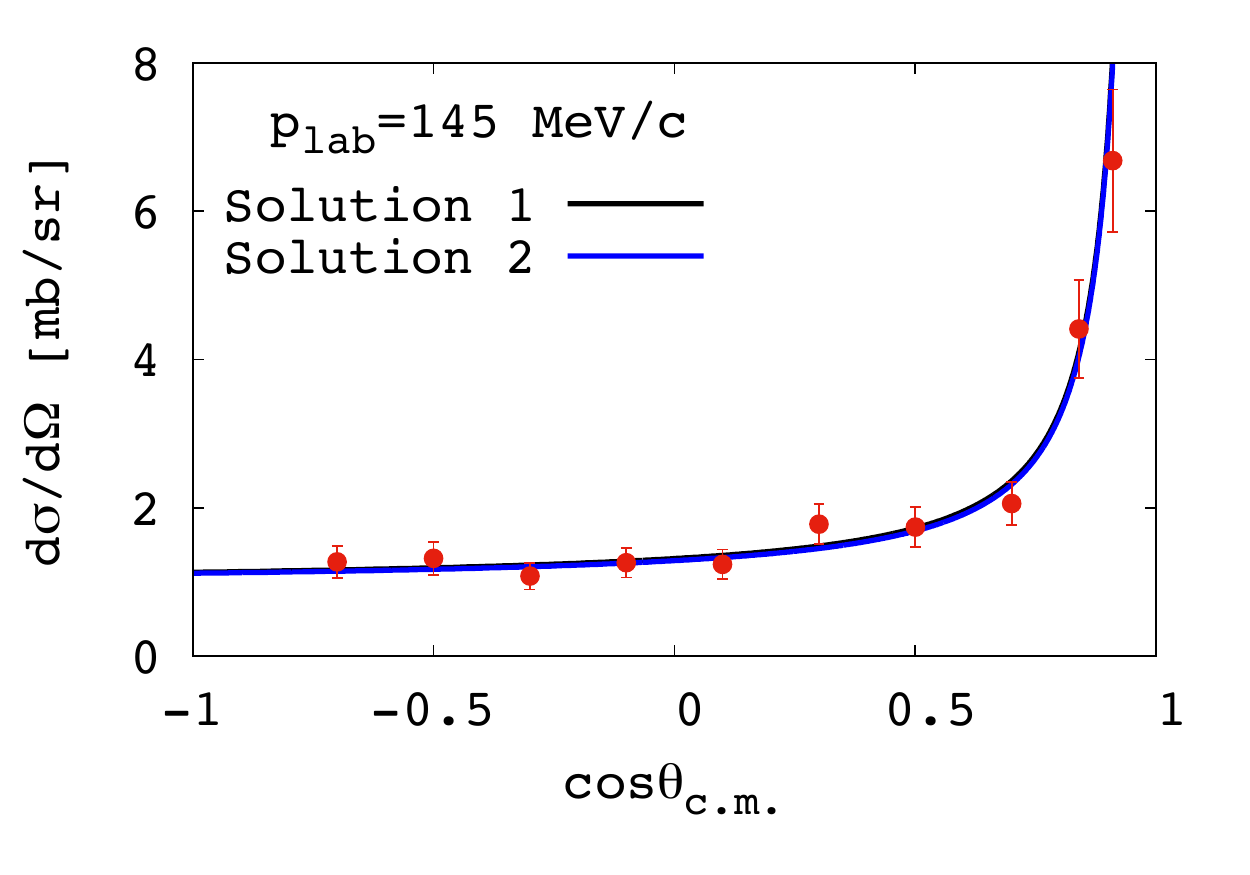}
  \end{center}
 \end{minipage}
 \begin{minipage}{0.5\hsize}
  \begin{center}
   \includegraphics[width=80mm,bb=0 0 360 252]{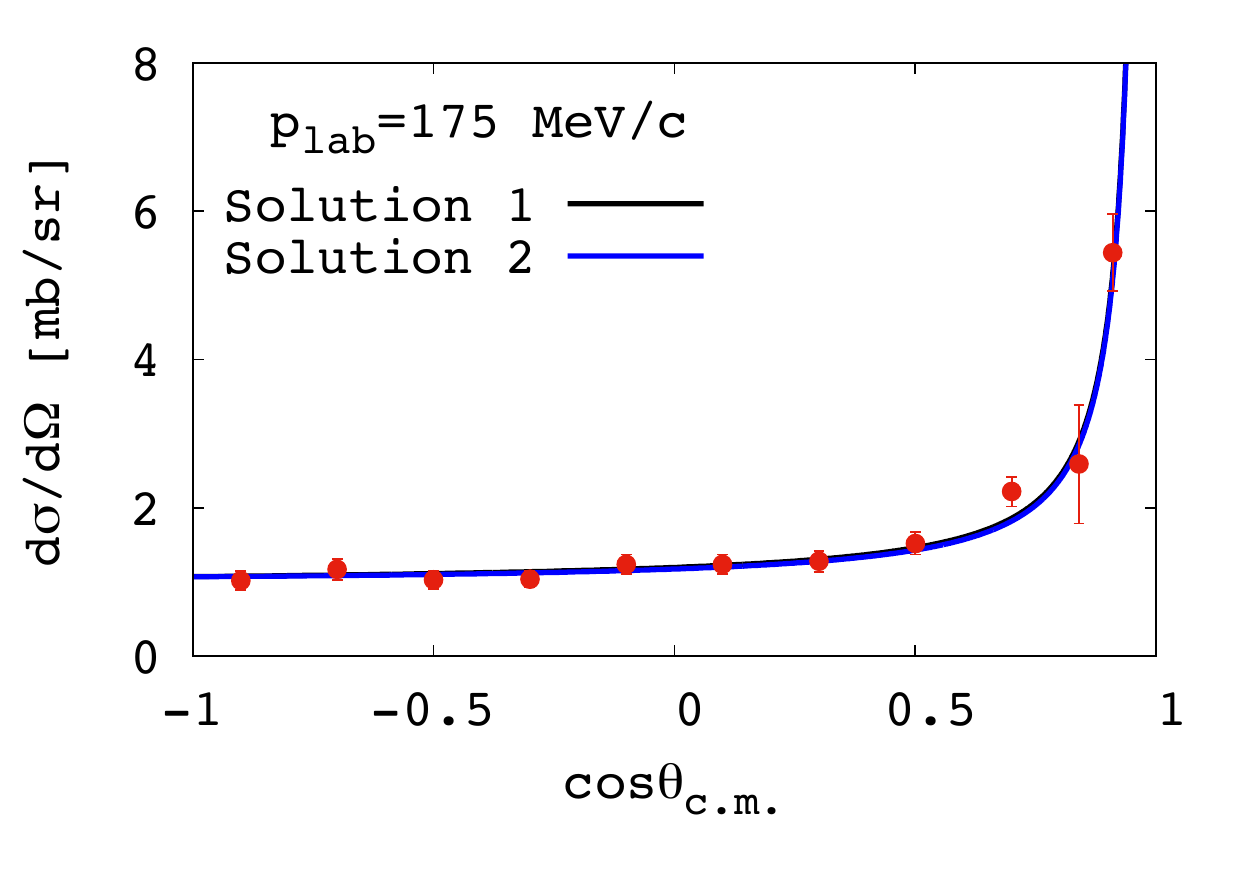}
  \end{center}
 \end{minipage}
\begin{minipage}{0.5\hsize}
  \begin{center}
   \includegraphics[width=80mm,bb=0 0 360 252]{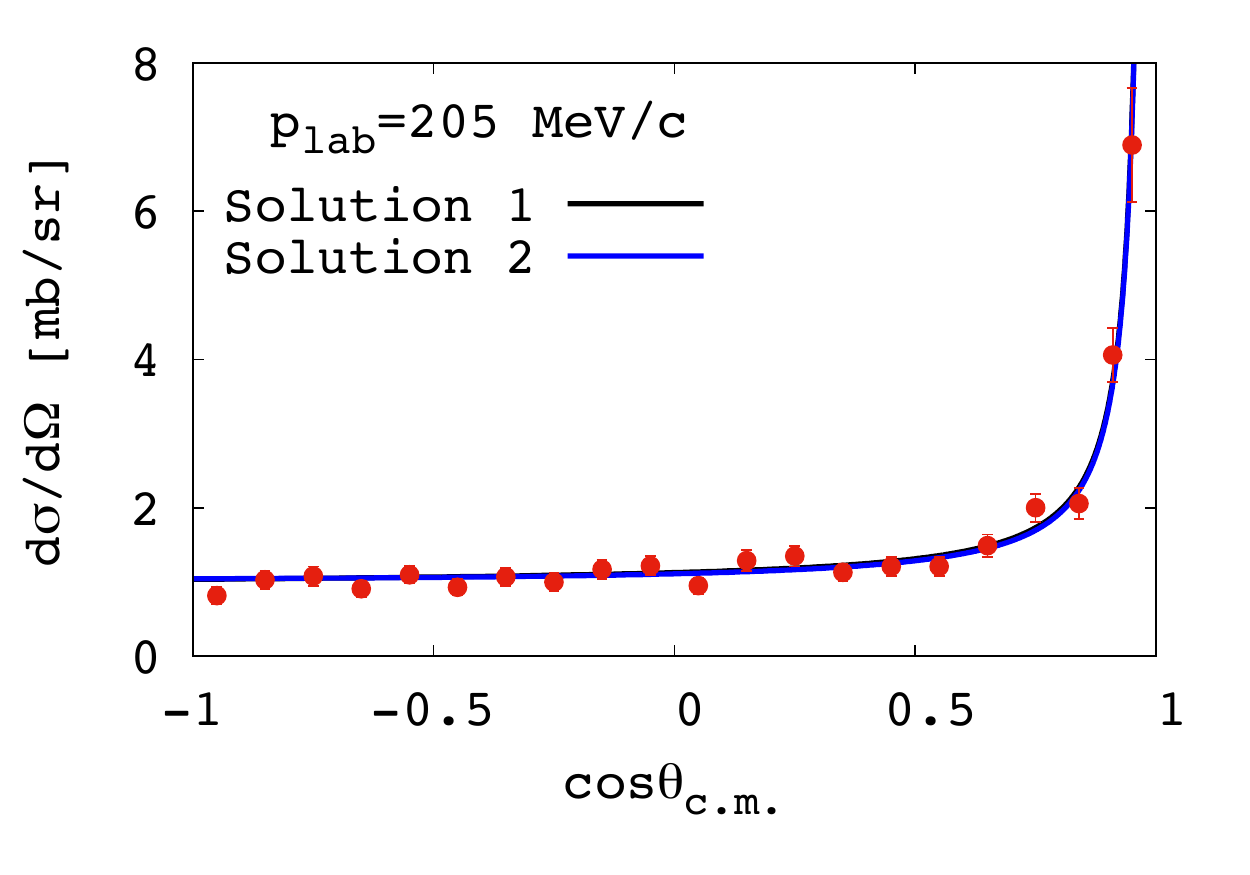}
  \end{center}
 \end{minipage}
\begin{minipage}{0.5\hsize}
  \begin{center}
   \includegraphics[width=80mm,bb=0 0 360 252]{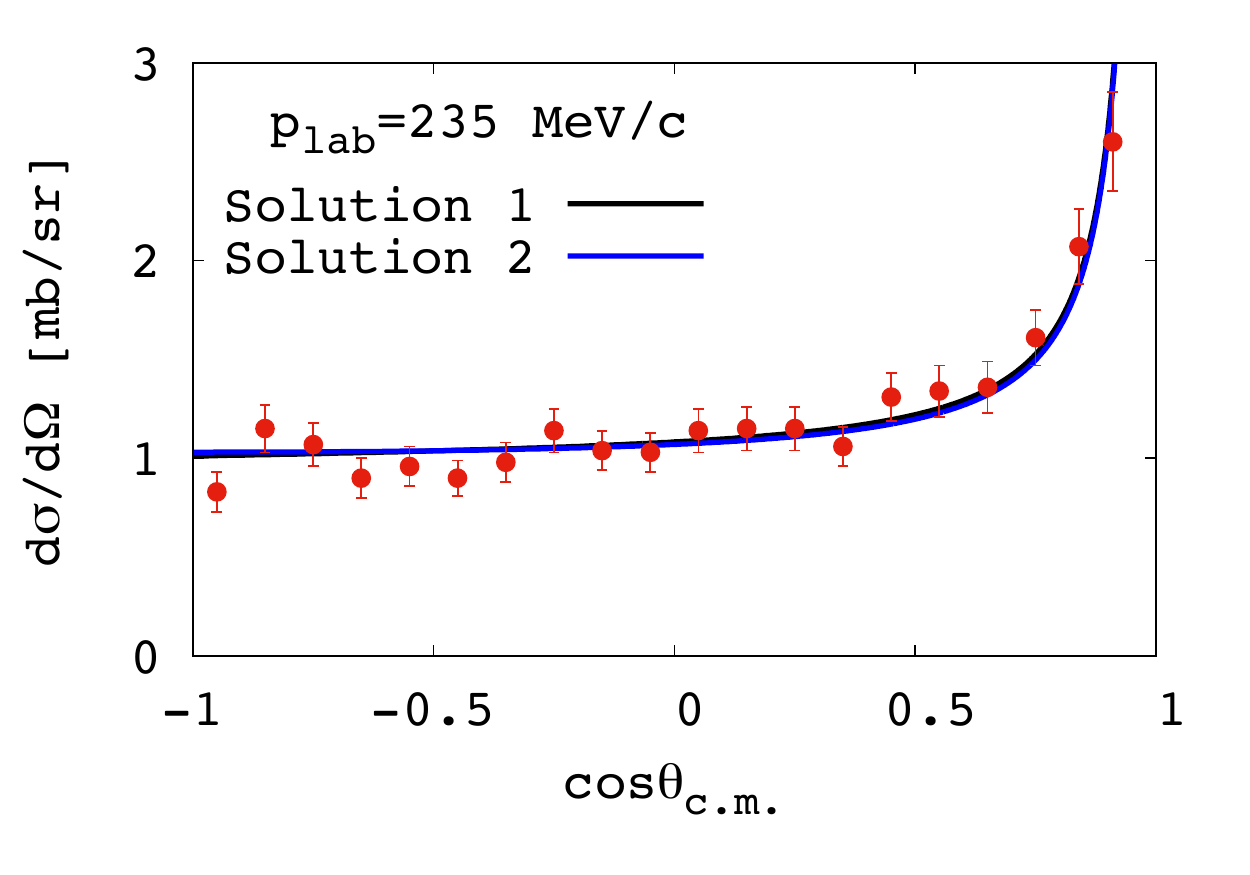}
  \end{center}
 \end{minipage}
 \begin{minipage}{0.5\hsize}
  \begin{center}
   \includegraphics[width=80mm,bb=0 0 360 252]{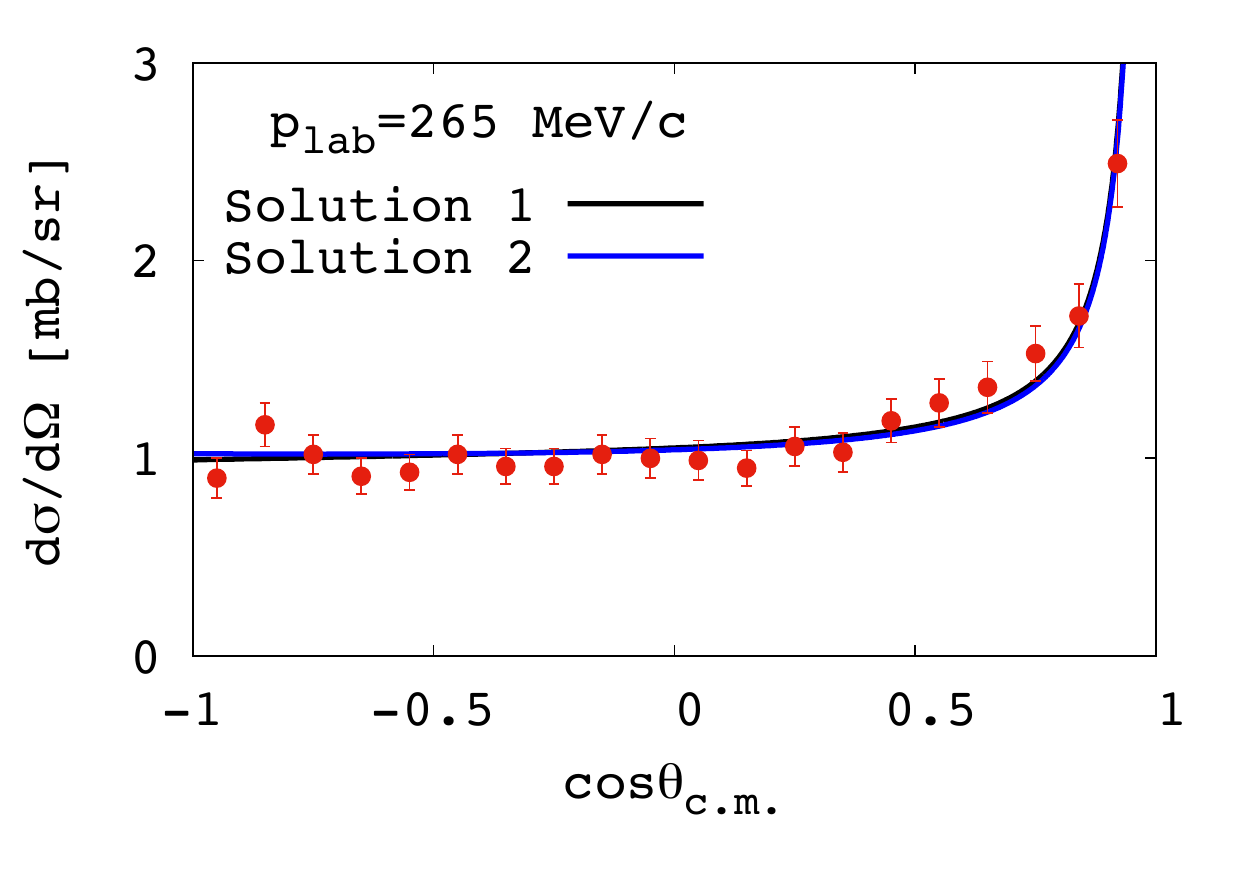}
  \end{center}
 \end{minipage}
\begin{minipage}{0.5\hsize}
  \begin{center}
   \includegraphics[width=80mm,bb=0 0 360 252]{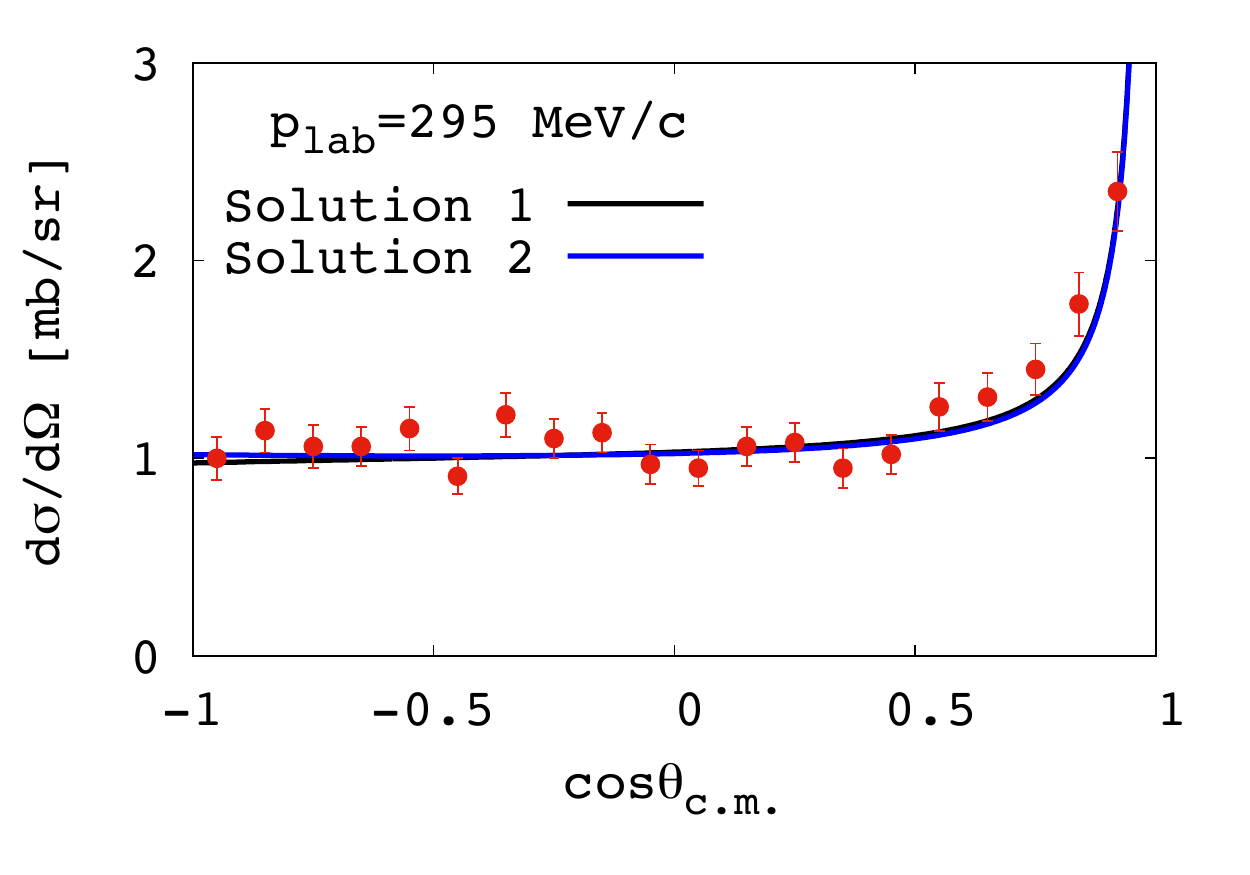}
  \end{center}
 \end{minipage}
\begin{minipage}{0.5\hsize}
  \begin{center}
   \includegraphics[width=80mm,bb=0 0 360 252]{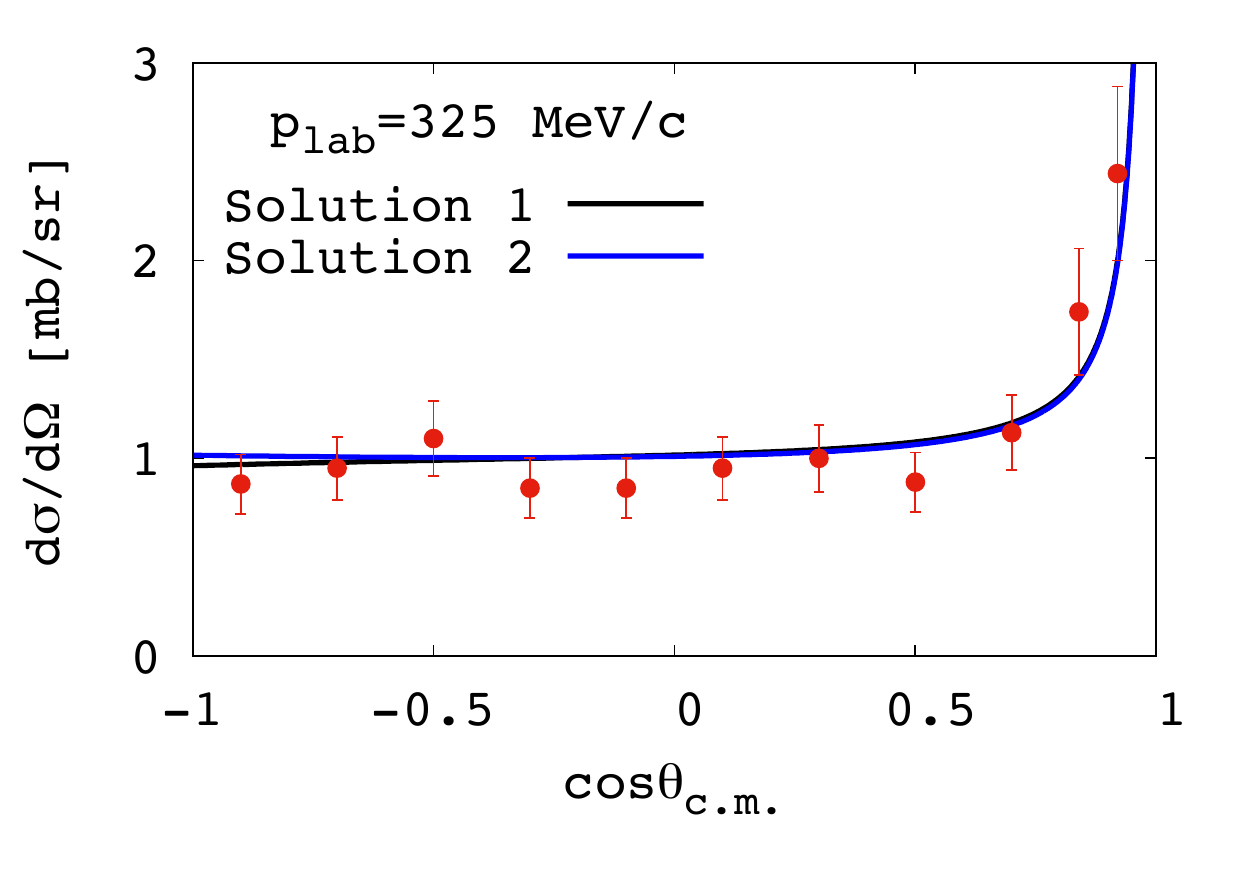}
  \end{center}
 \end{minipage}
\begin{minipage}{0.5\hsize}
  \begin{center}
   \includegraphics[width=80mm,bb=0 0 360 252]{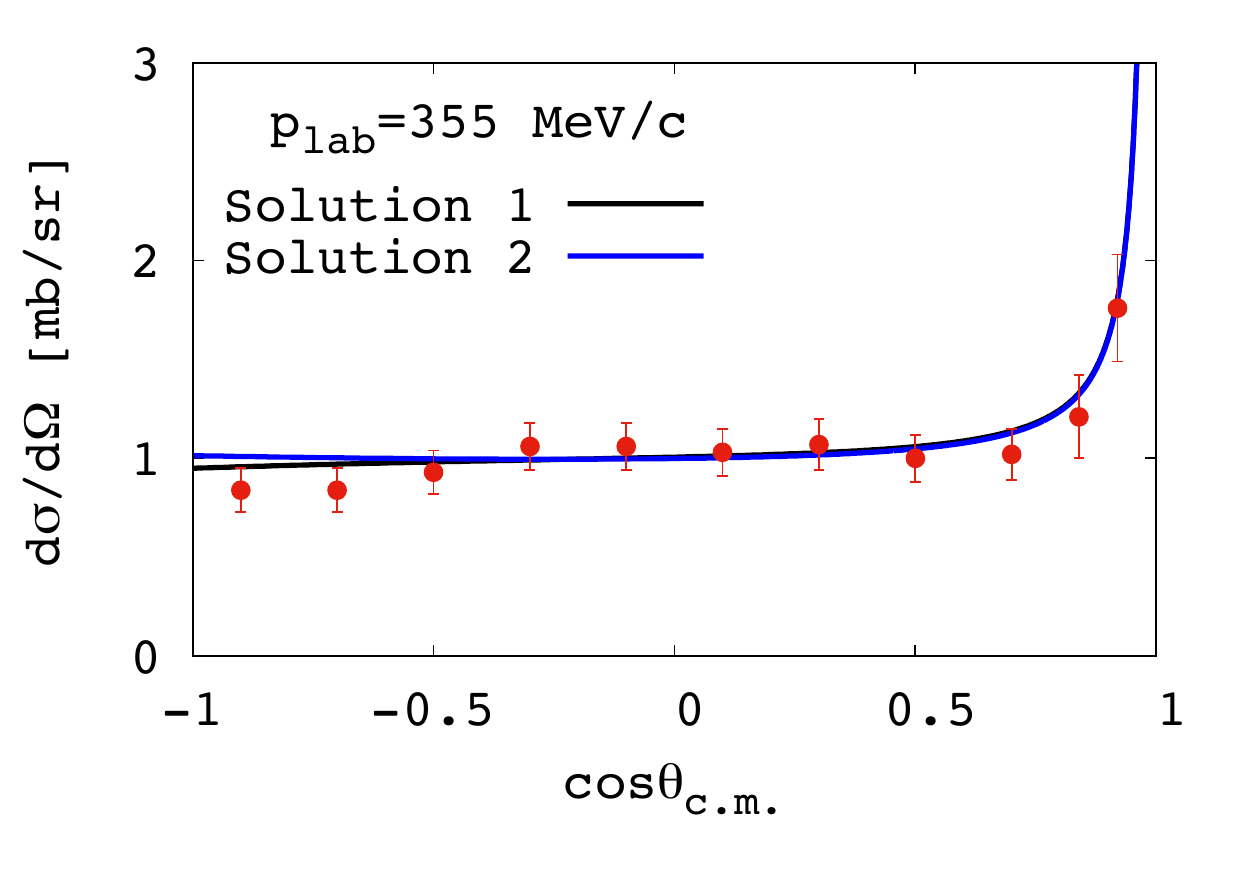}
  \end{center}
 \end{minipage}
 \end{figure}

\begin{figure}[]
    \begin{tabular}{c}
      \begin{minipage}[t]{1.0\hsize}
        \centering
        \includegraphics[keepaspectratio, scale=0.8, bb=0 0 360 252]{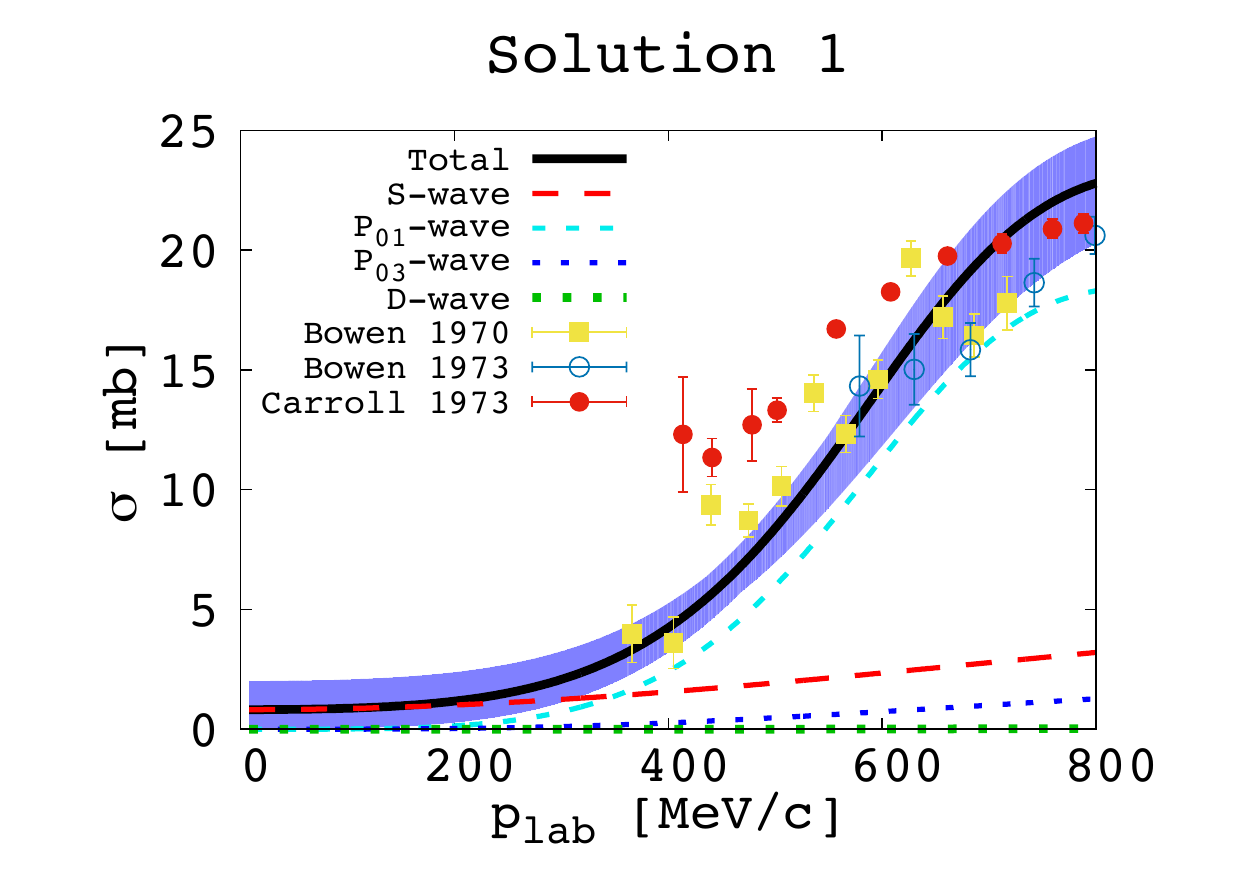}
      \end{minipage} \\
      \begin{minipage}[t]{1.0\hsize}
        \centering
        \includegraphics[keepaspectratio, scale=0.8,bb=0 0 360 252]{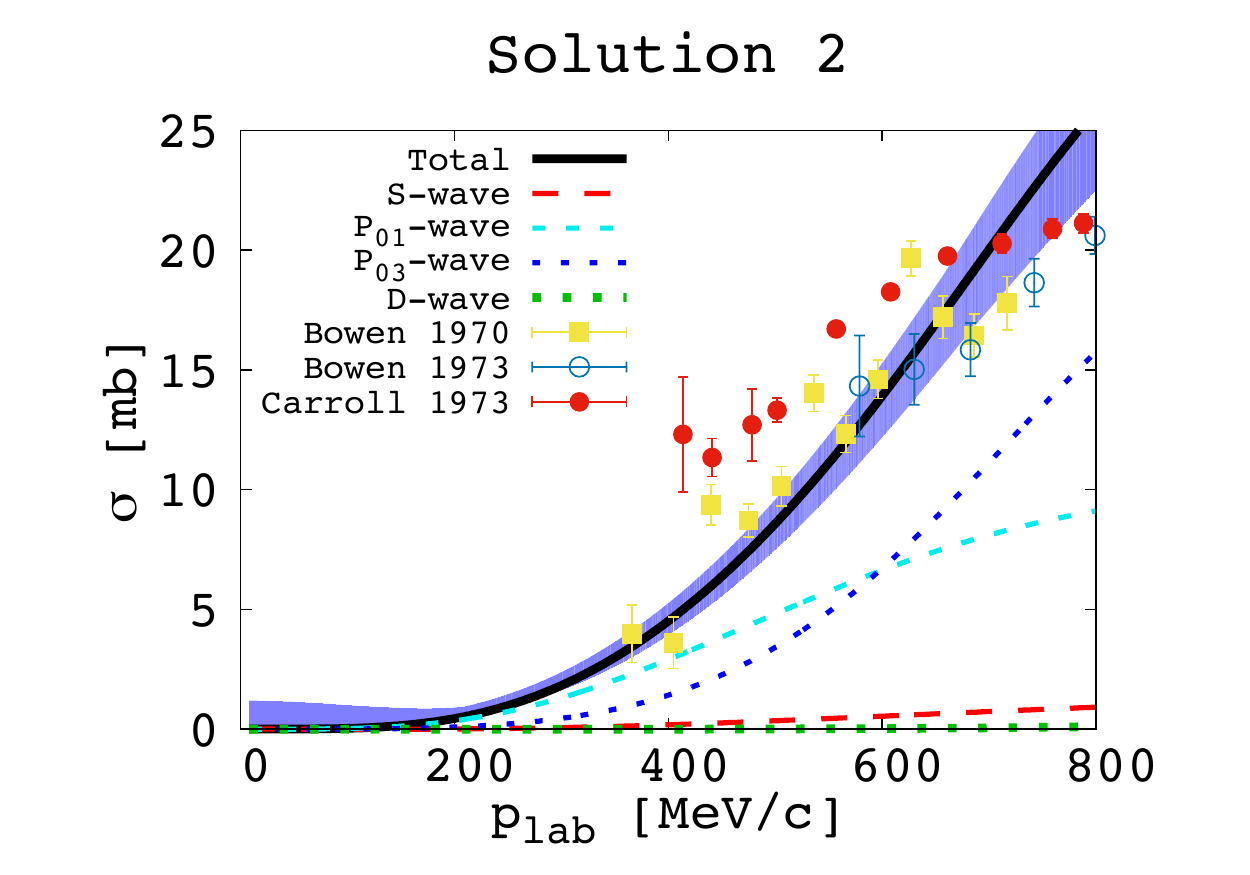}
      \end{minipage} \\
    \end{tabular}
     \caption{
   The $I=0$ total cross sections calculated using Solutions 1 and 2 in comparison
  with the experimental data \cite{bowen1970, bowen1973, carroll1973}.
  The solid line shows that the best-fit solution.
  The shaded area shows that the allowable region of the parameter around
  the vicinity of the best-fit solution.
  }
\label{fig:i0_tot}
  \end{figure}

\begin{figure}[]
 \begin{minipage}{0.5\hsize}
  \begin{center}
   \includegraphics[width=80mm, bb=0 0 360 252]{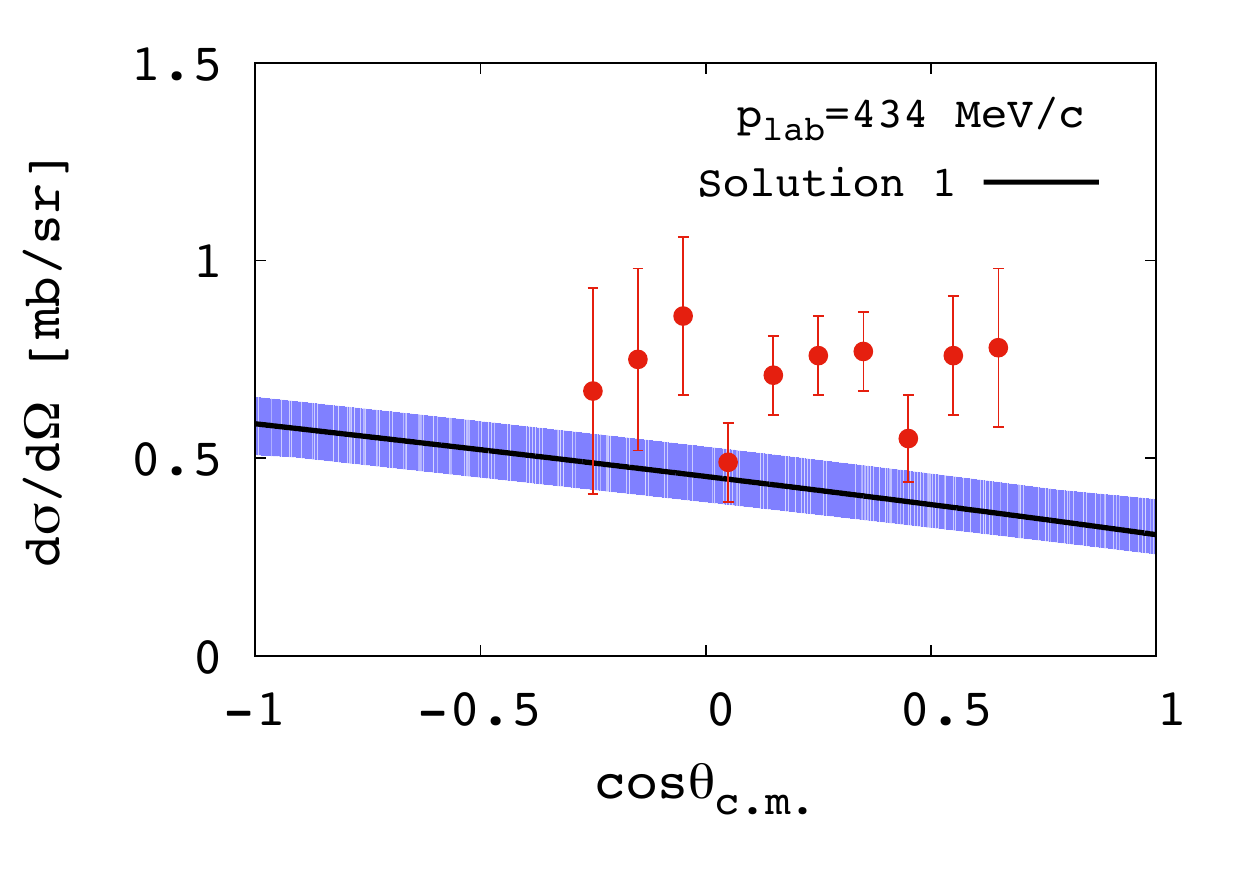}
  \end{center}
 \end{minipage}
 \begin{minipage}{0.5\hsize}
  \begin{center}
   \includegraphics[width=80mm, bb=0 0 360 252]{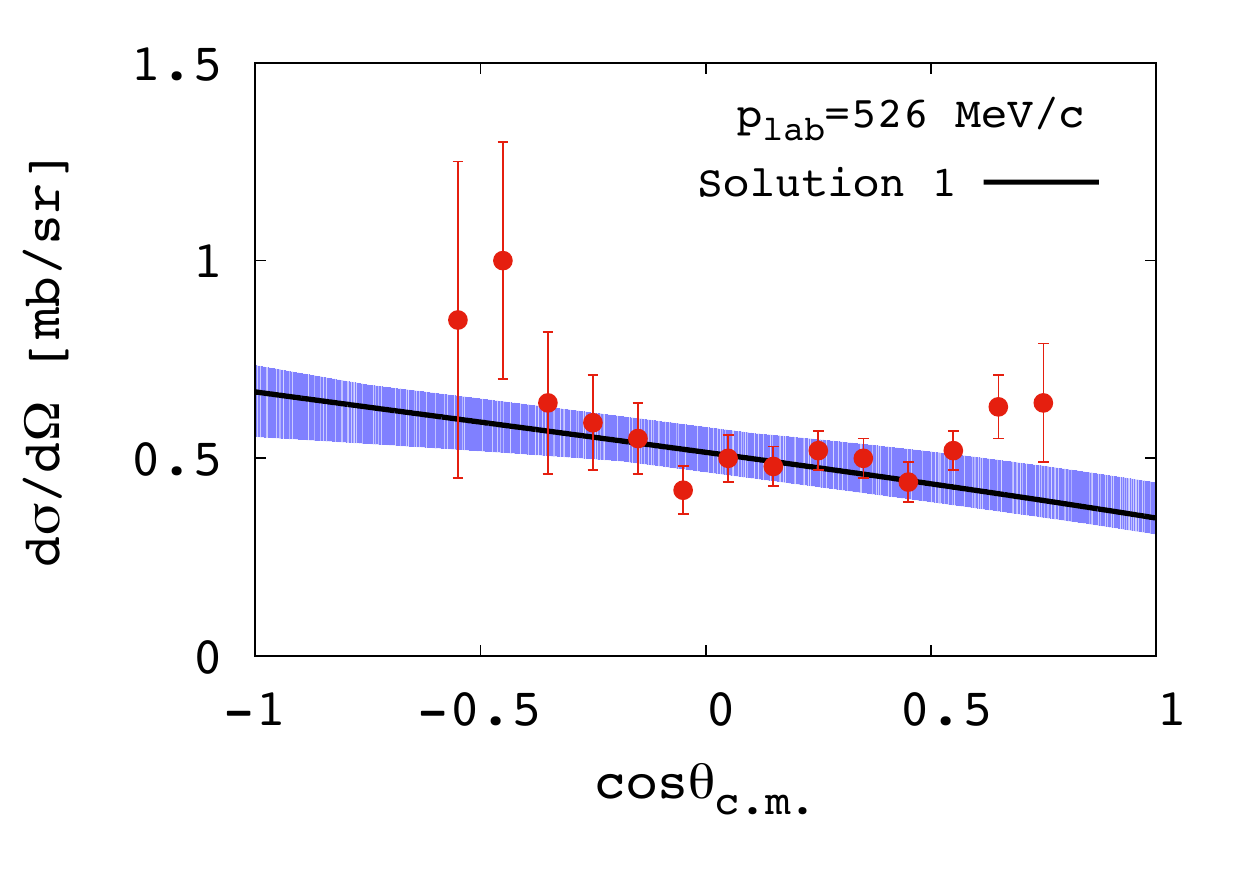}
  \end{center}
 \end{minipage}
\begin{minipage}{0.5\hsize}
  \begin{center}
   \includegraphics[width=80mm, bb=0 0 360 252]{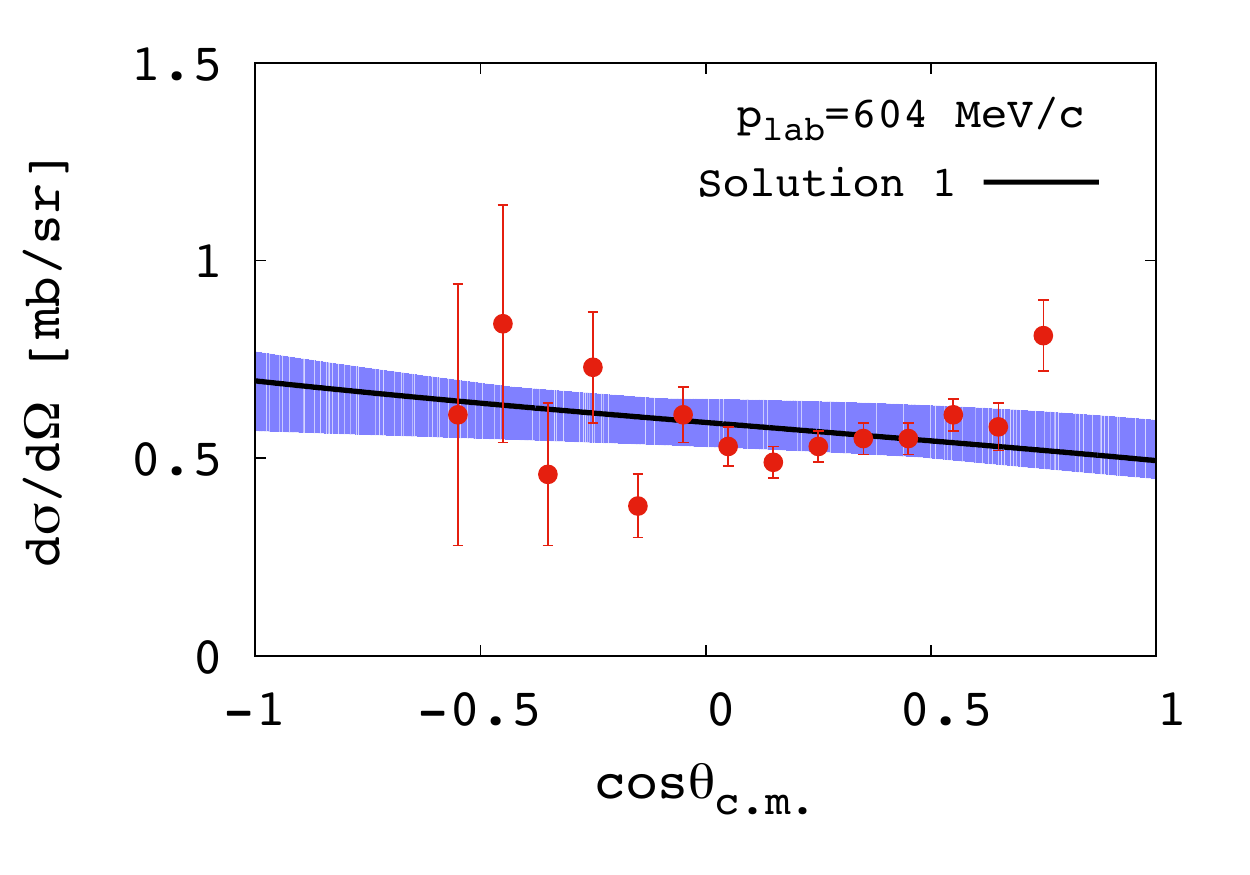}
  \end{center}
 \end{minipage}
\begin{minipage}{0.5\hsize}
  \begin{center}
   \includegraphics[width=80mm, bb=0 0 360 252]{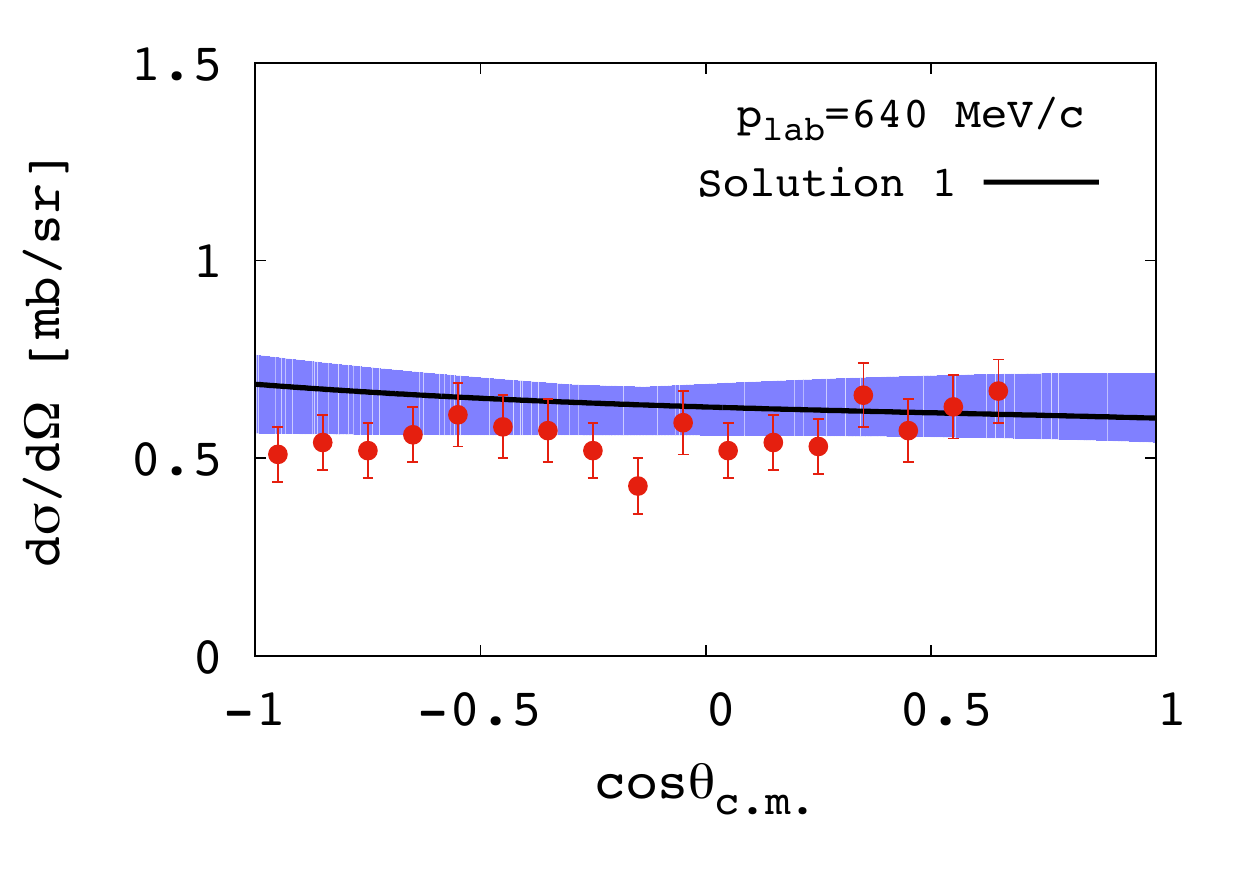}
  \end{center}
 \end{minipage}
 \begin{minipage}{0.5\hsize}
  \begin{center}
   \includegraphics[width=80mm, bb=0 0 360 252]{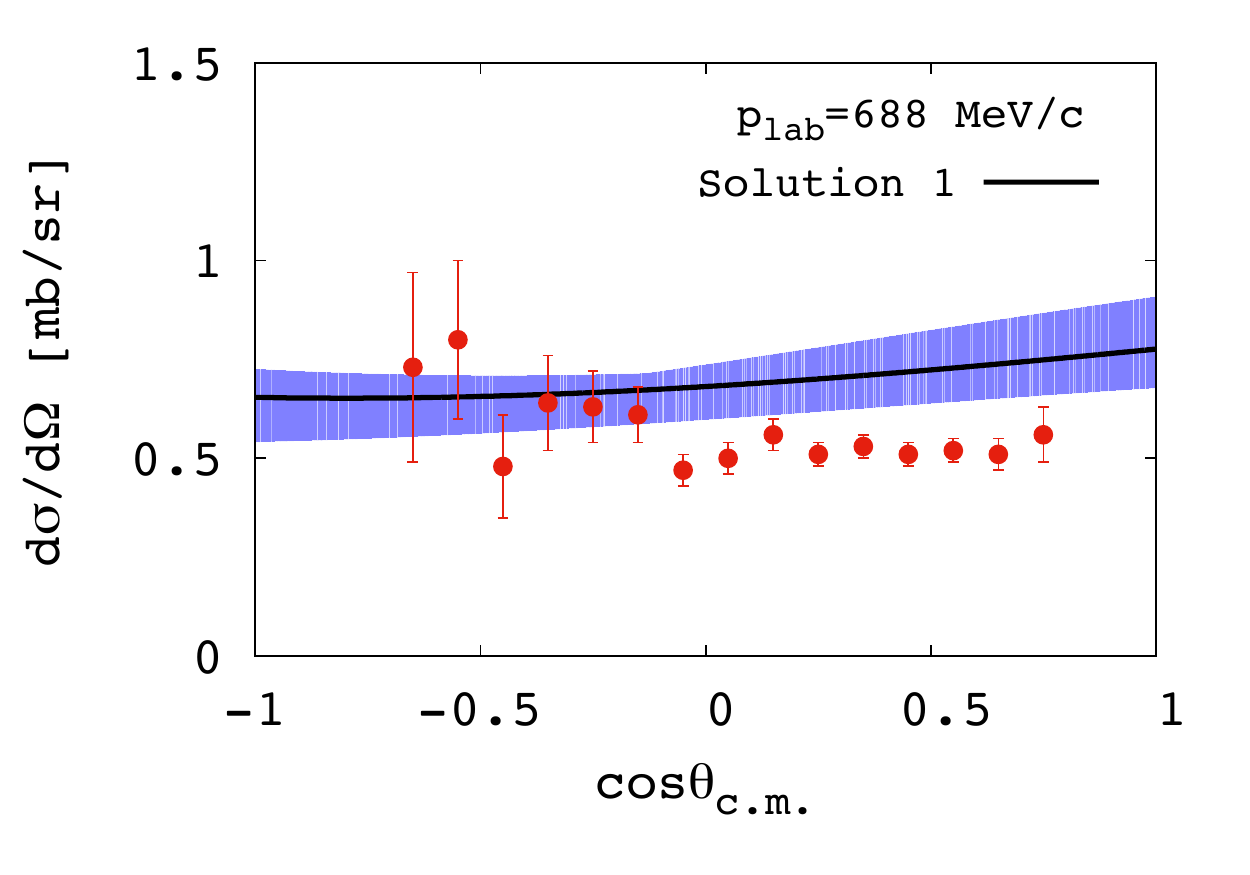}
  \end{center}
 \end{minipage}
\begin{minipage}{0.5\hsize}
  \begin{center}
   \includegraphics[width=80mm, bb=0 0 360 252]{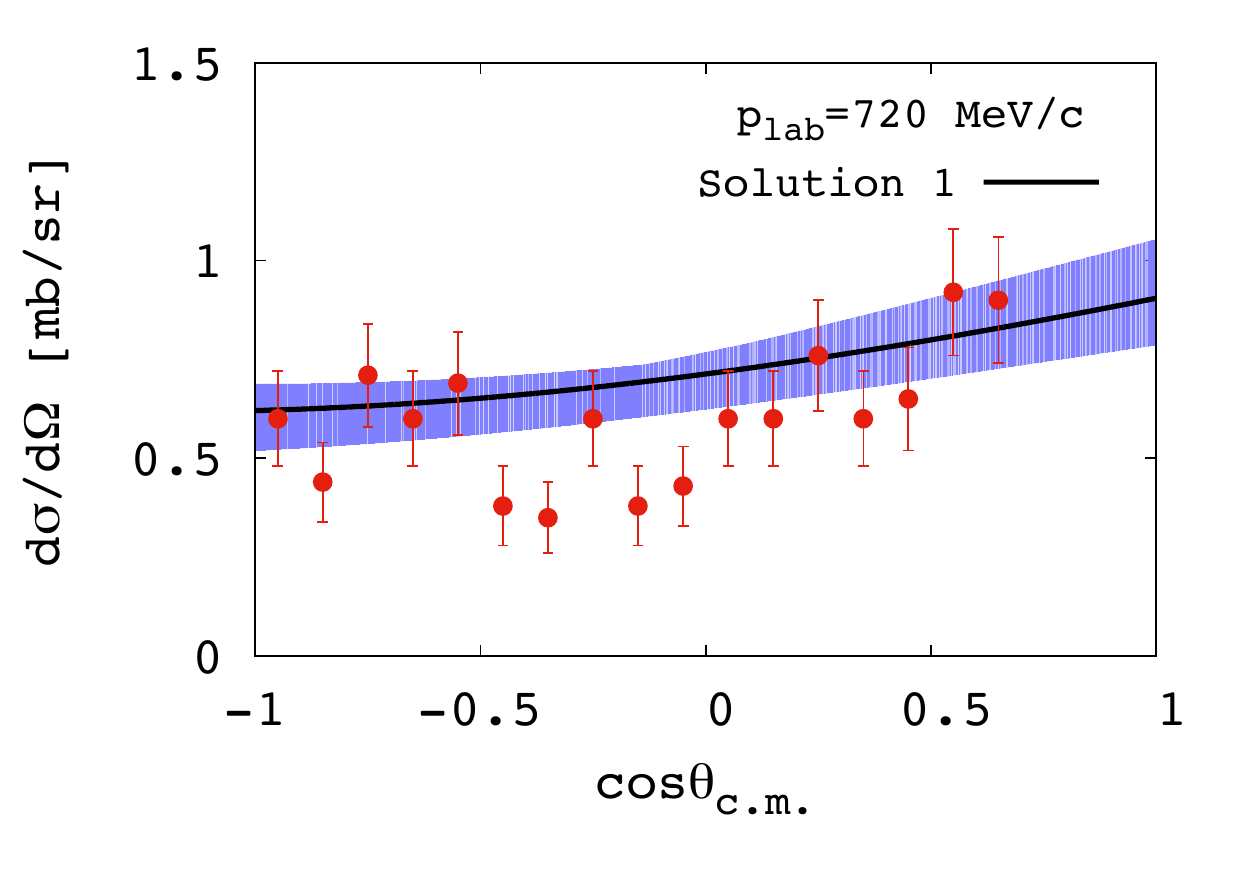}
  \end{center}
 \end{minipage}
\begin{minipage}{0.5\hsize}
  \begin{center}
   \includegraphics[width=80mm, bb=0 0 360 252]{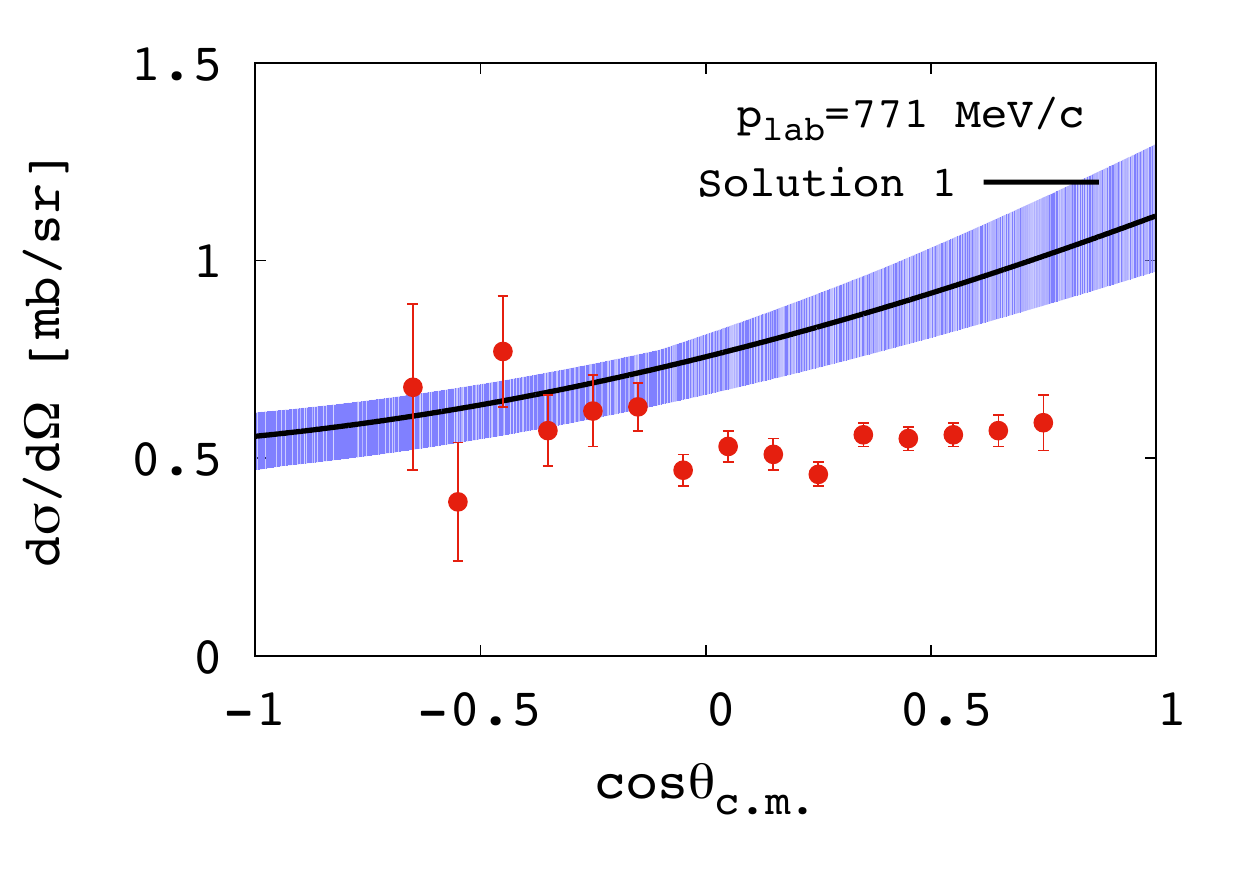}
  \end{center}
 \end{minipage}
\begin{minipage}{0.5\hsize}
  \begin{center}
   \includegraphics[width=80mm, bb=0 0 360 252]{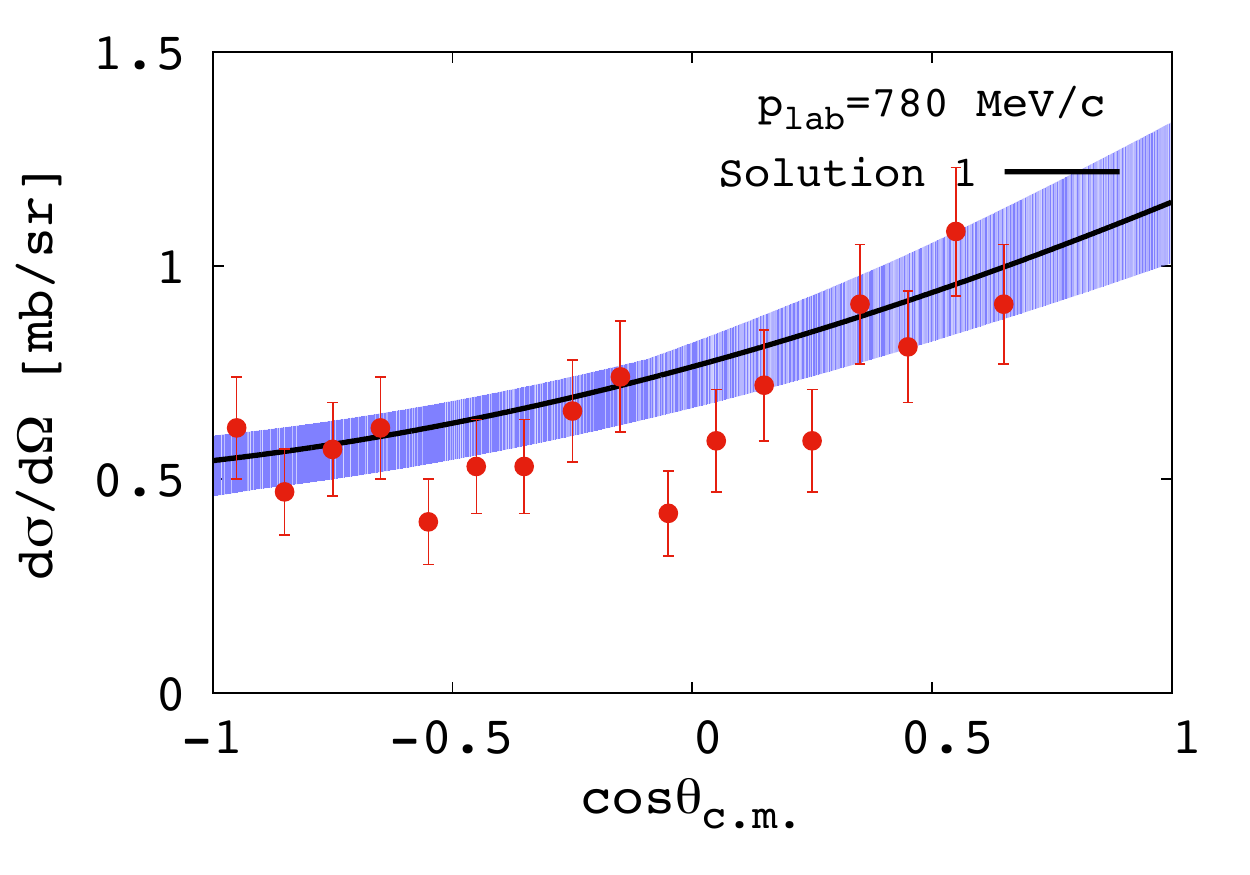}
  \end{center}
 \end{minipage}
 \vspace{-0.5cm}
 \caption{
The differential cross sections of $K^{+}n$ elastic scattering
 using Solution 1
 in comparison with the experimental data of Ref. \cite{Giacomelli:1973ed, dam1975}.
The momenta at the $p_{{\rm lab}}$=640, 720 and 780 MeV/c are the data from Ref. \cite{Giacomelli:1973ed}.
The others are the data from Ref. \cite{dam1975}.}
\label{fig:kn_diff_sol1}
 \end{figure}

\begin{figure}[]
 \begin{minipage}{0.5\hsize}
  \begin{center}
   \includegraphics[width=80mm,bb=0 0 360 252]{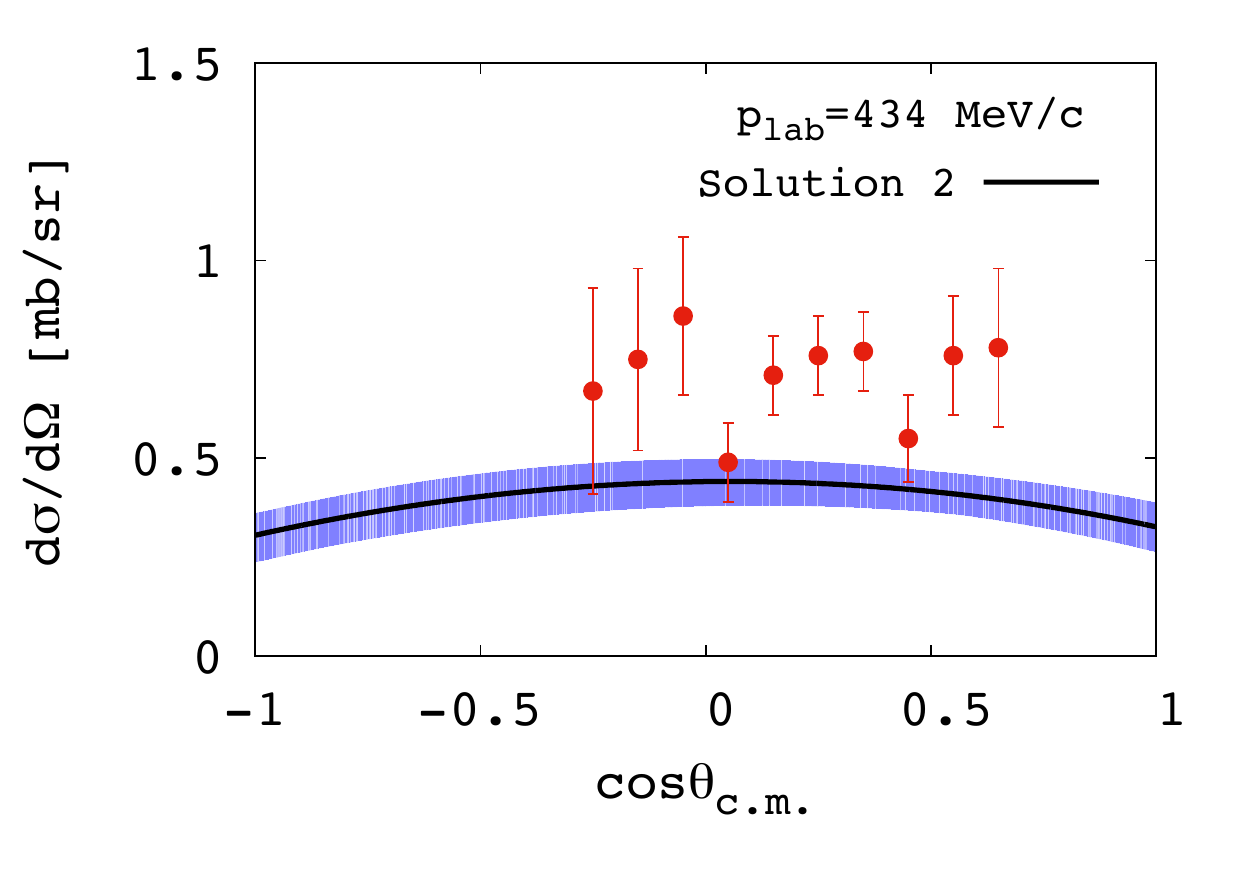}
  \end{center}
 \end{minipage}
 \begin{minipage}{0.5\hsize}
  \begin{center}
   \includegraphics[width=80mm,bb=0 0 360 252]{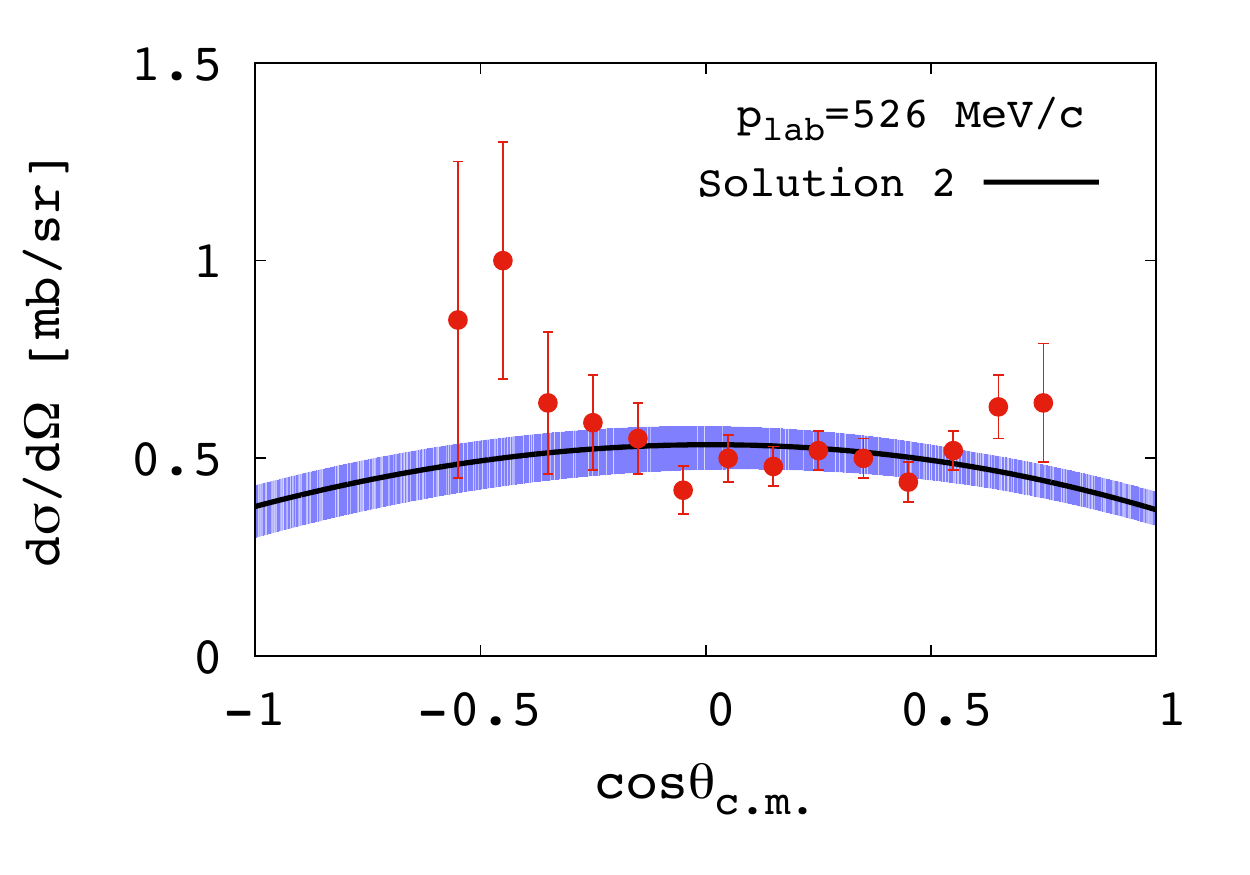}
  \end{center}
 \end{minipage}
\begin{minipage}{0.5\hsize}
  \begin{center}
   \includegraphics[width=80mm,bb=0 0 360 252]{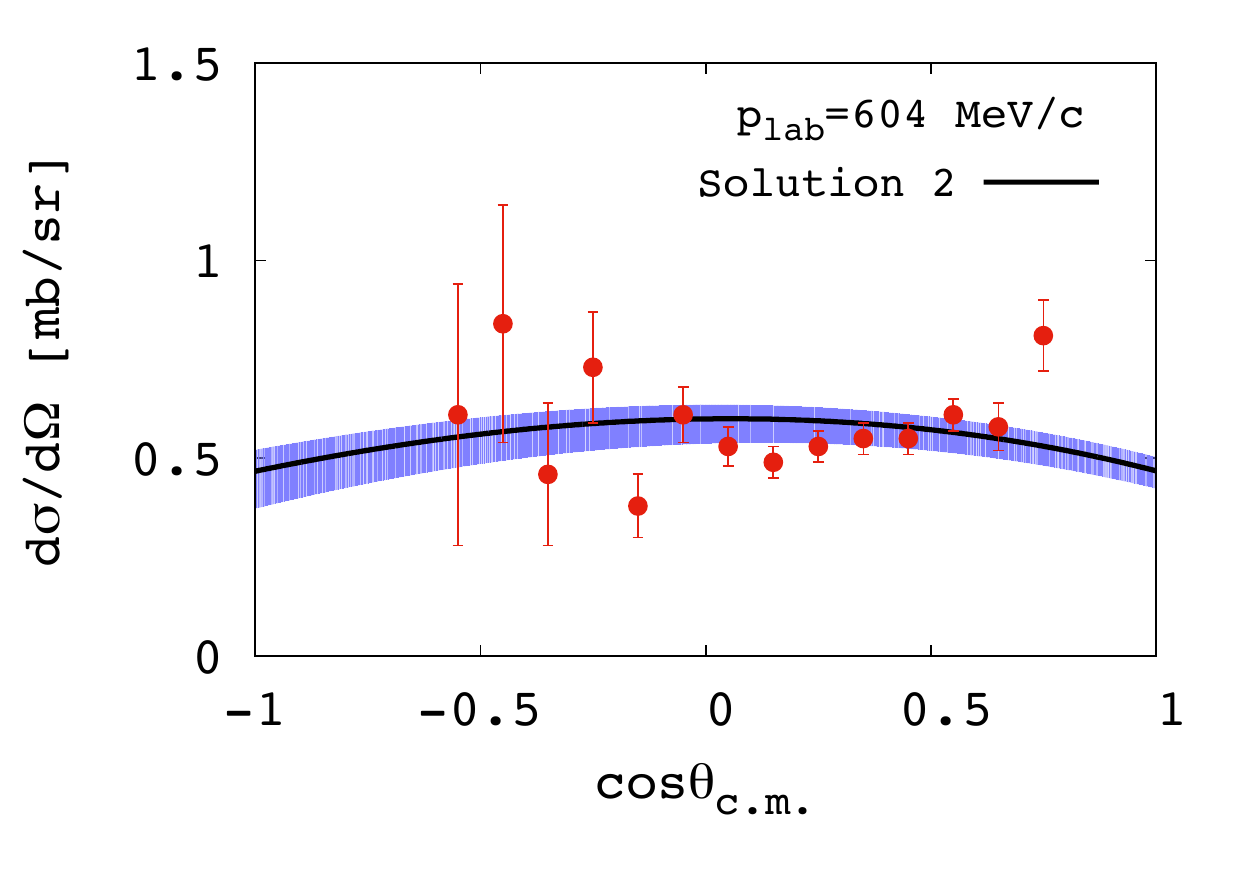}
  \end{center}
 \end{minipage}
\begin{minipage}{0.5\hsize}
  \begin{center}
   \includegraphics[width=80mm,bb=0 0 360 252]{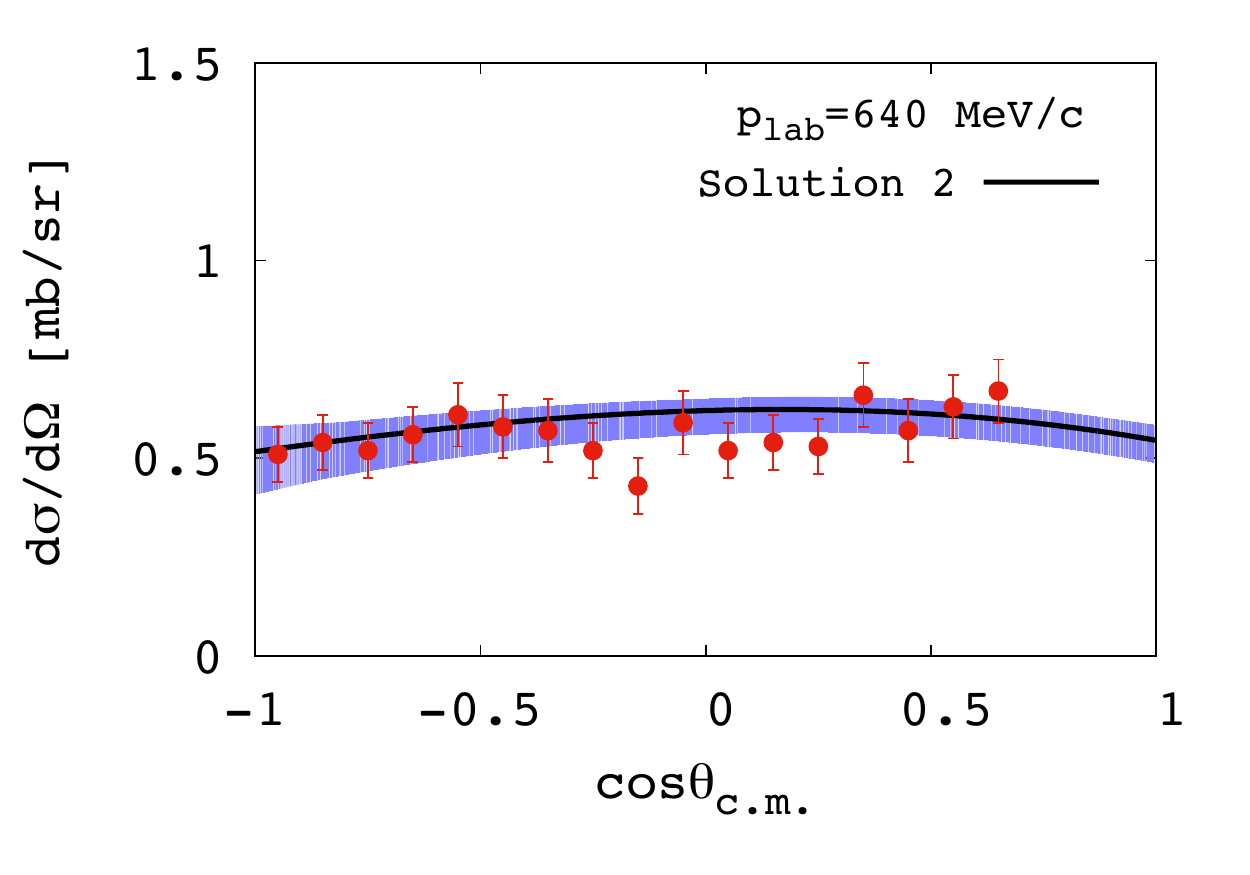}
  \end{center}
 \end{minipage}
 \begin{minipage}{0.5\hsize}
  \begin{center}
   \includegraphics[width=80mm,bb=0 0 360 252]{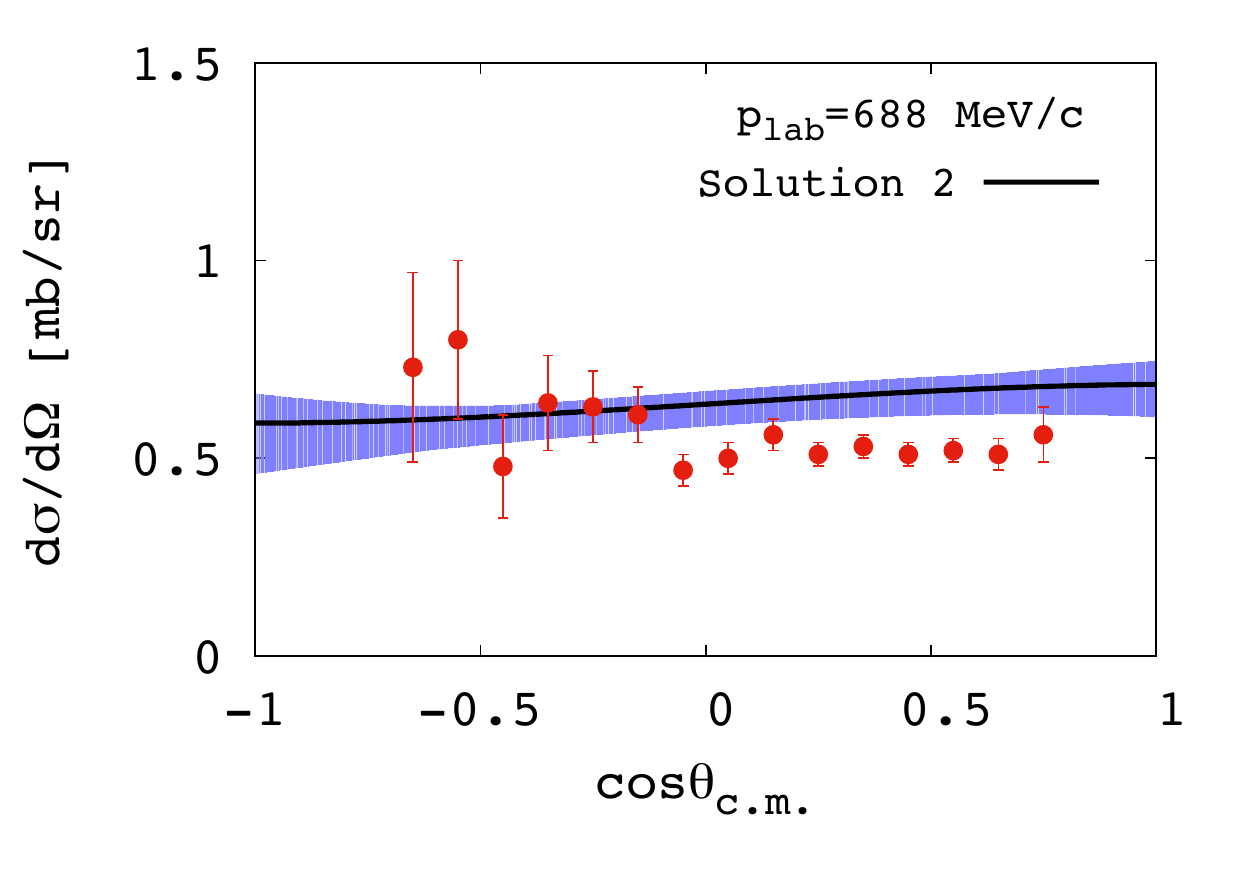}
  \end{center}
 \end{minipage}
\begin{minipage}{0.5\hsize}
  \begin{center}
   \includegraphics[width=80mm,bb=0 0 360 252]{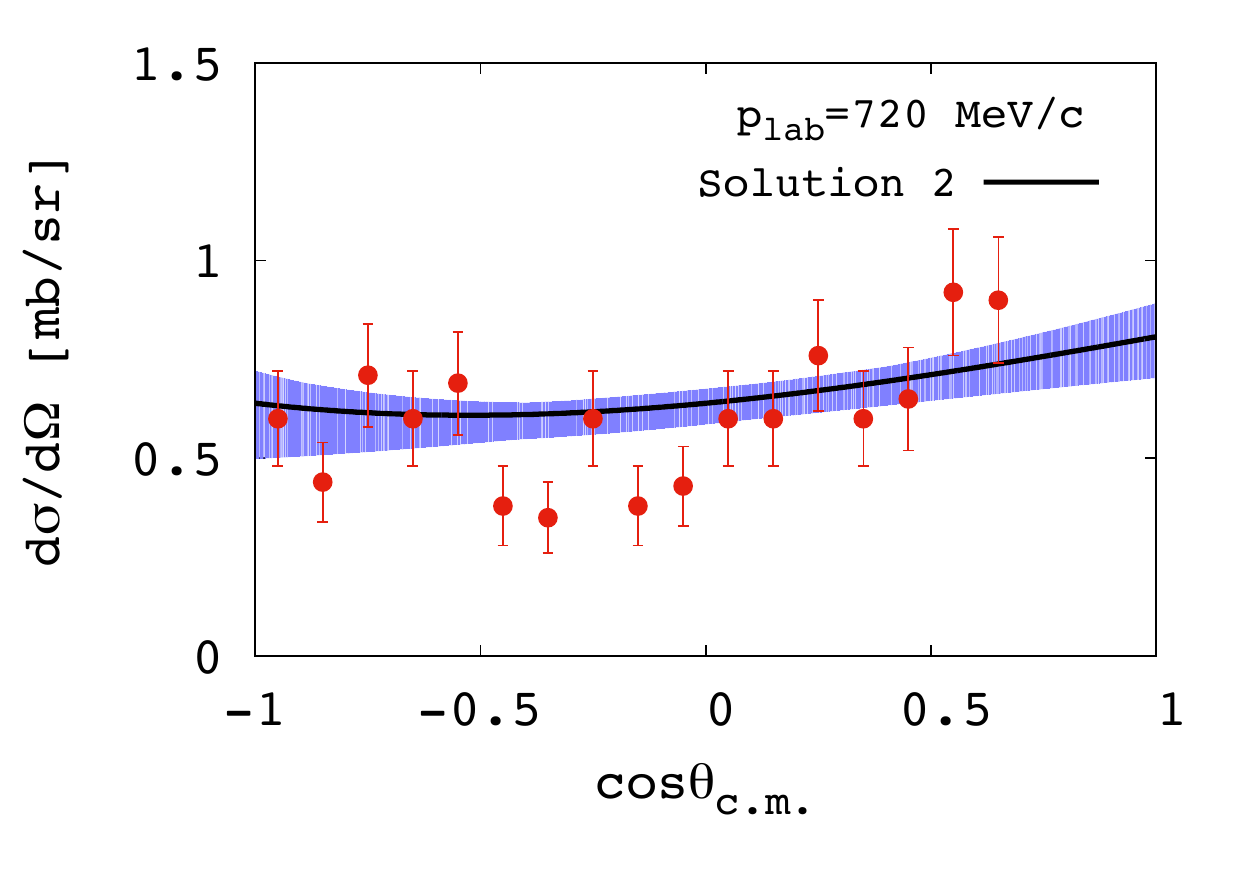}
  \end{center}
 \end{minipage}
\begin{minipage}{0.5\hsize}
  \begin{center}
   \includegraphics[width=80mm,bb=0 0 360 252]{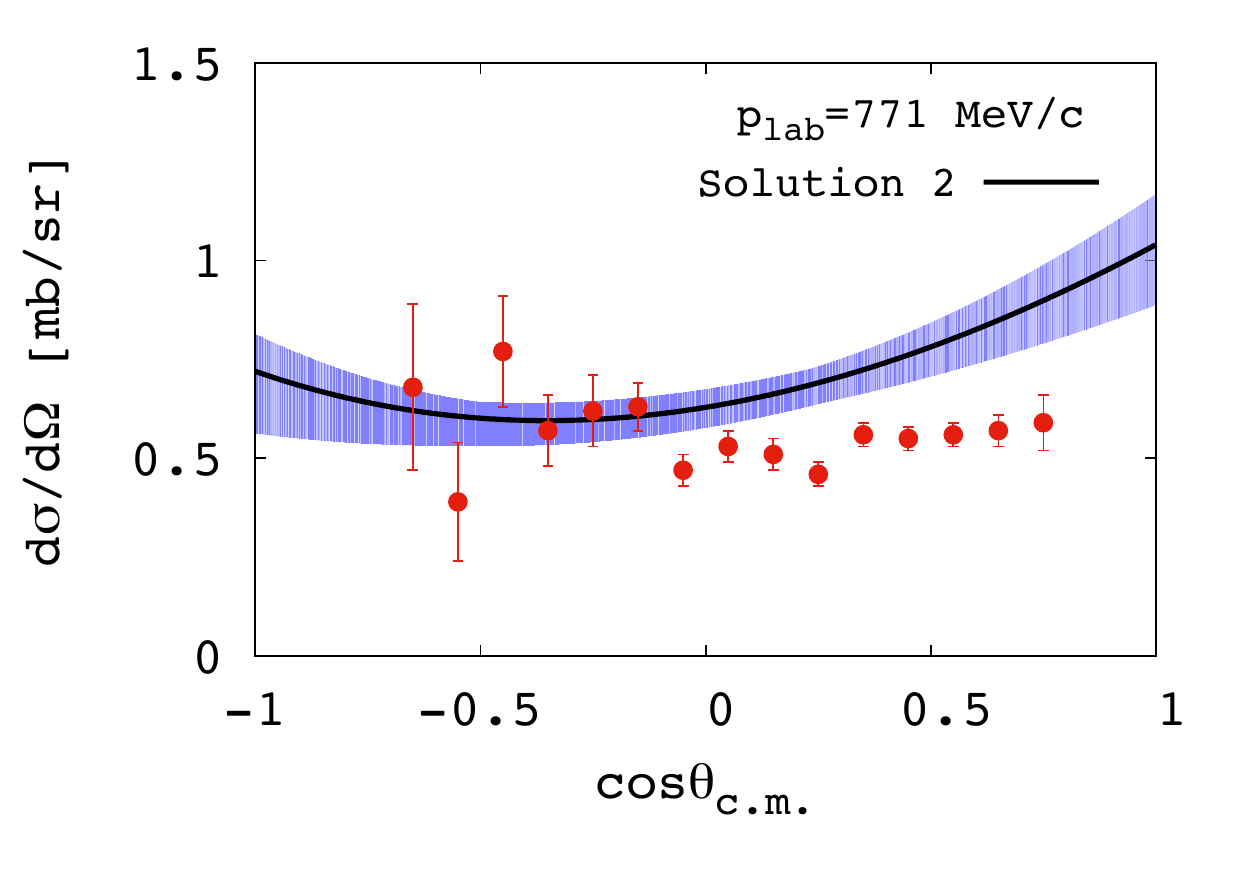}
  \end{center}
 \end{minipage}
\begin{minipage}{0.5\hsize}
  \begin{center}
   \includegraphics[width=80mm,bb=0 0 360 252]{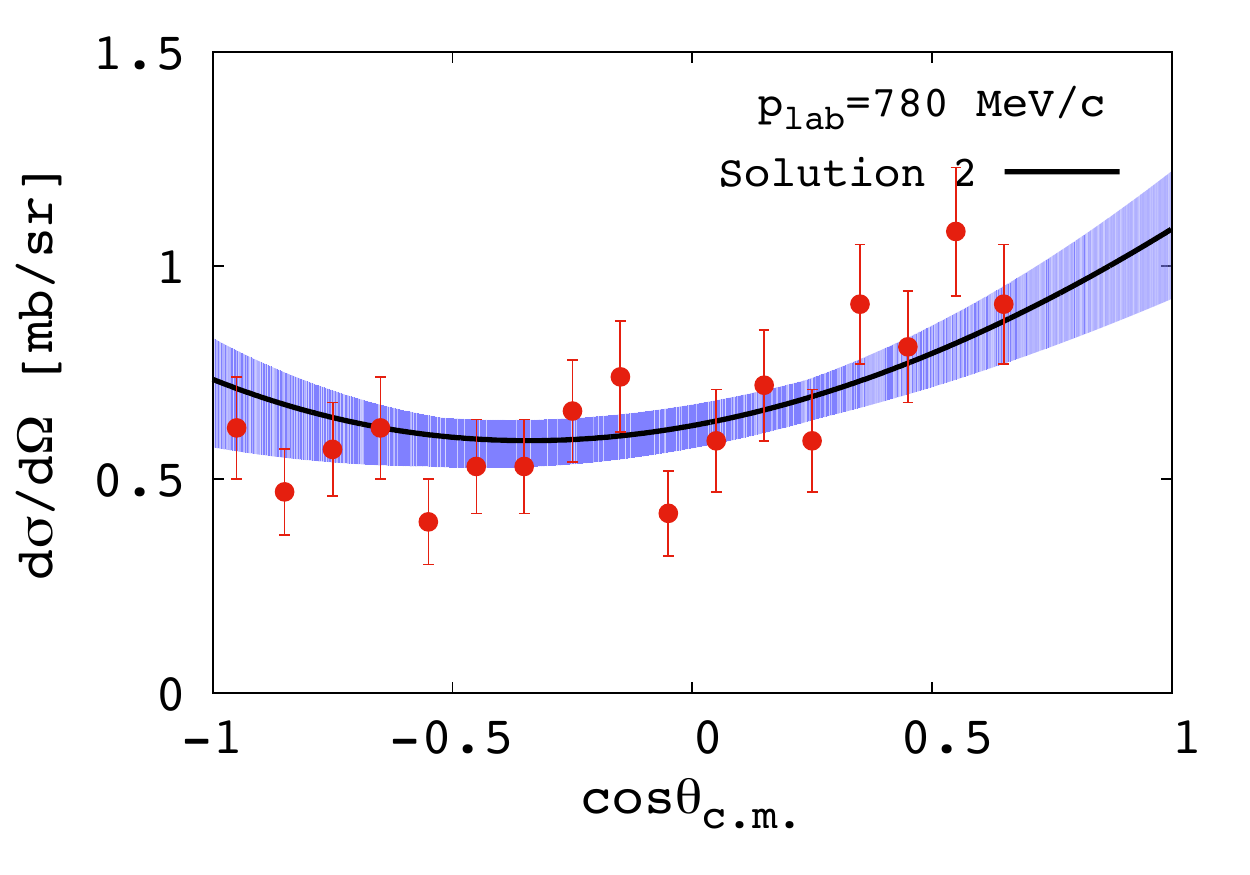}
  \end{center}
 \end{minipage}
 \vspace{-0.5cm}
 \caption{
The differential cross sections of $K^{+}n$ elastic scattering
 using Solution 2.}
\label{fig:kn_diff_sol2}
 \end{figure}

\begin{figure}[]
 \begin{minipage}{0.5\hsize}
  \begin{center}
   \includegraphics[width=80mm,bb=0 0 360 252]{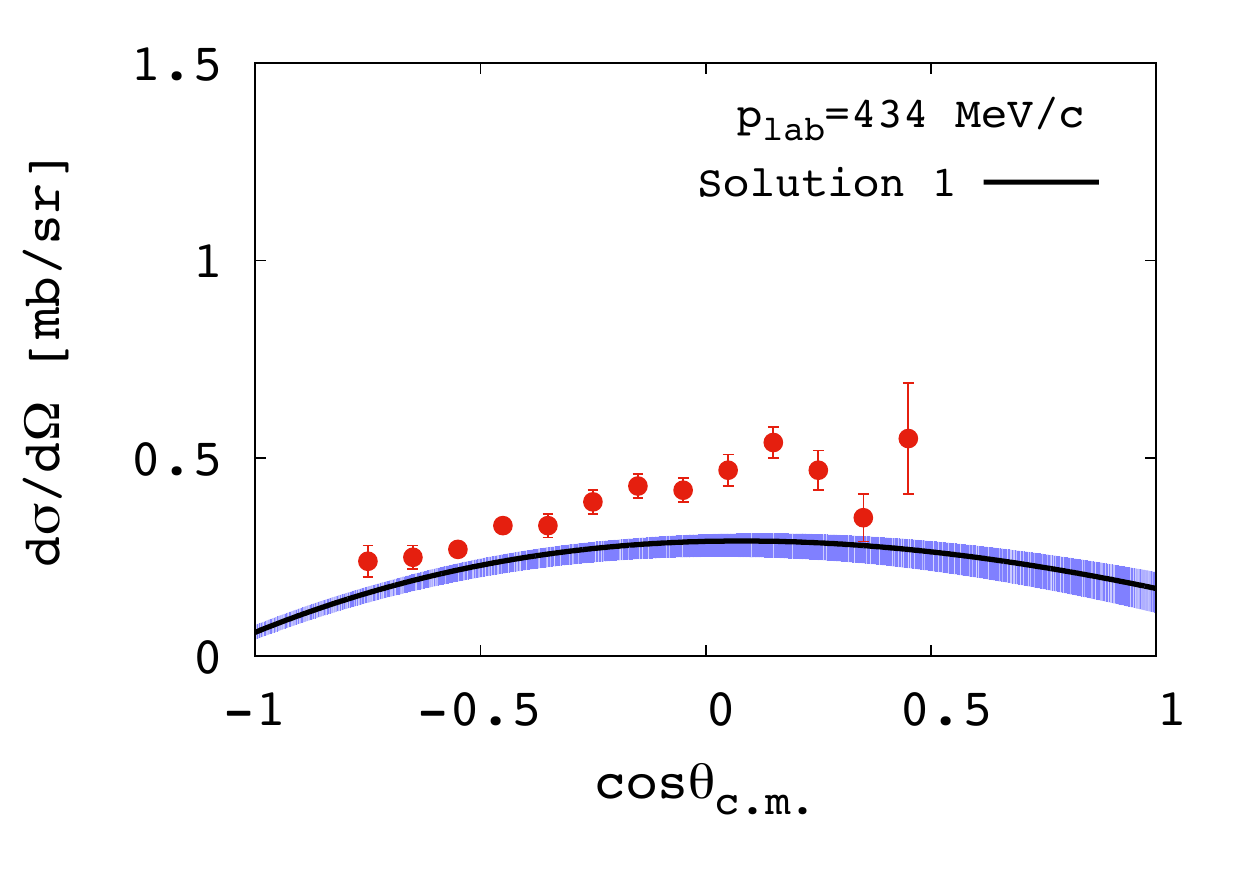}
  \end{center}
 \end{minipage}
 \begin{minipage}{0.5\hsize}
  \begin{center}
   \includegraphics[width=80mm,bb=0 0 360 252]{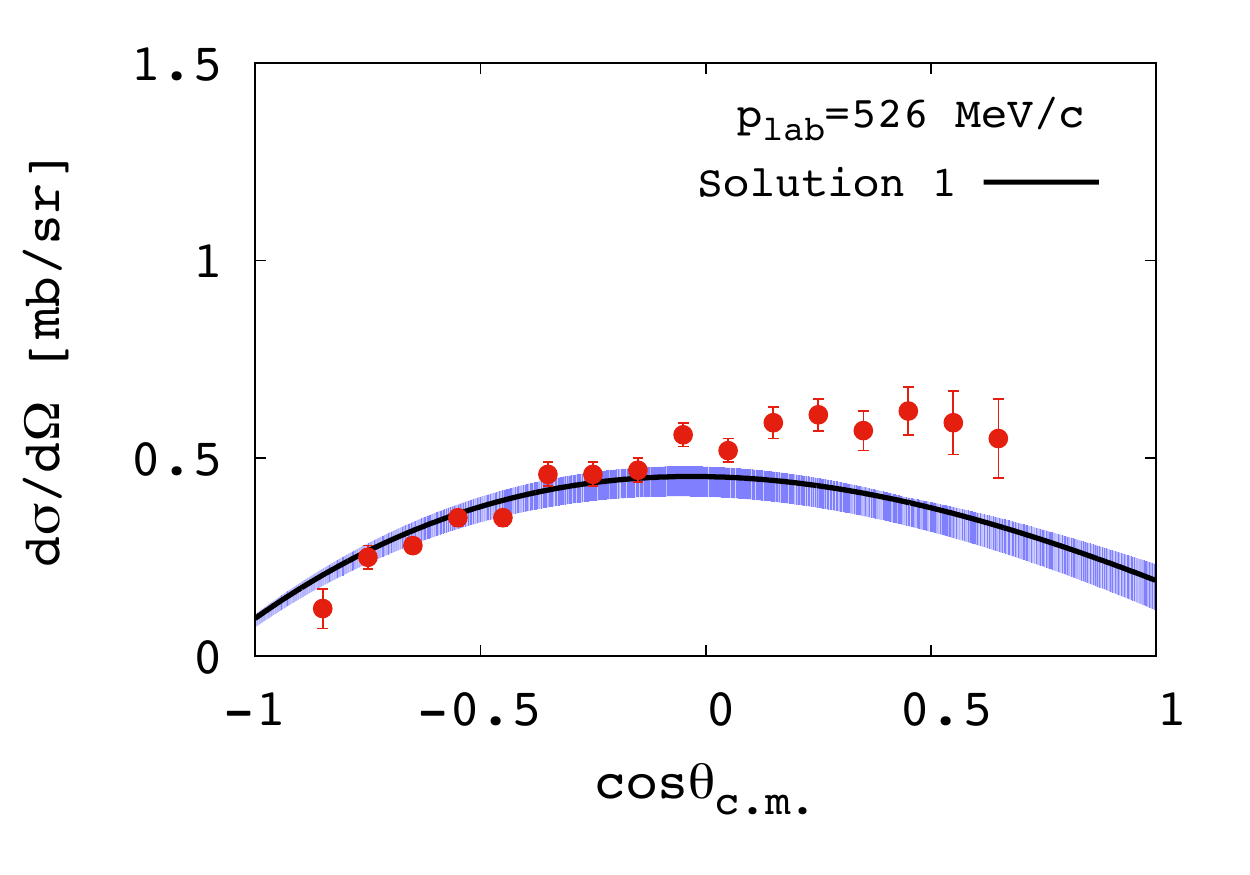}
  \end{center}
 \end{minipage}
\begin{minipage}{0.5\hsize}
  \begin{center}
   \includegraphics[width=80mm,bb=0 0 360 252]{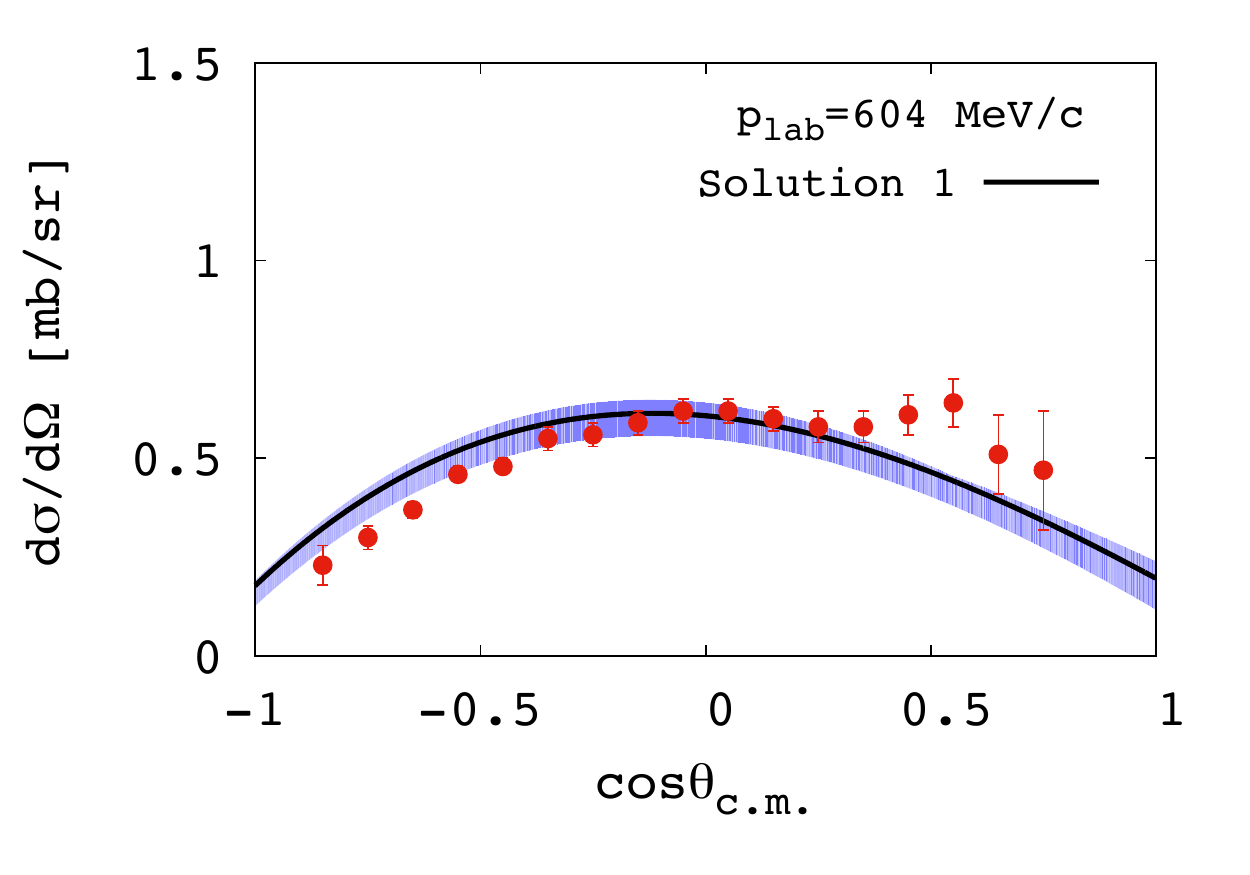}
  \end{center}
 \end{minipage}
\begin{minipage}{0.5\hsize}
  \begin{center}
   \includegraphics[width=80mm,bb=0 0 360 252]{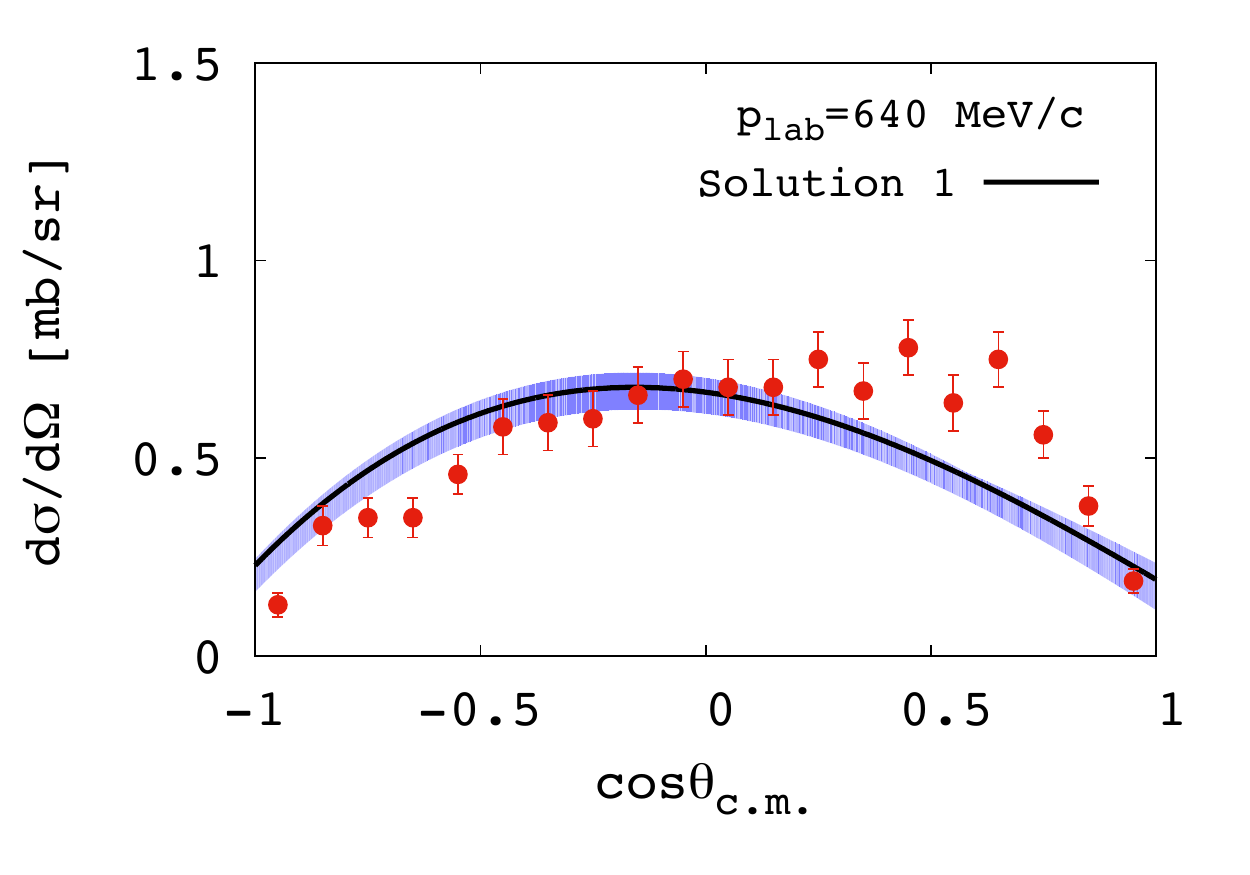}
  \end{center}
 \end{minipage}
 \begin{minipage}{0.5\hsize}
  \begin{center}
   \includegraphics[width=80mm,bb=0 0 360 252]{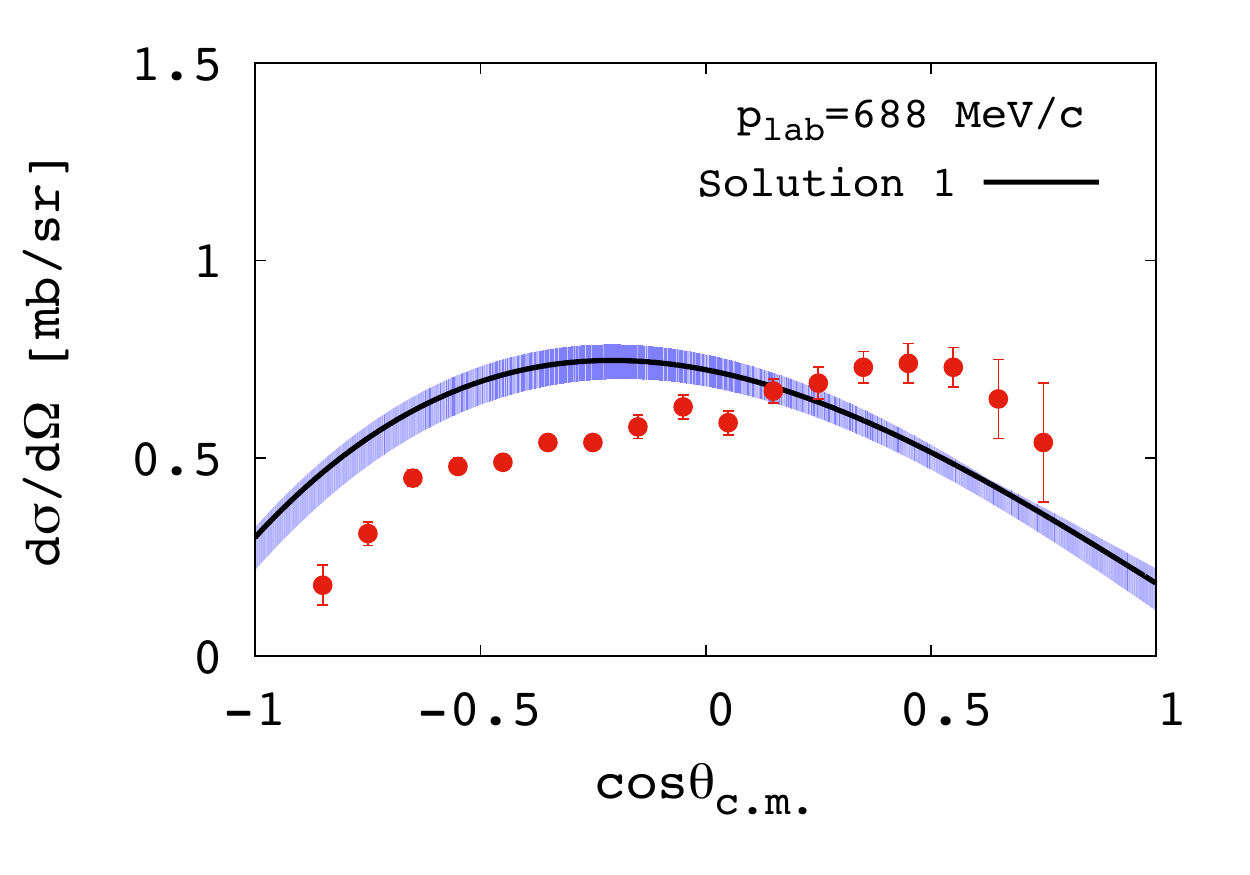}
  \end{center}
 \end{minipage}
\begin{minipage}{0.5\hsize}
  \begin{center}
   \includegraphics[width=80mm,bb=0 0 360 252]{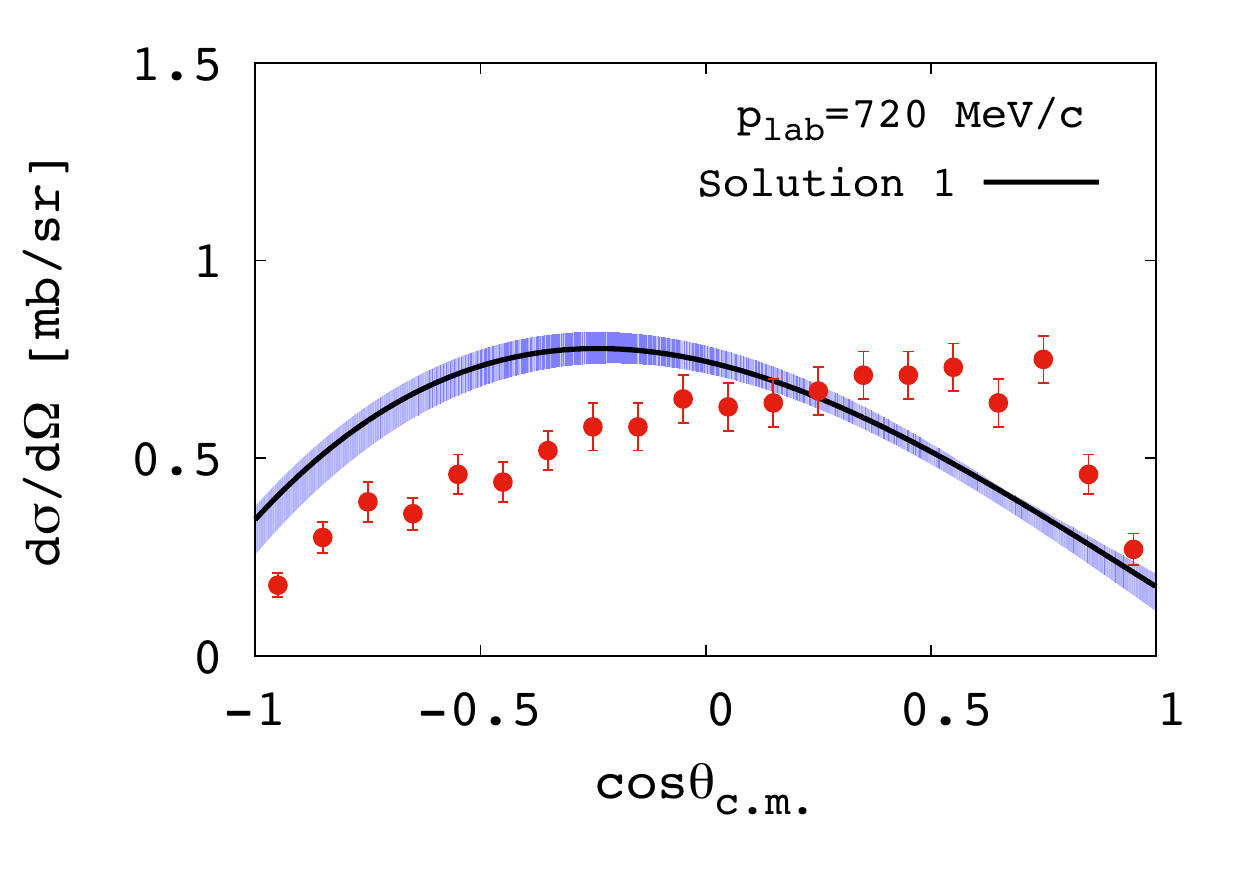}
  \end{center}
 \end{minipage}
\begin{minipage}{0.5\hsize}
  \begin{center}
   \includegraphics[width=80mm,bb=0 0 360 252]{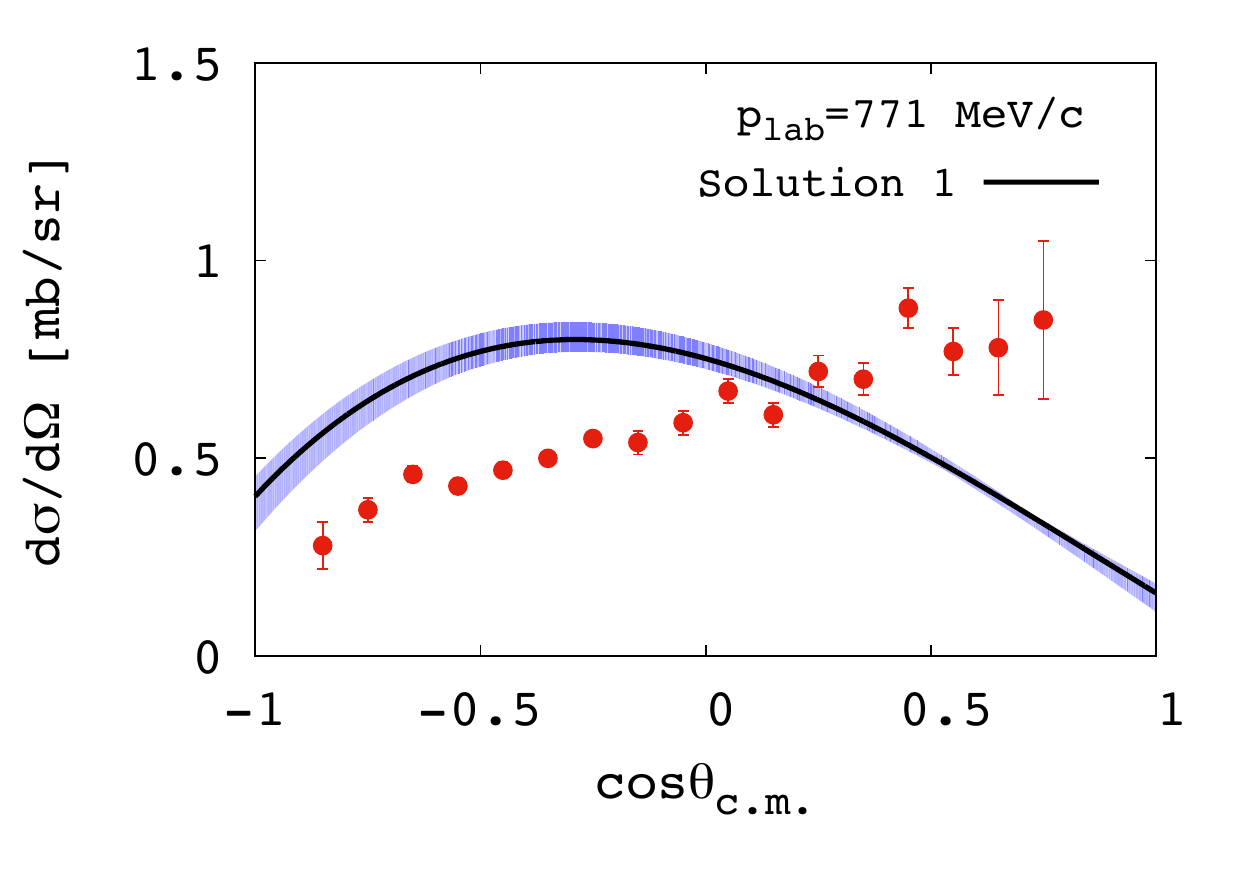}
  \end{center}
 \end{minipage}
\begin{minipage}{0.5\hsize}
  \begin{center}
   \includegraphics[width=80mm,bb=0 0 360 252]{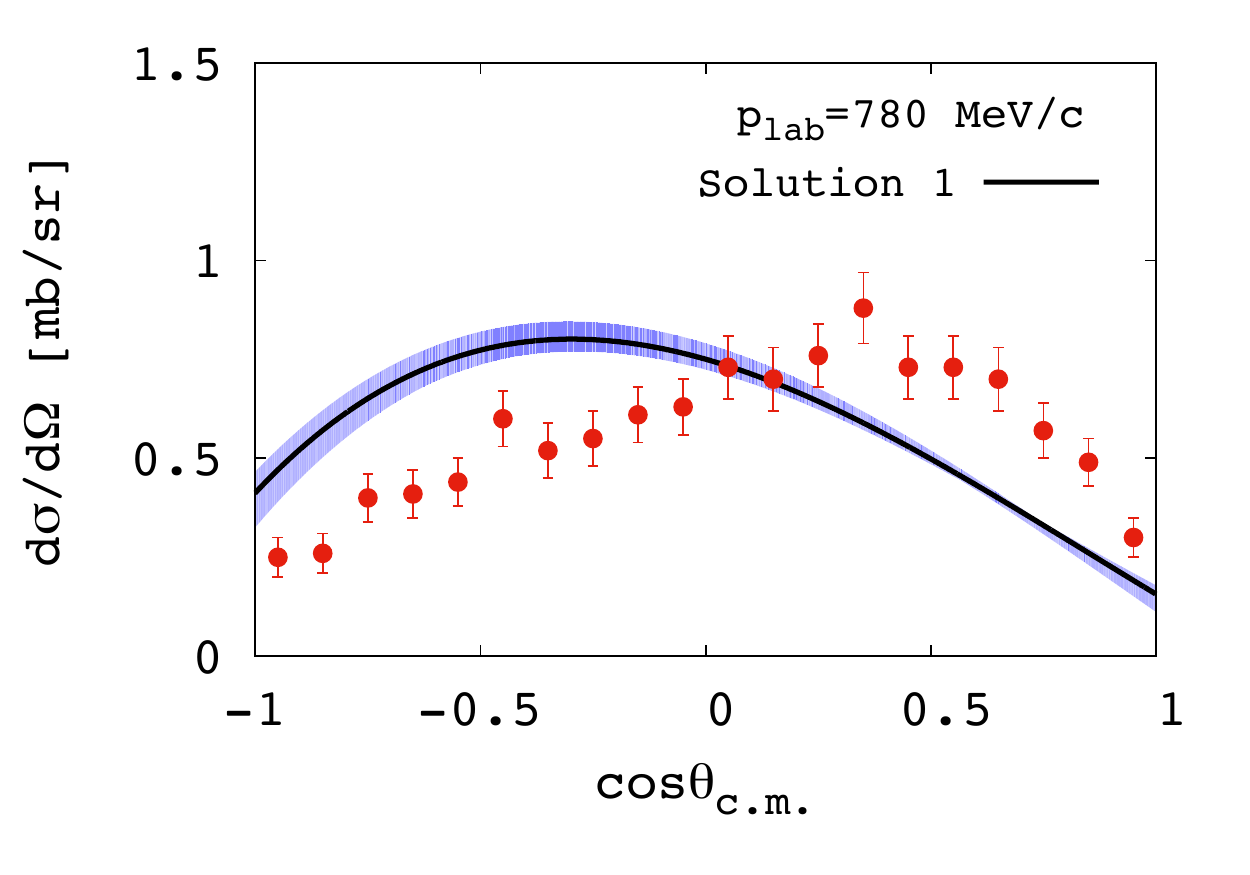}
  \end{center}
 \end{minipage}
 \vspace{-0.5cm}
 \caption{
 The differential cross sections of $K^{+}n$ charge exchange scattering 
 using Solution 1 in 
 comparison with the experimental data of Ref. \cite{Giacomelli:1972uj, dam1975}.
The momenta at the $p_{{\rm lab}}$=640, 720 and 780 MeV/c are the data from Ref.~\cite{Giacomelli:1972uj}.
The others are the data from Ref.~\cite{dam1975}. }
\label{fig:cex_diff_sol1}
 \end{figure}

\begin{figure}[]
 \begin{minipage}{0.5\hsize}
  \begin{center}
   \includegraphics[width=80mm,bb=0 0 360 252]{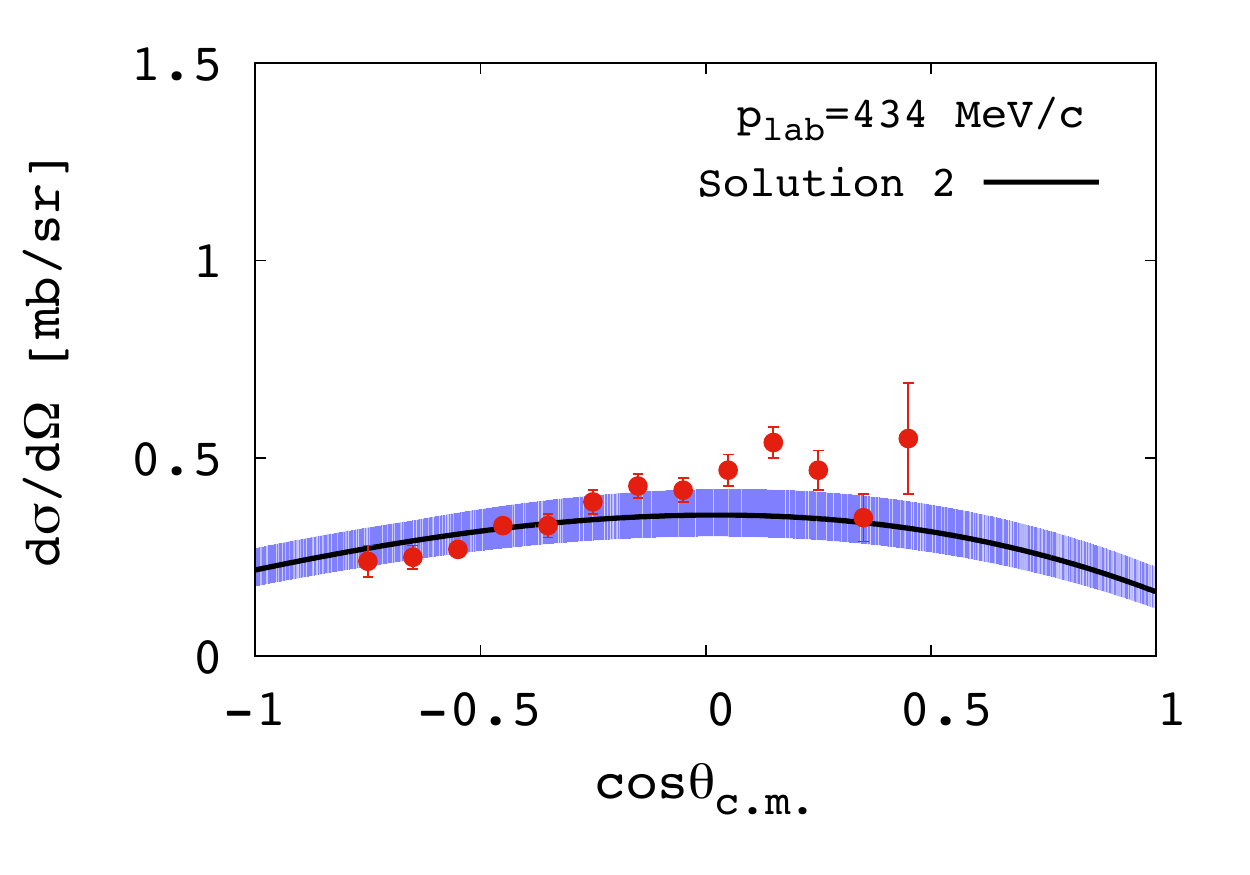}
  \end{center}
 \end{minipage}
 \begin{minipage}{0.5\hsize}
  \begin{center}
   \includegraphics[width=80mm,bb=0 0 360 252]{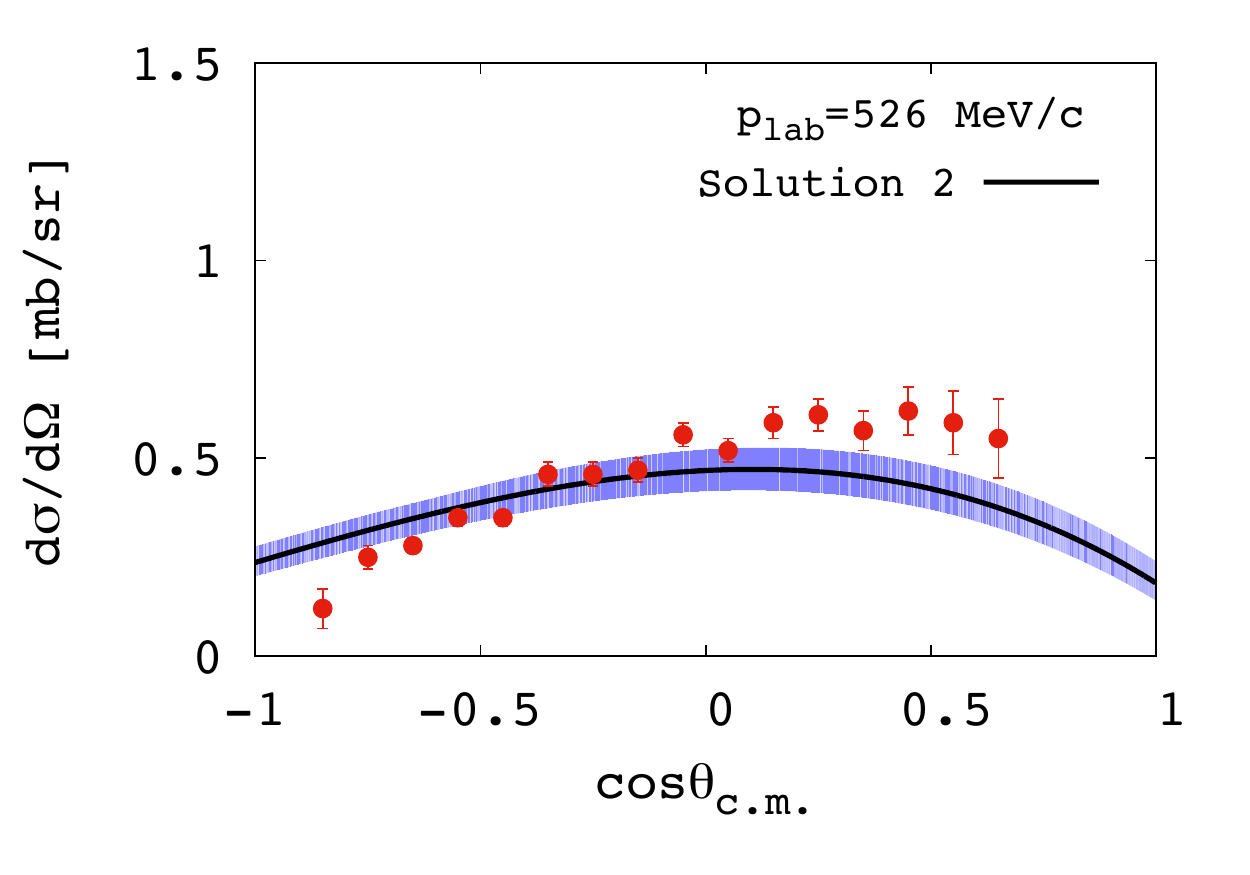}
  \end{center}
 \end{minipage}
\begin{minipage}{0.5\hsize}
  \begin{center}
   \includegraphics[width=80mm,bb=0 0 360 252]{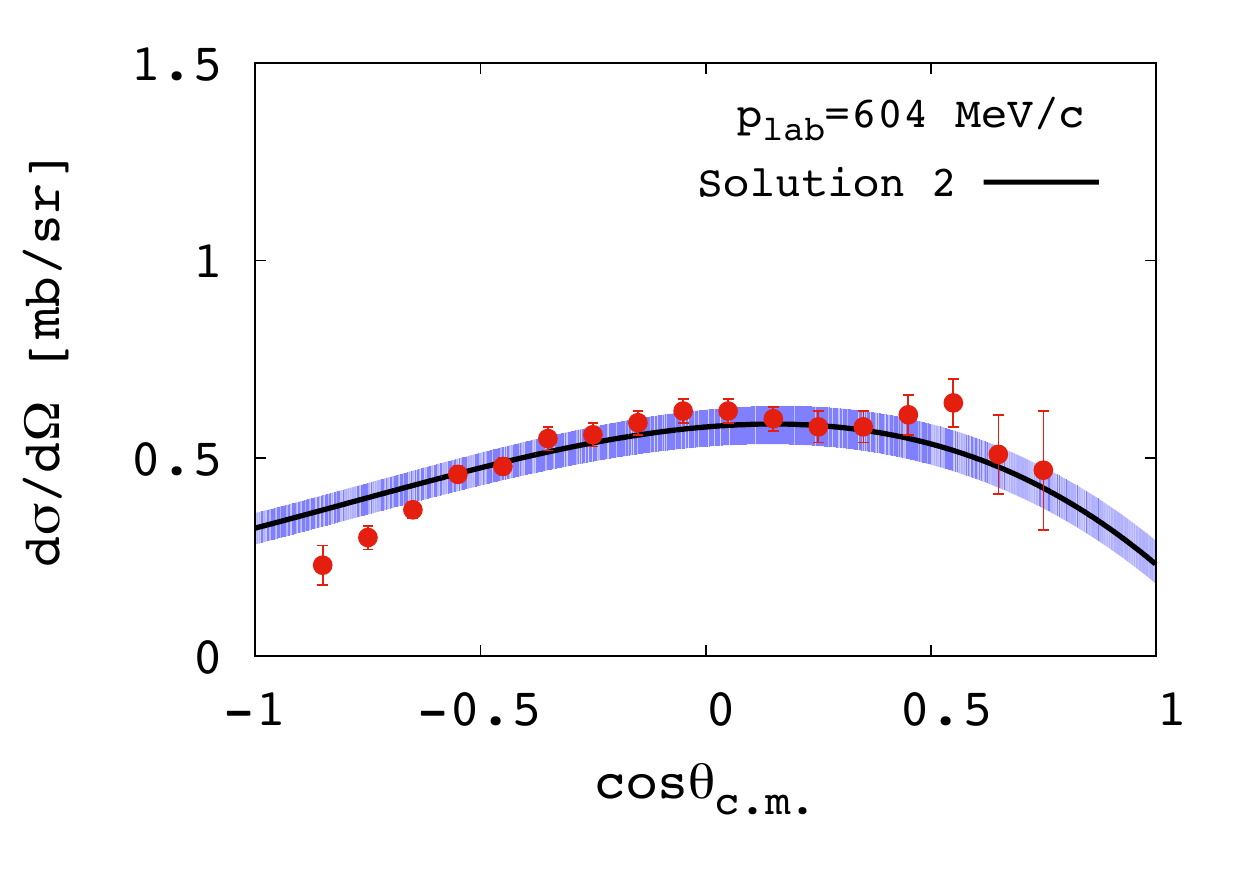}
  \end{center}
 \end{minipage}
\begin{minipage}{0.5\hsize}
  \begin{center}
   \includegraphics[width=80mm,bb=0 0 360 252]{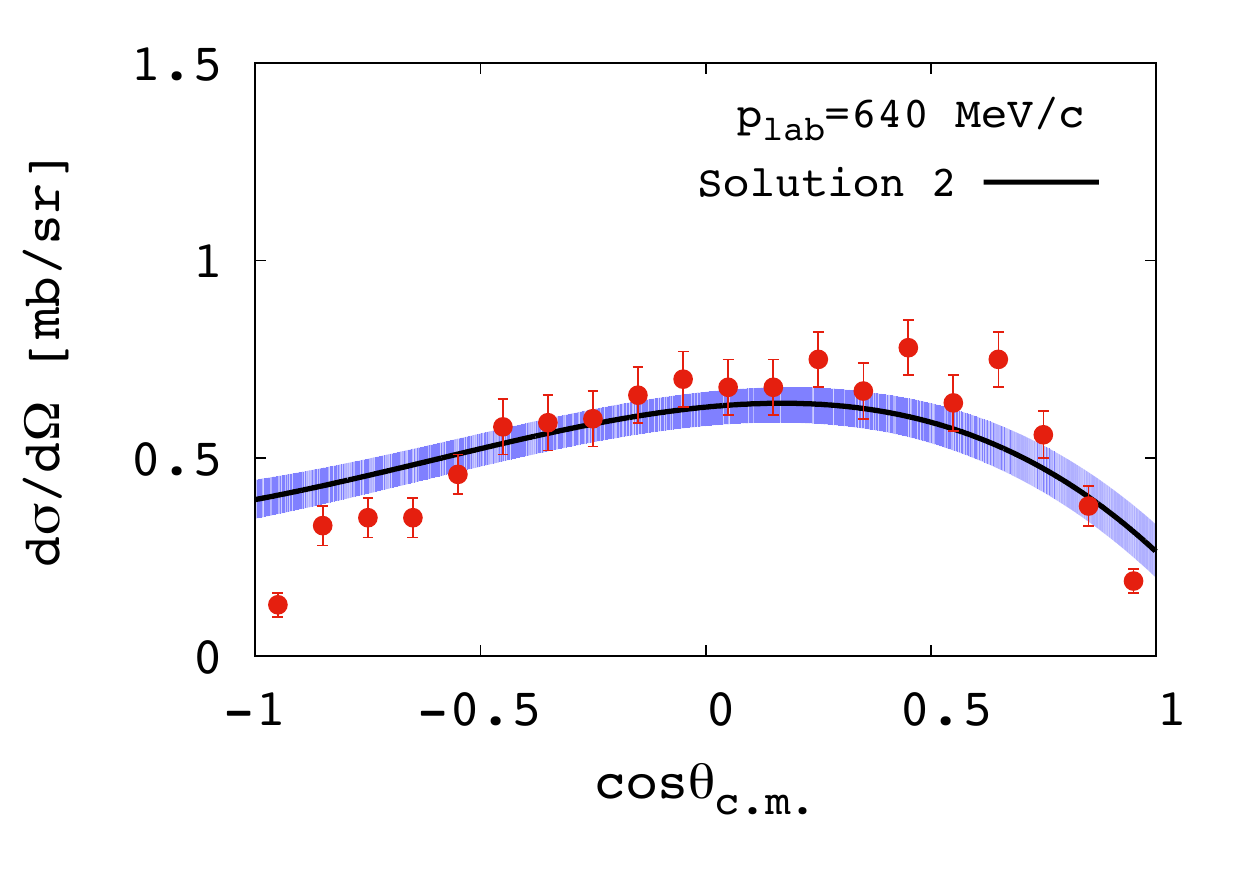}
  \end{center}
 \end{minipage}
 \begin{minipage}{0.5\hsize}
  \begin{center}
   \includegraphics[width=80mm,bb=0 0 360 252]{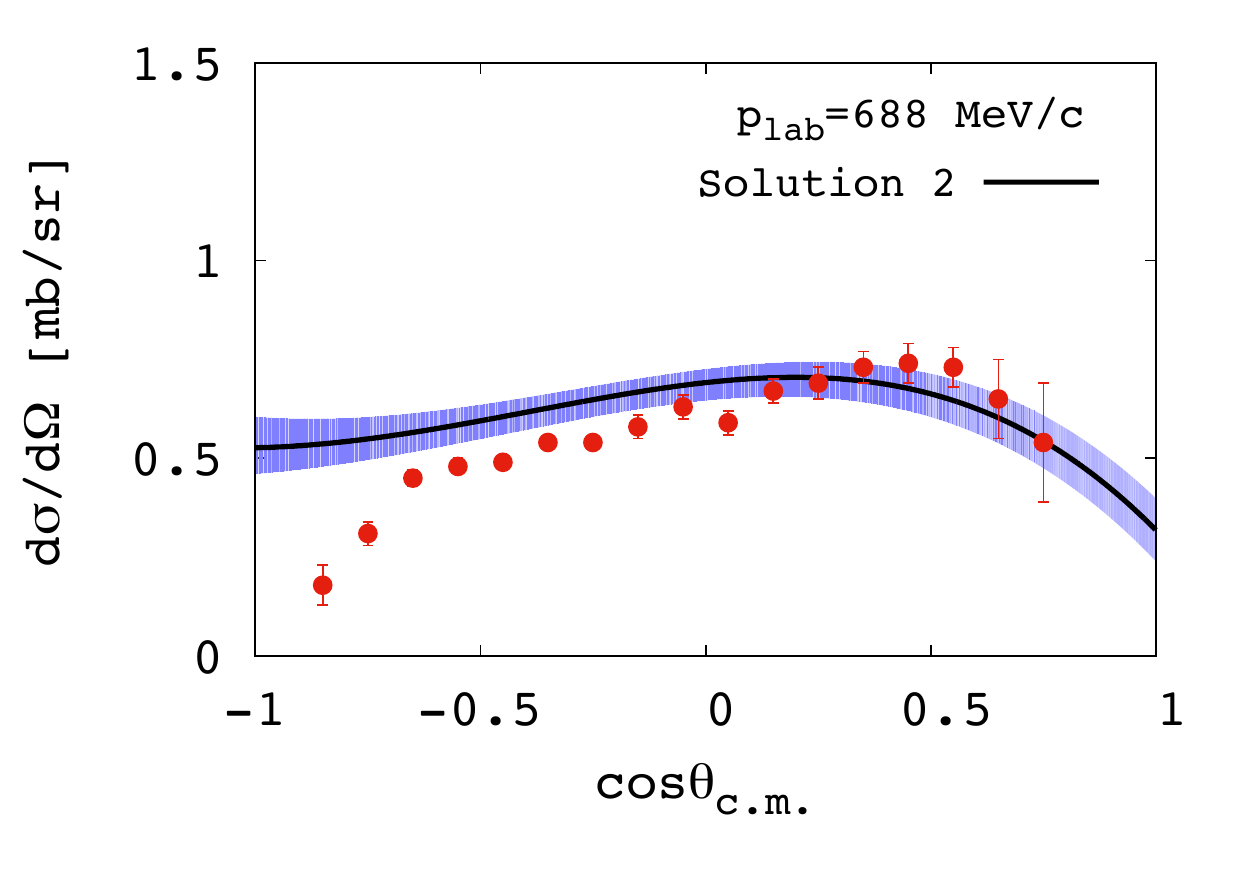}
  \end{center}
 \end{minipage}
\begin{minipage}{0.5\hsize}
  \begin{center}
   \includegraphics[width=80mm,bb=0 0 360 252]{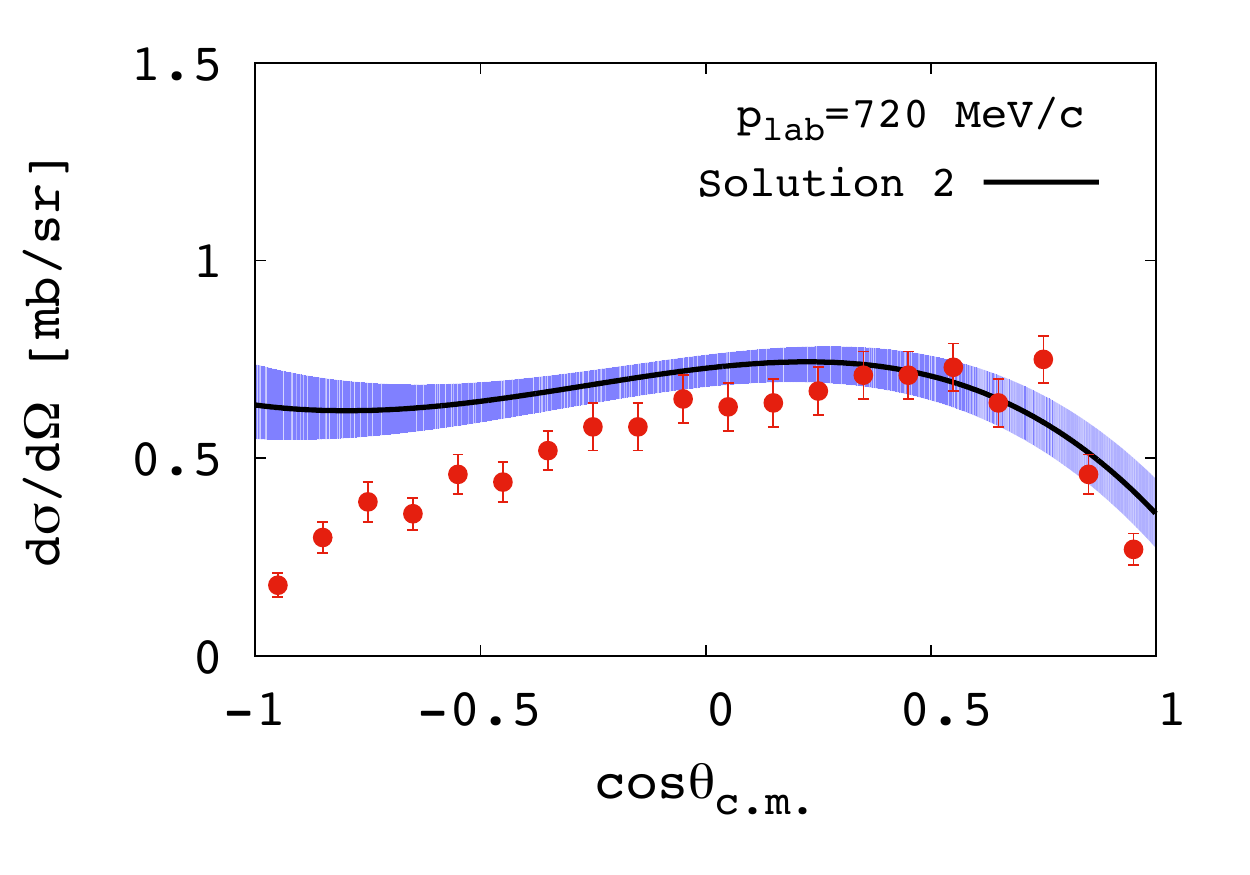}
  \end{center}
 \end{minipage}
\begin{minipage}{0.5\hsize}
  \begin{center}
   \includegraphics[width=80mm,bb=0 0 360 252]{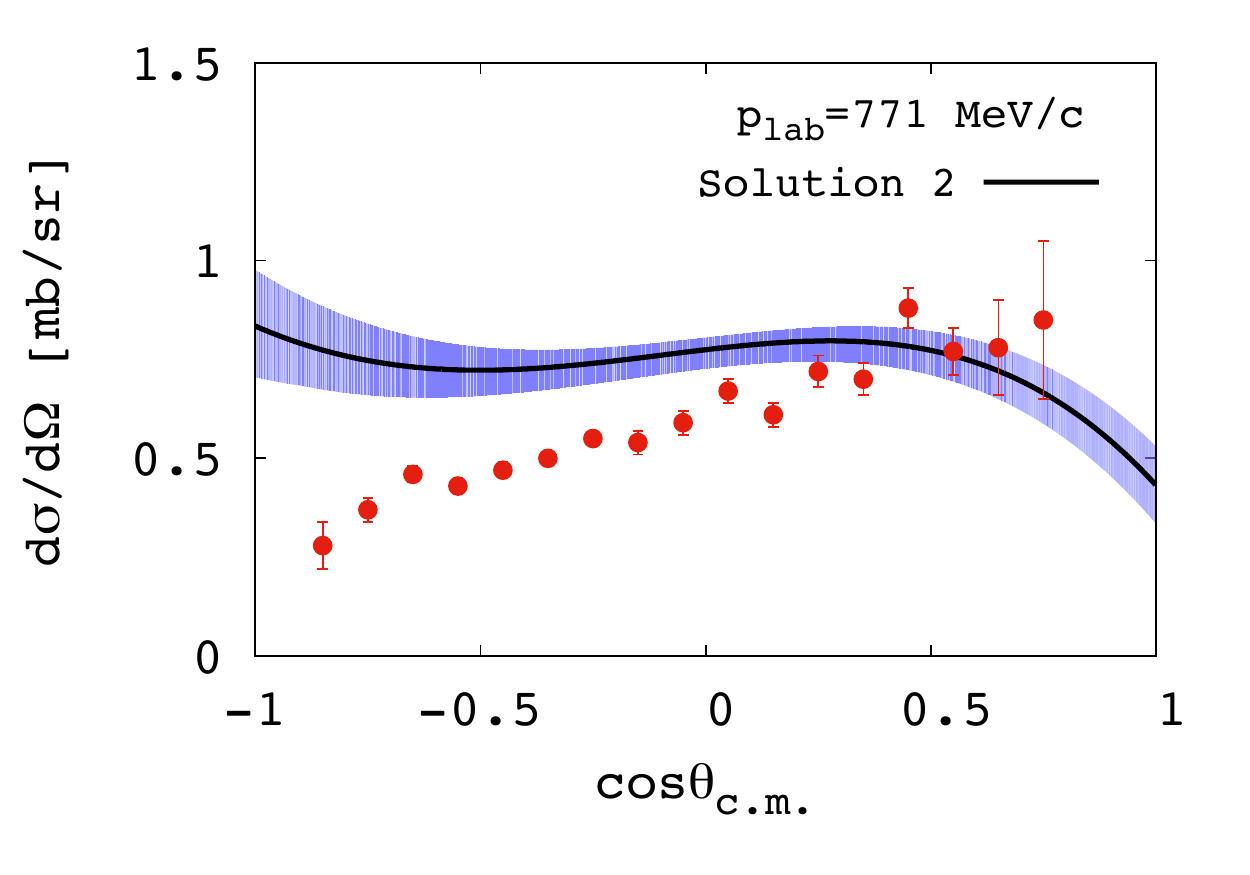}
  \end{center}
 \end{minipage}
\begin{minipage}{0.5\hsize}
  \begin{center}
   \includegraphics[width=80mm,bb=0 0 360 252]{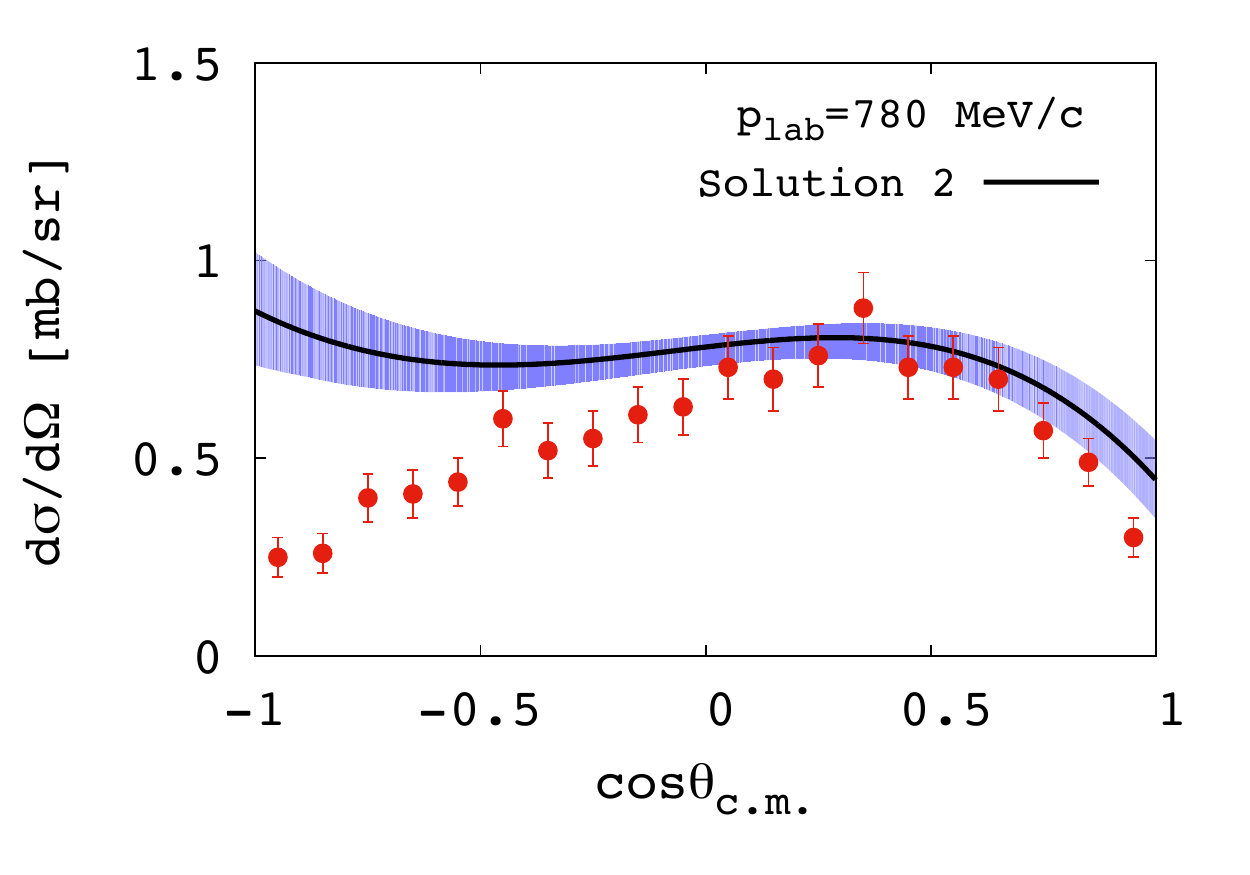}
  \end{center}
 \end{minipage}
 \vspace{-0.5cm}
 \caption{
The differential cross sections of $K^{+}n$ charge exchange scattering 
 using Solution~2.}
\label{fig:cex_diff_sol2}
 \end{figure}

Next, we determine the $I=0$ low-energy constants $b^{I=0}, d^{I=0}, g^{I=0}$ and $h^{I=0}$
using the data of the $K^{+}n \to K^{+}n$ and $K^{+}n \to K^{0} p$ differential cross sections 
at $p_{{\rm lab}} = 526$, 604 and 640 MeV/c given in Refs \cite{Giacomelli:1972uj, Giacomelli:1973ed, dam1975}
together with  the $I=0$ total cross section of Ref.~\cite{bowen1970}
between $p_{{\rm lab}} = 366$ to 717 MeV/c,
which is referred as Bowen 1970 in Fig. \ref{fig:i0_tot}.
We have confirmed that even if we include the $I=0$ total cross section
of Ref. \cite{carroll1973} referred as Carroll 1973 in Fig. \ref{fig:i0_tot},
we obtain similar parameter sets with much worse $\chi^{2}$ values.
It would imply that our model prefers the data of Bowen 1970 \cite{bowen1970} and 1973 \cite{bowen1973}. 
As a fine-tuning, we use the data of Bowen 1970 as the $I=0$ total cross section data.
The $K^{+}n$ elastic and charge exchange scattering amplitudes
are linear combinations of the $I=0$ and $I=1$ amplitudes
as shown in Eqs.~(\ref{eq:Kn}) and (\ref{eq:CE}).
The $I=1$ amplitude is already determined with the $K^{+}p$ elastic scattering.  
The $I=1$ parameters are fixed, when the $I=0$ parameters are determined. 
The fitted results for the $I=0$ parameters are summarized in the Table \ref{tab:parameter}.
Here we propose two solutions which have different character in the $I=0$ total cross section, 
as we will discuss in details later.
In Fig.~\ref{fig:i0_tot}, we show the $I=0$ total cross sections calculated with Solution 1, 2 and
find that these two solutions reproduce well the observed total cross section.
The band
shown in the figure show the allowable region of each solution around the
vicinity of the local minimum of $\chi^{2}$, that is
$4.54<\chi^{2}/N<5.53$ for Solution 1 
and  $4.06<\chi^{2}/N<5.01$ for Solution 2.
As one can see, the $I=0$ total cross section rapidly increases around $p_{\rm lab} = 500$ MeV/c.
In the two solutions, the partial wave which is responsible for the rapid increase of the cross section 
is different. 
Actually, as we shall see later, this feature links to the property of 
a possible resonance appearing in $I=0$ with a large width. 
In Solution 1, the $P_{01}$ amplitude\footnote{
We use the partial wave convention $L_{I\, 2J}$ with orbital angular momentum $L$, 
isospin $I$ and total angular momentum $J = L \pm 1/2$.} 
dominantly contributes, and thus 
the rapid increases is caused by the $P_{01}$ amplitude. In Solution 2, 
both $P_{01}$ and $P_{03}$ amplitudes provide the contribution for the $I=0$ total 
cross section and the $P_{03}$ amplitude is responsible for the rapid increase. 
In this way, these two solutions have 
their own characteristic features in the $I=0$ total cross section. 
In summary, we would say that Solution 1 is ``$P_{01}$ dominant solution",
Solution 2  is ``$P_{03}$ dominant solution".
%
In the following, we show the result of the differential cross sections used $I=0$ and 1 amplitudes.
Similar to the $I=0$ total cross section, we show the allowable region of solutions as the band.
Figures \ref{fig:kn_diff_sol1} and \ref{fig:kn_diff_sol2} show 
the $K^{+}n$ elastic differential cross section for Solutions 1 and 2.
Solutions 1 and 2  are mostly consistent with the experimental data.
Figures \ref{fig:cex_diff_sol1} and \ref{fig:cex_diff_sol2} show the $K^{+}n$ charge exchange differential cross section.
Solutions 1 and 2 show relatively good reproduction except for the forward and backward scattering.
We cannot find sizable contradictions for Solutions 1 and 2 with the experimental data.
As seen later, other analyses support Solution 1.

%

%
%

\subsection{Possible broad resonances}
\begin{figure}[]
 \begin{center}
   \includegraphics[width=100mm,bb=0 0 3946 2705]{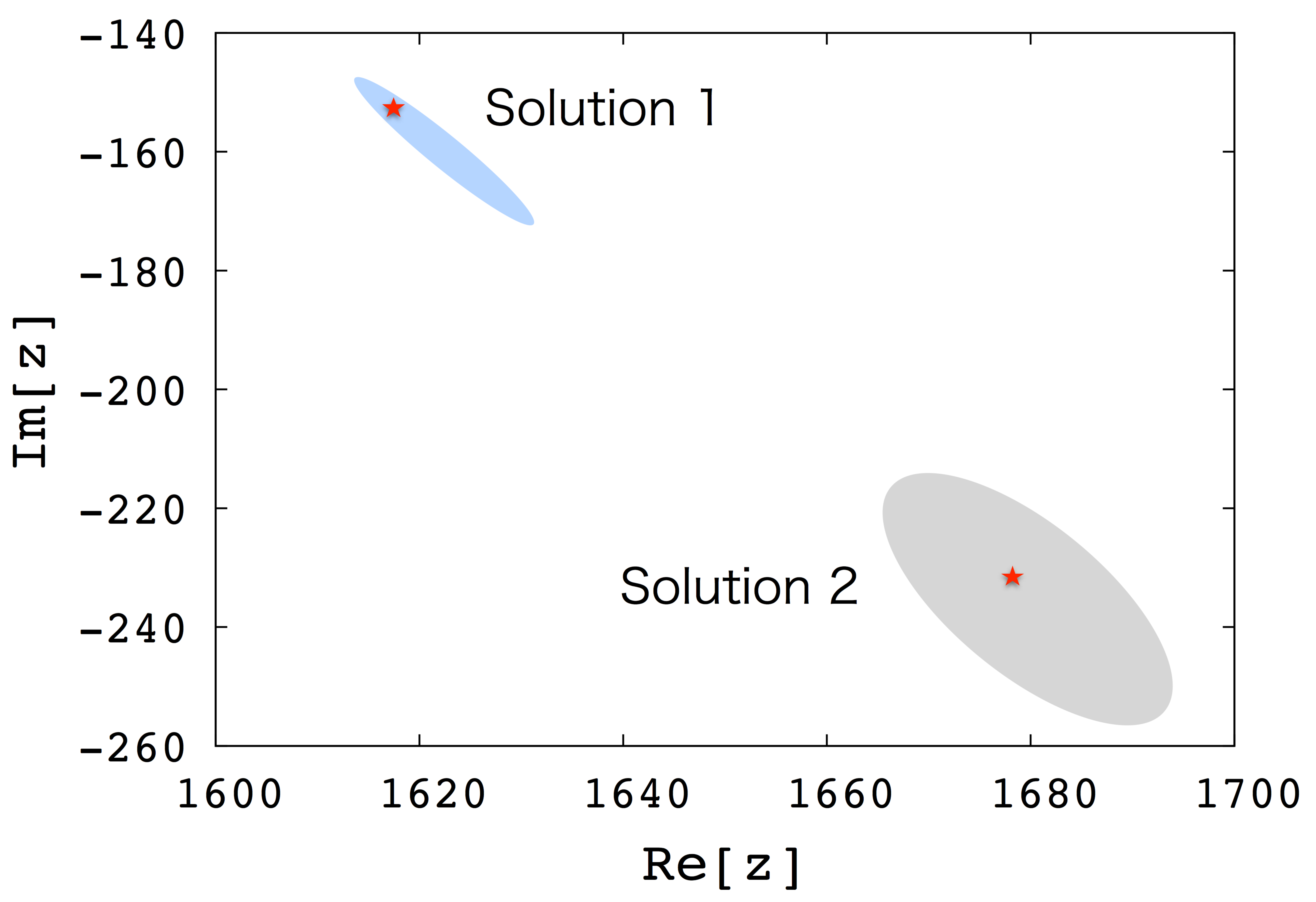}
 \end{center}
 \caption{The distribution of the poles of the amplitude in the complex energy plane $z$.
 The red star stands for the best-fit value.}
 \label{fig:poles}
\end{figure}

We have constructed the $KN$ amplitudes which reproduce the experimental data well.
In the following, we concentrate on the $KN$ partial wave amplitudes with $I=0$ and 
discuss the outcome from the obtained amplitude. 
First of all, we look for poles of the scattering amplitude in the complex energy plane.
Having the $KN$ scattering amplitude
in an analytic form, we can perform analytic continuation of the scattering amplitude into 
the complex energy plane. 
We find a pole in the $P_{01}$ amplitude of Solution 1 at $z = 1617 - 153 i$~MeV,
which corresponds to a resonance state with mass 1617~MeV/$\rm c^{2}$,  
width 305~MeV and $J^{p} = (1/2)^{+}$.  
The resonance has a quite large width and it could be hard to pin down the resonance
in production experiments.
Similarly, we find a pole of the $P_{03}$ amplitude of Solution 2 in the complex energy plane at 
$z=1678 - 232 i $~MeV corresponding to a resonance state with mass 
1678~MeV/${\rm c^{2}}$, width 463~MeV and $J^{p} = (3/2)^{+}$.
Since this resonance state
is located far from the real axis, it 
is not constrained well by experimental observation 
appearing in the real axis and theoretical uncertainty should be large and this solution 
could be unstable against small deviation of experimental data. 
These results are summarized in Table \ref{tab:amp}.
These resonances could be compared with the state found in the chiral soliton model \cite{weigel}
with around 1700 MeV/${\rm c^{2}}$ mass even though it has a narrow width.
In Fig. \ref{fig:poles}, we show the distribution of the poles in the vicinity of the best-fit value.


\begin{table}[]
\begin{center}
\caption{The resonance states of Solutions 1 and 2.}
  \begin{tabular}{l c c} \hline
    amplitude ($J^{P}$) & mass [MeV] &  width [MeV]  \\ \hline \hline
    Solution 1 \enspace $P_{01}$ ($ \frac{1}{2}^{+}$)  & 1617 & 305  \\ \hline 
    Solution 2 \enspace $P_{03}$  ($ \frac{3}{2}^{+}$) & 1678  & 463  \\ \hline
  \end{tabular}
  \label{tab:amp}
  \end{center}
\end{table}

\begin{figure}[]
 \begin{minipage}{0.24\hsize}
  \begin{center}
   \includegraphics[width=53mm,bb=0 0 360 252]{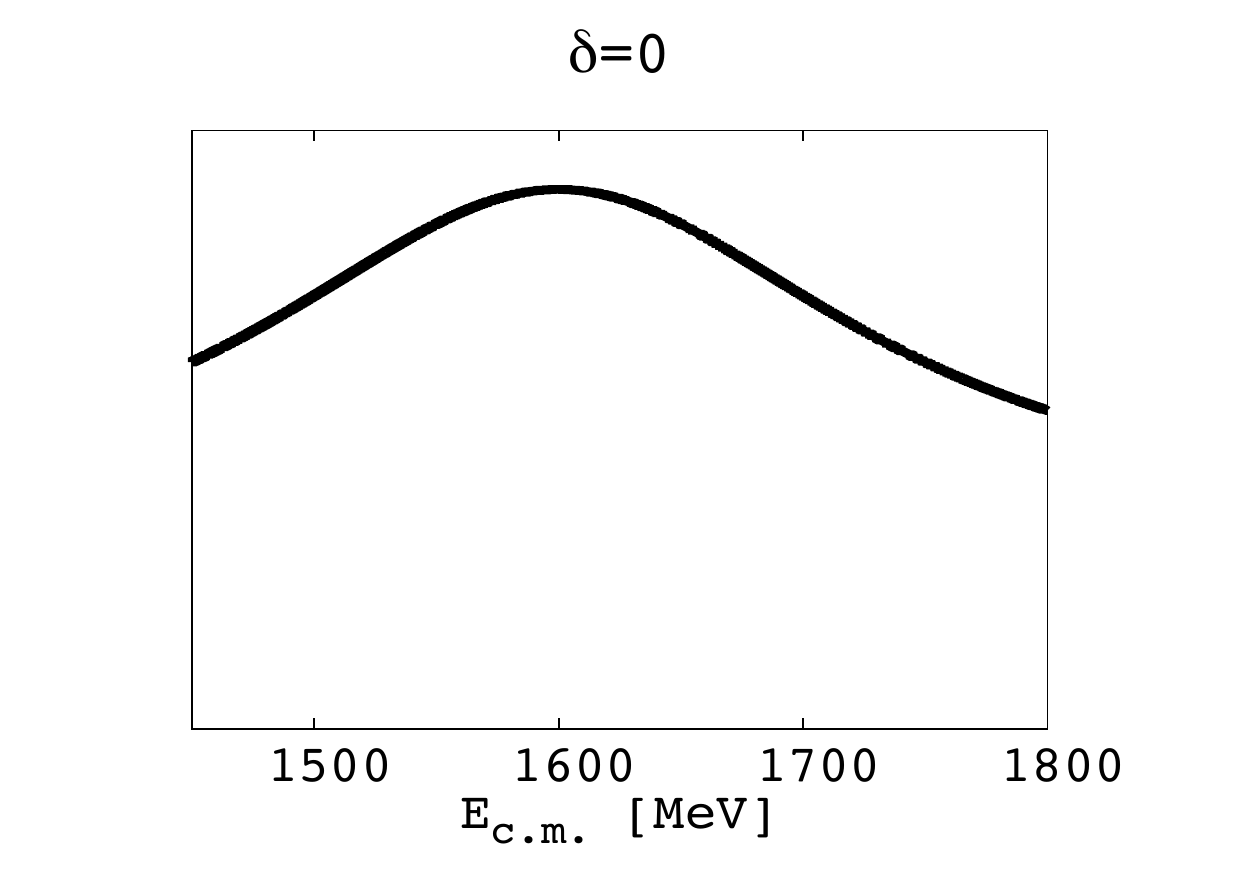}
  \end{center}
 \end{minipage}
 \begin{minipage}{0.24\hsize}
  \begin{center}
   \includegraphics[width=53mm,bb=0 0 360 252]{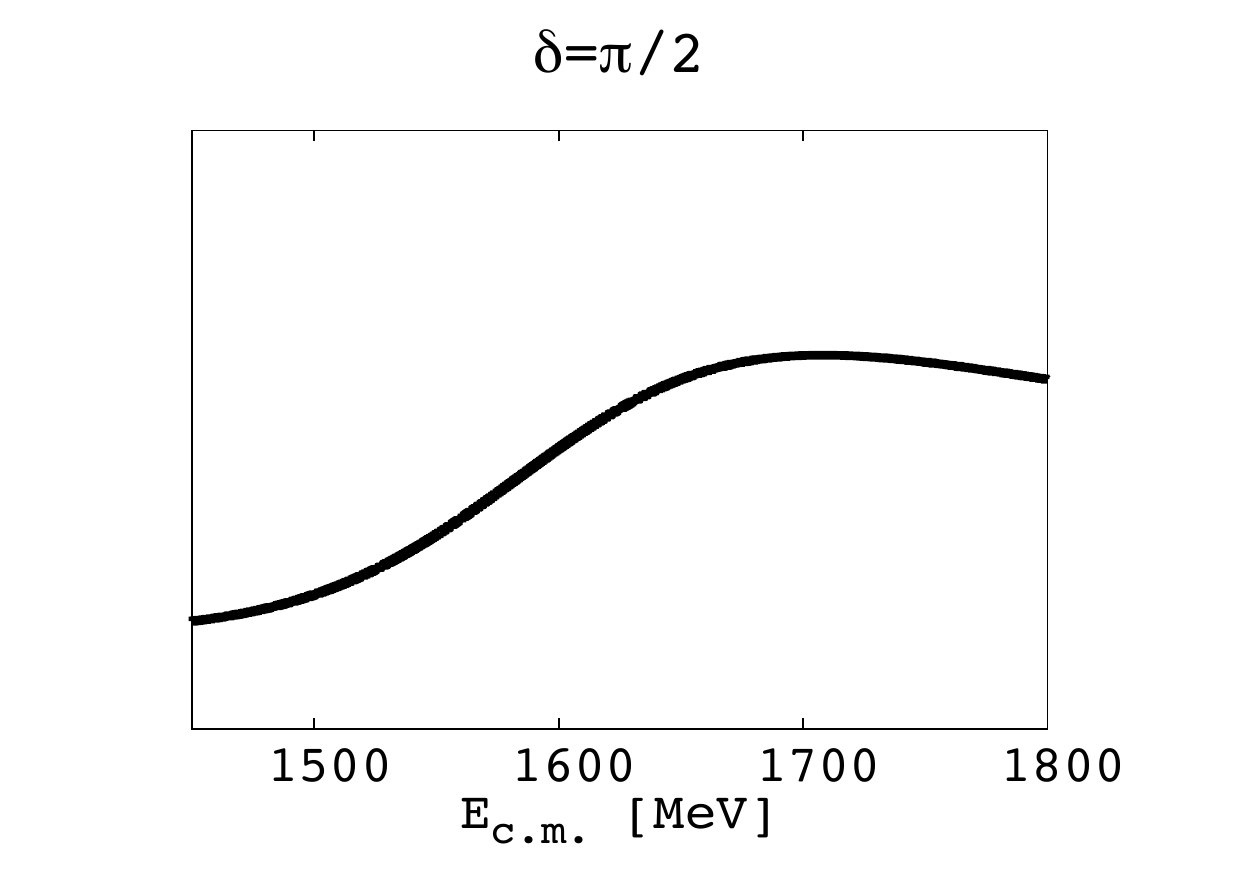}
  \end{center}
 \end{minipage}
  \begin{minipage}{0.24\hsize}
  \begin{center}
   \includegraphics[width=53mm,bb=0 0 360 252]{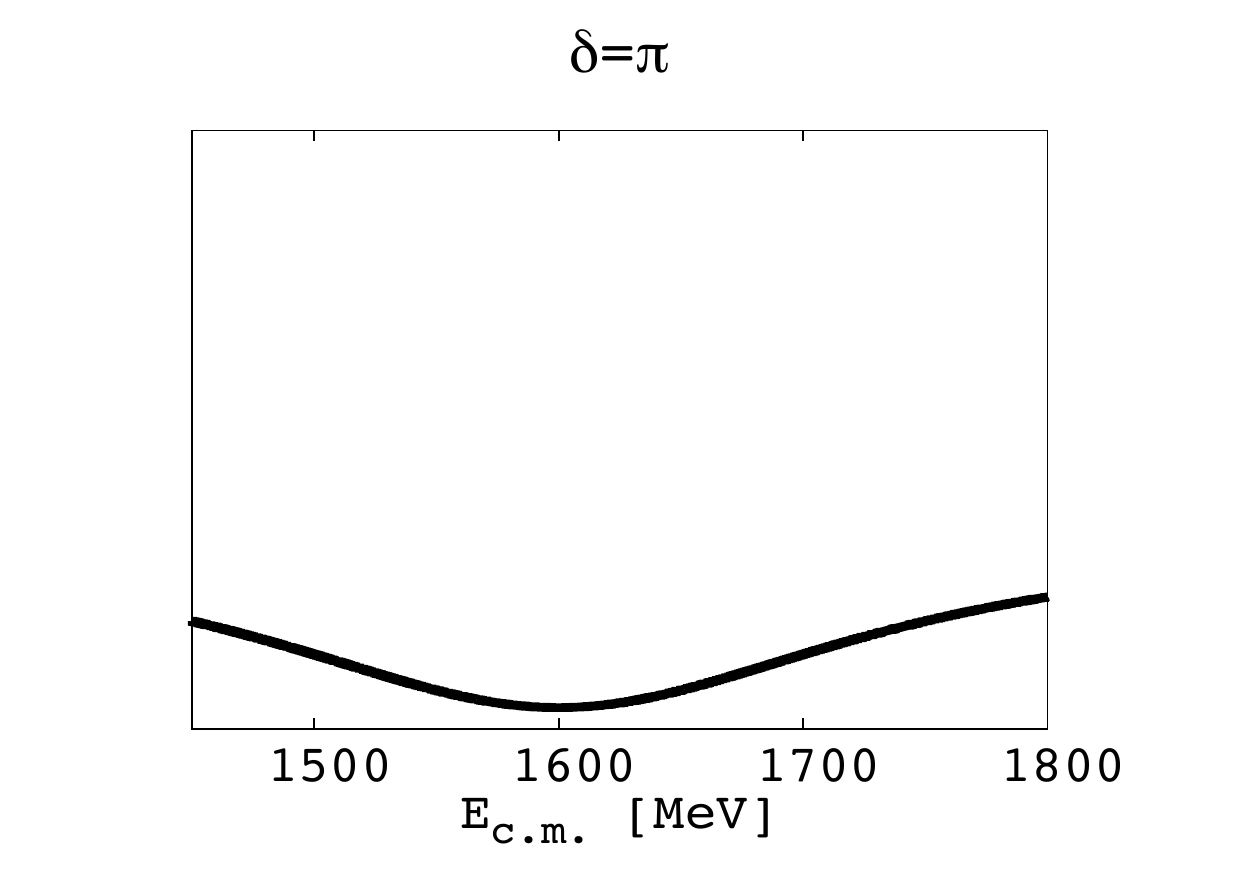}
  \end{center}
 \end{minipage}
   \begin{minipage}{0.24\hsize}
  \begin{center}
   \includegraphics[width=53mm,bb=0 0 360 252]{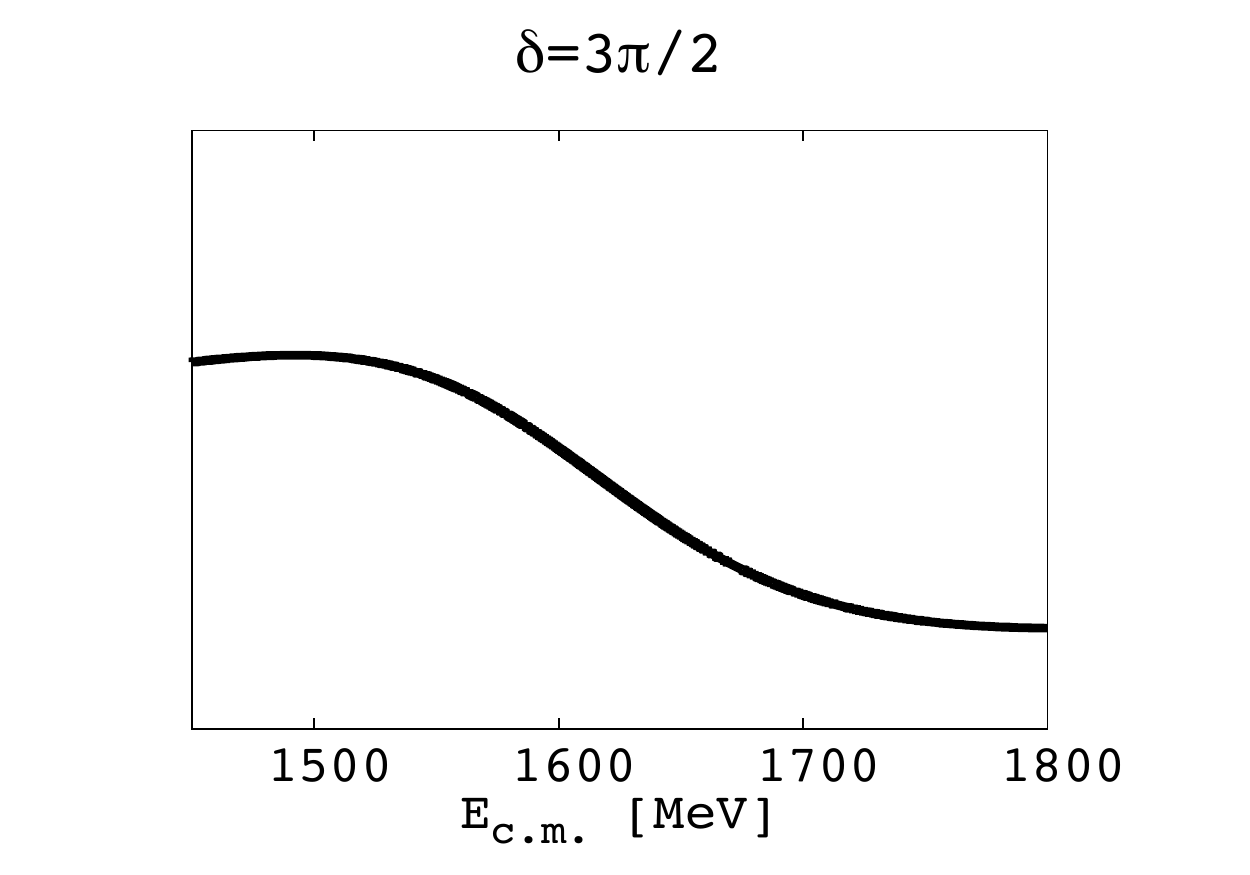}
  \end{center}
 \end{minipage}
\caption{Fano resonance. Cross sections of amplitudes composed of a resonance and 
 a continuum background with relative phase $\delta$ are shown. 
 The resonance is assumed to have 1600~MeV mass and 300 MeV width. 
 The resonance shape is dependent on the relative phase $\delta$. The vertical axis is in arbitrary unit. }
\label{fig:fano}
\end{figure}

{Even though we find the resonance state as a pole of the scattering amplitude, 
there are no peak structure in the scattering amplitude around the resonance energy.}
One usually expects 
that resonance states should appear as a peak in the cross section. It is not necessarily true
when the resonance has a large width and substantial coupling to non-resonance background.
We demonstrate this situation by using a simple amplitude in which a resonance pole is embedded 
in a constant background with a relative phase $\delta$:
\begin{equation}
  f(E) = \frac{i}{E - M + i \Gamma/2} + b e^{i\delta}. \label{eq:BWamp}
\end{equation}
In Fig.~\ref{fig:fano}, we show the cross sections of the amplitudes (\ref{eq:BWamp}) with 
$\delta= 0, \frac{\pi}{2}, \pi, \frac{3\pi}{2}$ for $M=1600$ MeV, $\Gamma=300$ MeV and 
$b=0.01$ MeV$^{-1}$.
As one can see in the figure, the resonance shape depends on the relative phase.
For $\delta=0$, the resonance and background contributions are interfered constructively 
a resonance peak appears in the cross section, while for $\delta = \pi$, the resonance 
and background contribute deconstructively and the resonance is seen as a dip.  
It is very interesting to see that, for the case of $\delta=\pi/2$, a rapid increase 
takes place at the resonance energy. This is also one of the resonance shapes.
These kinds of resonances are known as Fano resonance~\cite{Fano:1961zz}.
The resonance structure in the $KN$ $I=0$ channel might be one of
the example of Fano resonance.

\begin{figure}[]
    \begin{tabular}{c}
      \begin{minipage}[t]{1.0\hsize}
        \centering
        \includegraphics[keepaspectratio, scale=0.8,bb=0 0 360 252]{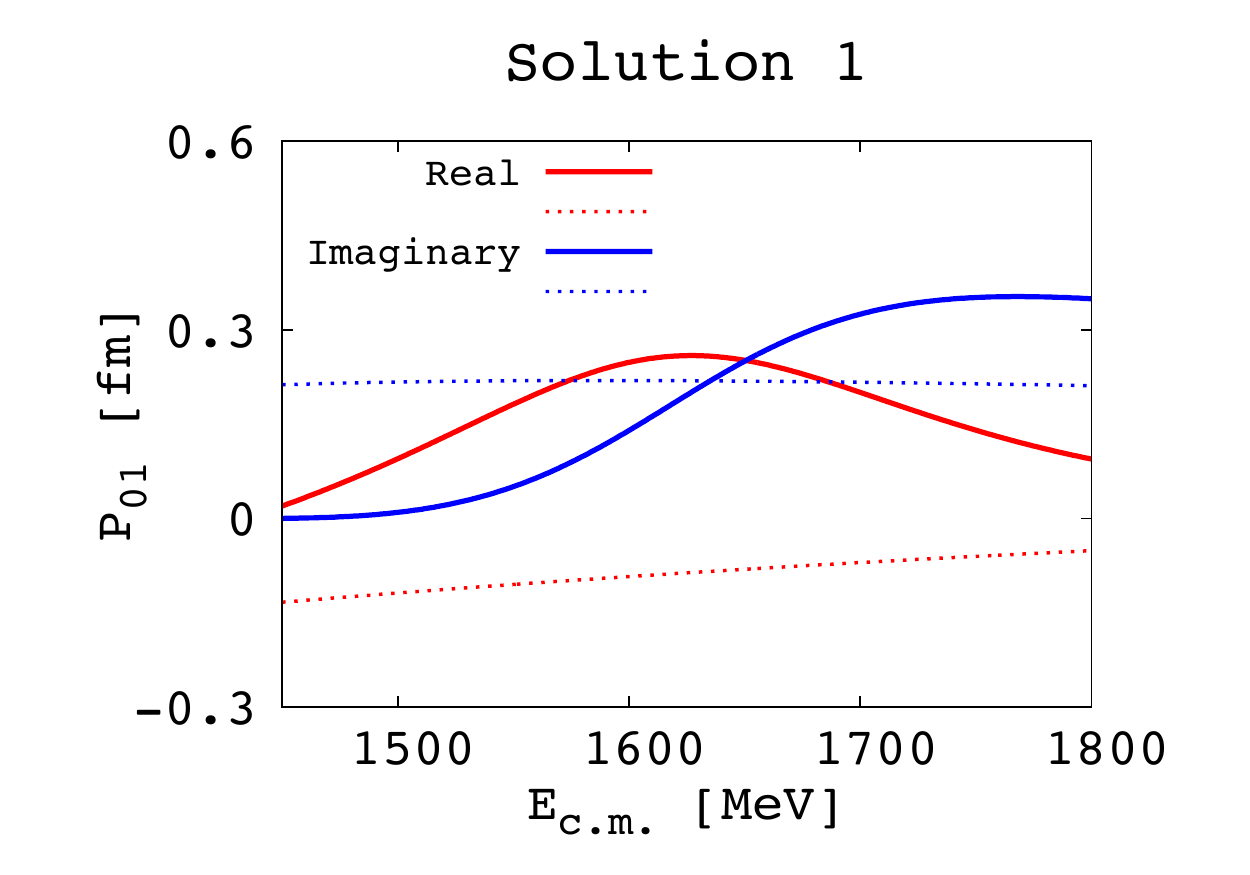}
      \end{minipage} \\
      \begin{minipage}[t]{1.0\hsize}
        \centering
        \includegraphics[keepaspectratio, scale=0.8,bb=0 0 360 252]{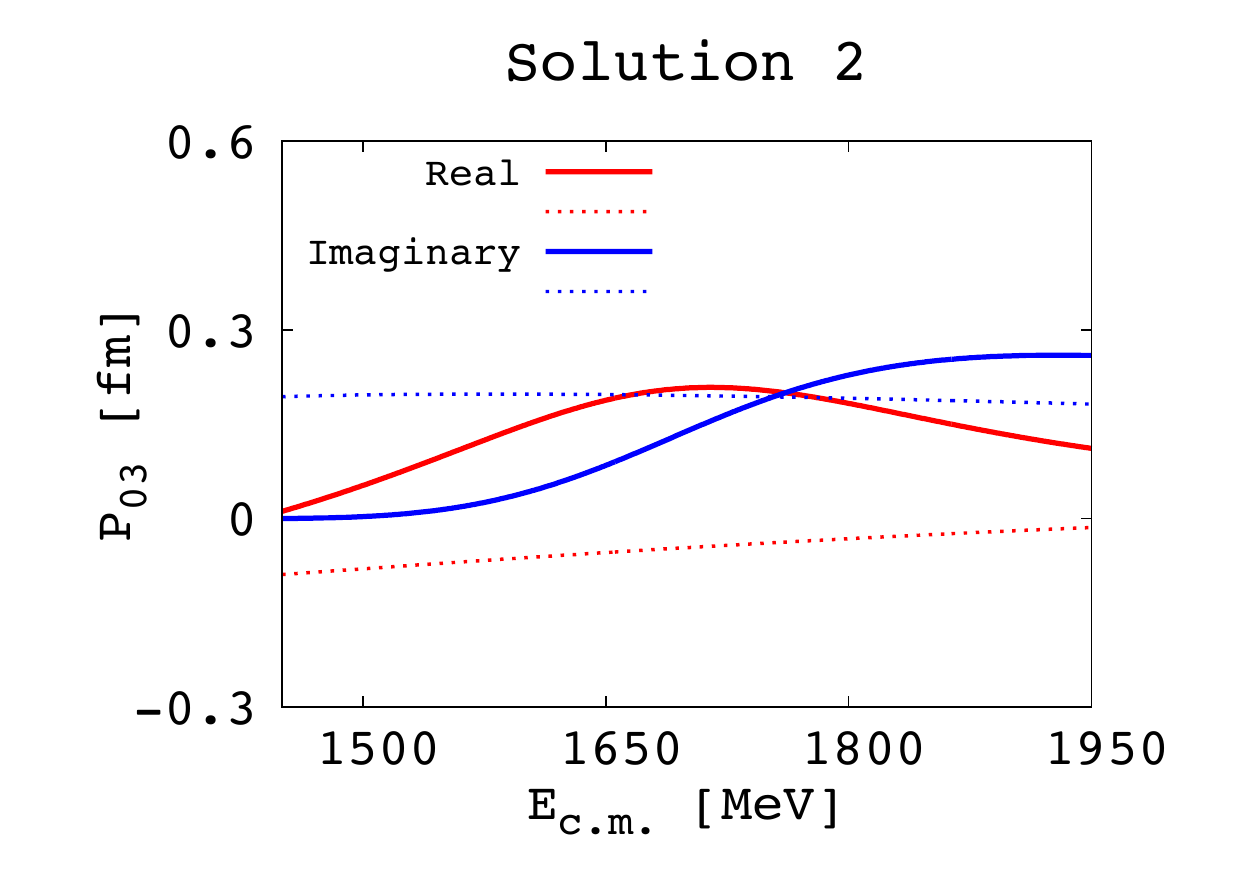}
      \end{minipage} \\
    \end{tabular}
     \caption{
     The real and imaginary parts of the $I=0$ amplitudes of $P_{01}$ for Solution 1,
     $P_{03}$ for Solution 2 around the resonance energy. 
     The solid lines stand for the original amplitudes, while 
     the dotted lines stand for the amplitudes obtained by subtracting the resonance pole.}
\label{fig:amp_i0}
  \end{figure}

In Fig.~\ref{fig:amp_i0}, we show the real and imaginary parts of 
the $I=0$ scattering amplitudes of $P_{01}$ for Solution 1 and
$P_{03}$ for Solution 2, where the resonances are found. As seen in figure, 
a typical resonance structure is seen in the amplitudes, but the role of the real and imaginary 
parts is interchanged. (Usually the imaginary part has a peak structure, while the real part
increases around the resonance point.) This is due to strong coupling of the resonance 
to the continuum background with some relative phase. In order to confirm whether 
the structure in the amplitude comes from the resonance state, we subtract
the resonance contribution from the amplitude.  We express the resonance contribution  
as the Breit-Wigner form, of which the numerator is obtained 
by calculating the residue of the amplitude at the resonance pole.
The subtracted amplitudes are shown as dotted lines in Fig,~\ref{fig:amp_i0}.
It implies that the subtracted amplitudes are almost constant without significant structure. 
Thus, the structure appearing in the amplitudes is caused by the resonance state. 

As we have mentioned above, the imaginary part of the amplitude rapidly 
increases around the resonance energy. According to the optical theorem,
the total cross section is proportional to the imaginary part. Therefore,
we conclude that the rapid increase seen in the $I=0$ total cross section around 
$p_{\rm lab}=500$ MeV/c can be a sign of the possible existence of a 
resonance with a large width. In addition, the spin-parity of the resonance 
can be learned by knowing which partial wave is responsible for the rapid increase 
of the $I=0$ total cross section. 
Here we have proposed 
two solutions; in Solution 1, the rapid increase appears in the $P_{01}$-wave
and the resonance should have $J^{p} = (1/2)^{+}$. In Solution 2, it does 
in the $P_{03}$-wave and the resonance should have $J^{p} = (3/2)^{+}$.
It would be very interesting if one could understand the feature of the
$I=0$ total cross section around $p_{\rm lab}=500$ MeV/c with more accurate 
experimental data. 

\begin{figure}[t]
 \begin{minipage}{0.5\hsize}
  \begin{center}
   \includegraphics[width=80mm,bb=0 0 360 252]{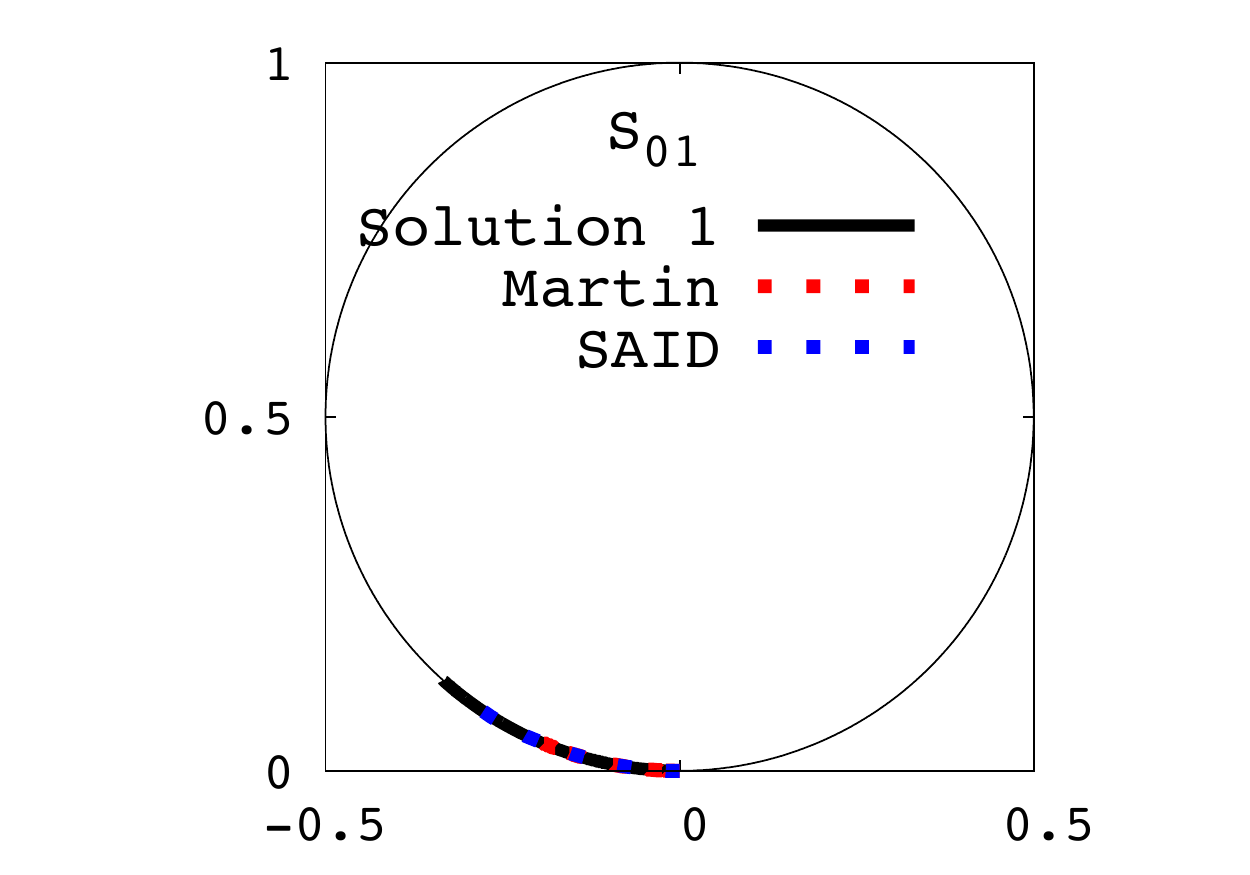}
  \end{center}
 \end{minipage}
 \begin{minipage}{0.5\hsize}
  \begin{center}
   \includegraphics[width=80mm,bb=0 0 360 252]{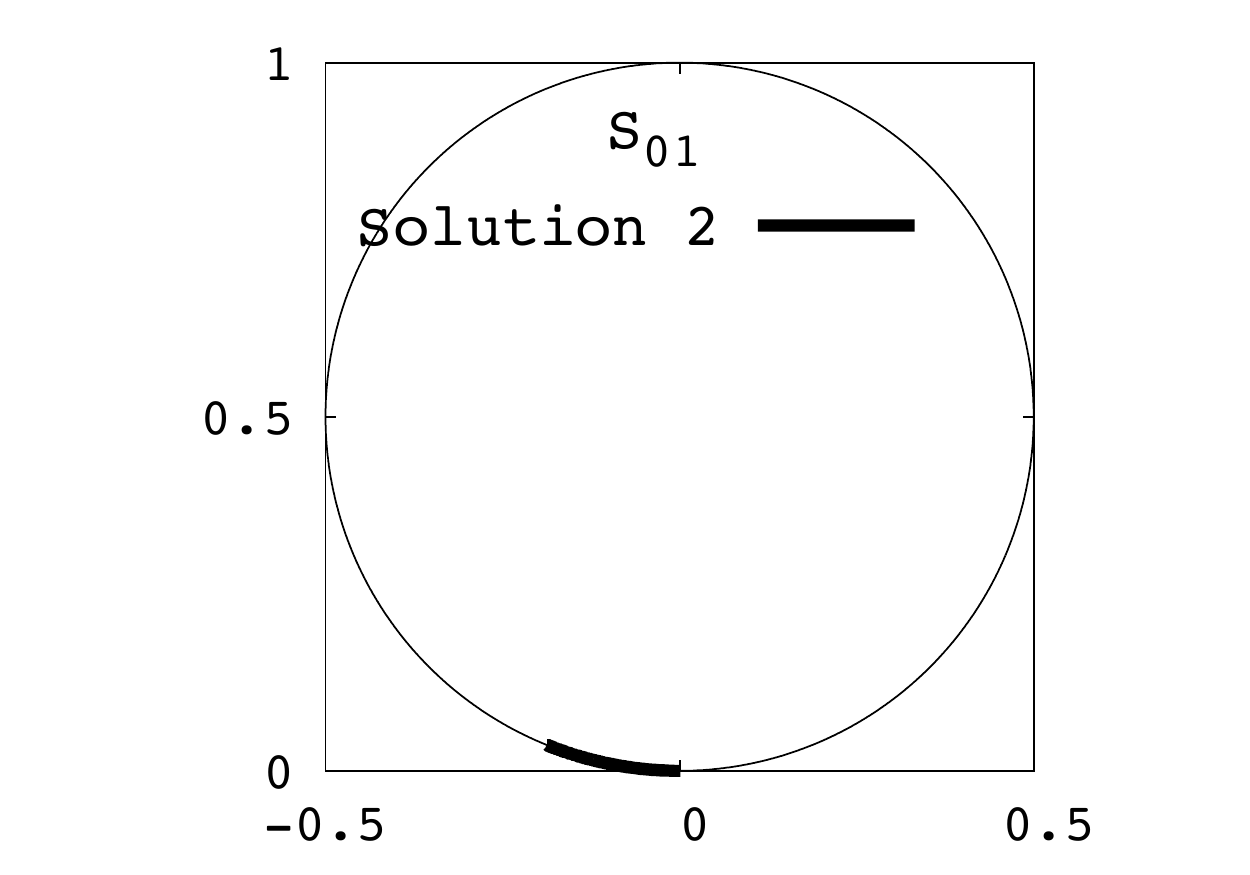}
  \end{center}
 \end{minipage}
 \begin{minipage}{0.5\hsize}
  \begin{center}
   \includegraphics[width=80mm,bb=0 0 360 252]{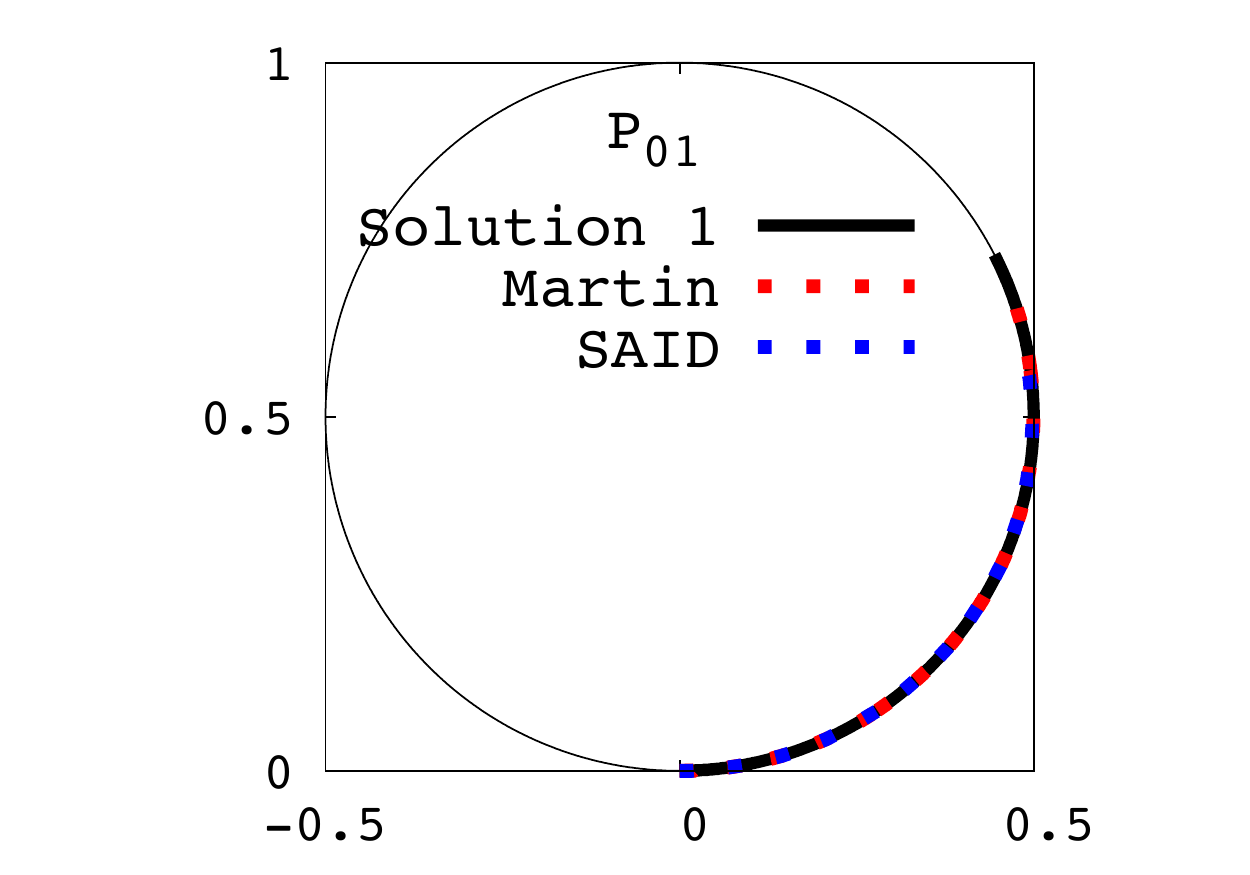}
  \end{center}
 \end{minipage}
 \begin{minipage}{0.5\hsize}
  \begin{center}
   \includegraphics[width=80mm,bb=0 0 360 252]{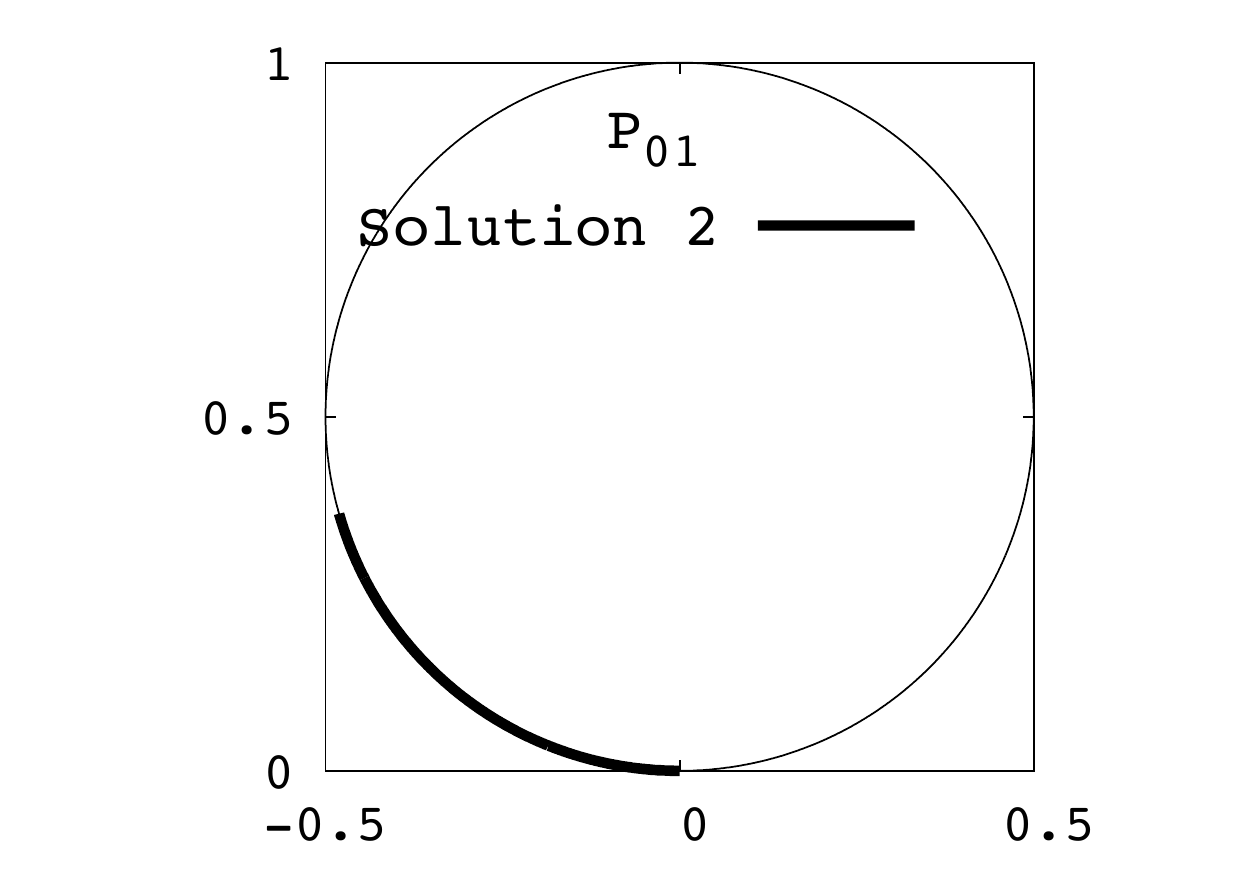}
  \end{center}
 \end{minipage}
   \begin{minipage}{0.50\hsize}
  \begin{center}
   \includegraphics[width=80mm,bb=0 0 360 252]{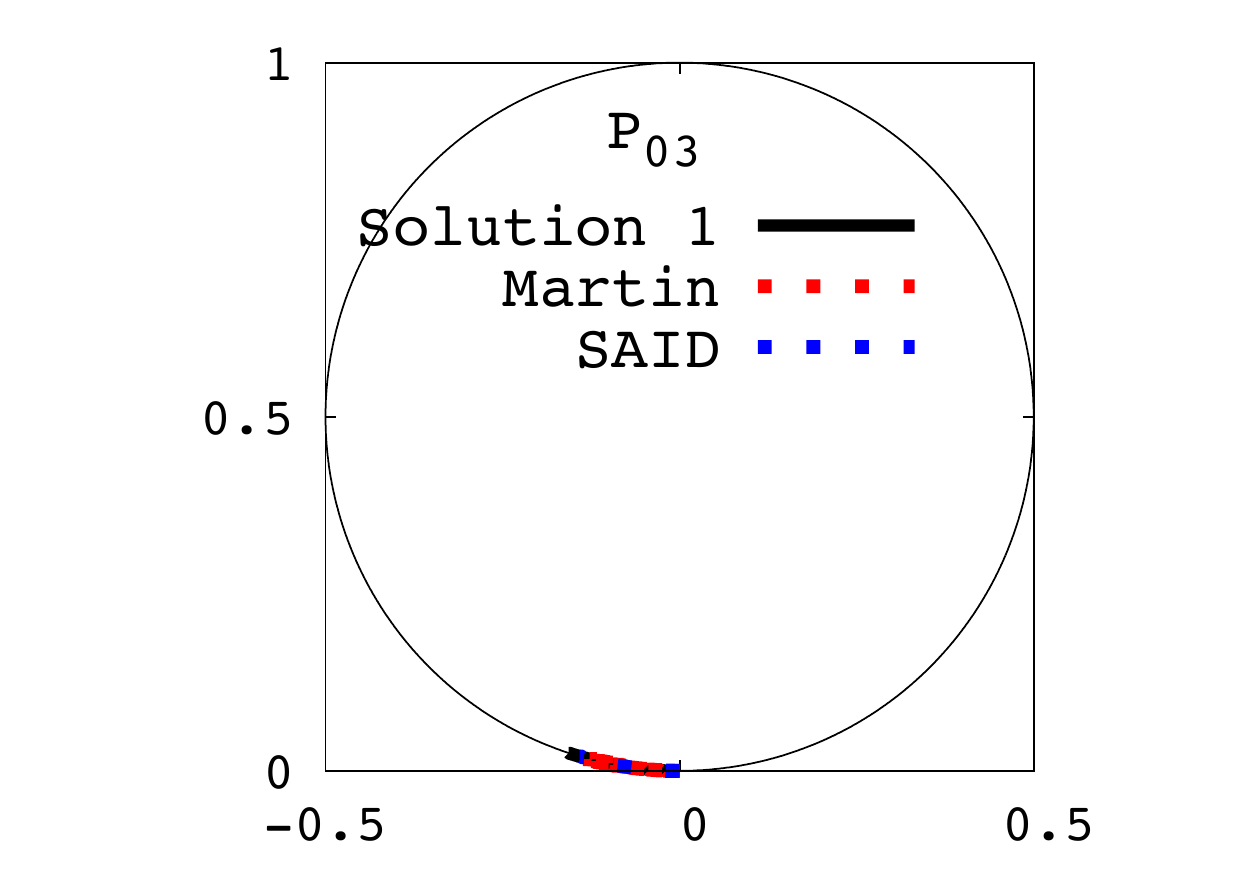}
  \end{center}
 \end{minipage}
 \begin{minipage}{0.5\hsize}
  \begin{center}
   \includegraphics[width=80mm,bb=0 0 360 252]{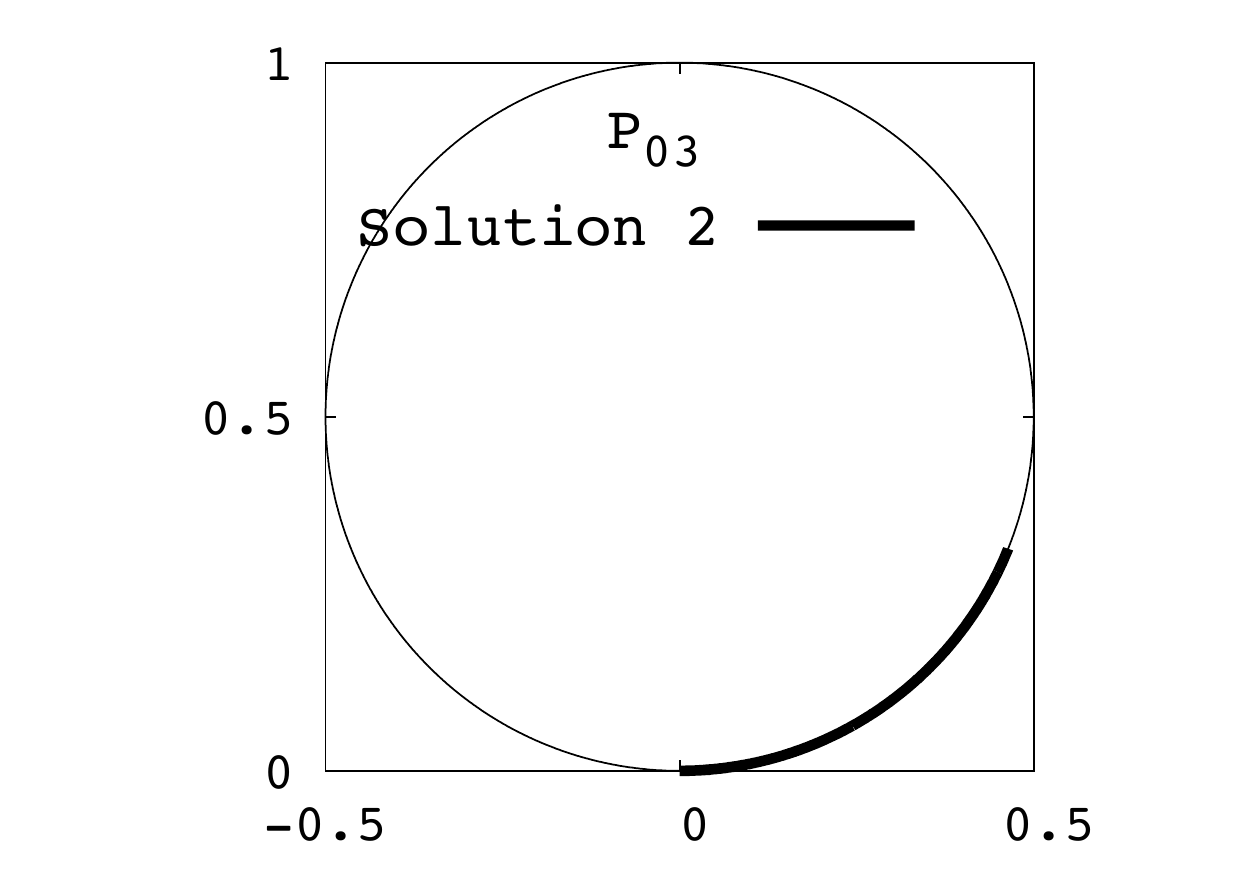}
  \end{center}
 \end{minipage}
  \vspace{-0.5cm}
 \caption{The Argand diagrams of Solutions 1 and 2 in the $I=0$ states up to the momentum $p_{{\rm lab}}=$800 MeV/c.
 Solution 1 is compared with the Martin's amplitude \cite{martin1975} and SAID amplitude 
  \cite{said} shown in the dotted line. It shows that Solution 1 is consistent with existing partial wave amplitude.}
\label{fig:arg_iso0}
 \end{figure}

\begin{figure}[t]
 \begin{minipage}{0.5\hsize}
  \begin{center}
   \includegraphics[width=80mm,bb=0 0 360 252]{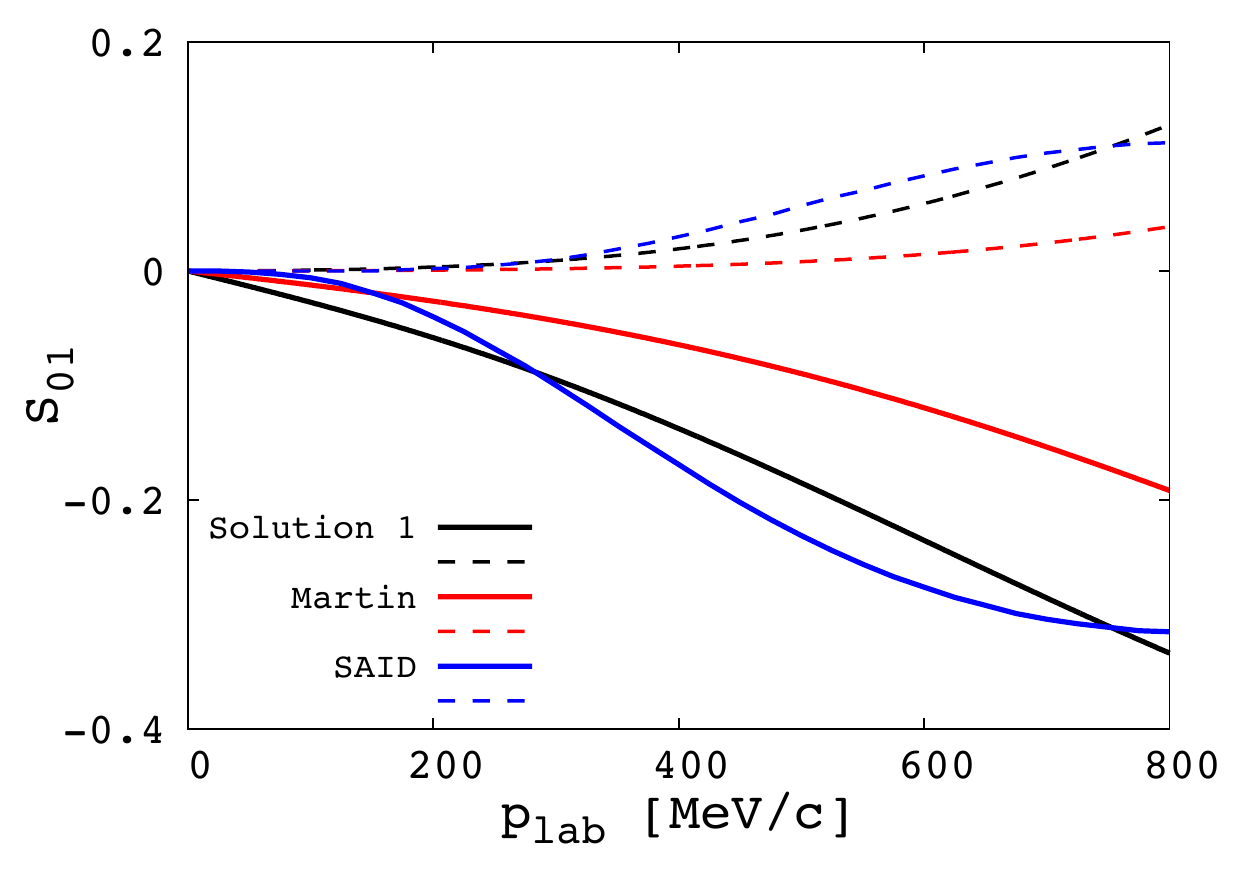}
  \end{center}
 \end{minipage}
 \begin{minipage}{0.5\hsize}
  \begin{center}
   \includegraphics[width=80mm,bb=0 0 360 252]{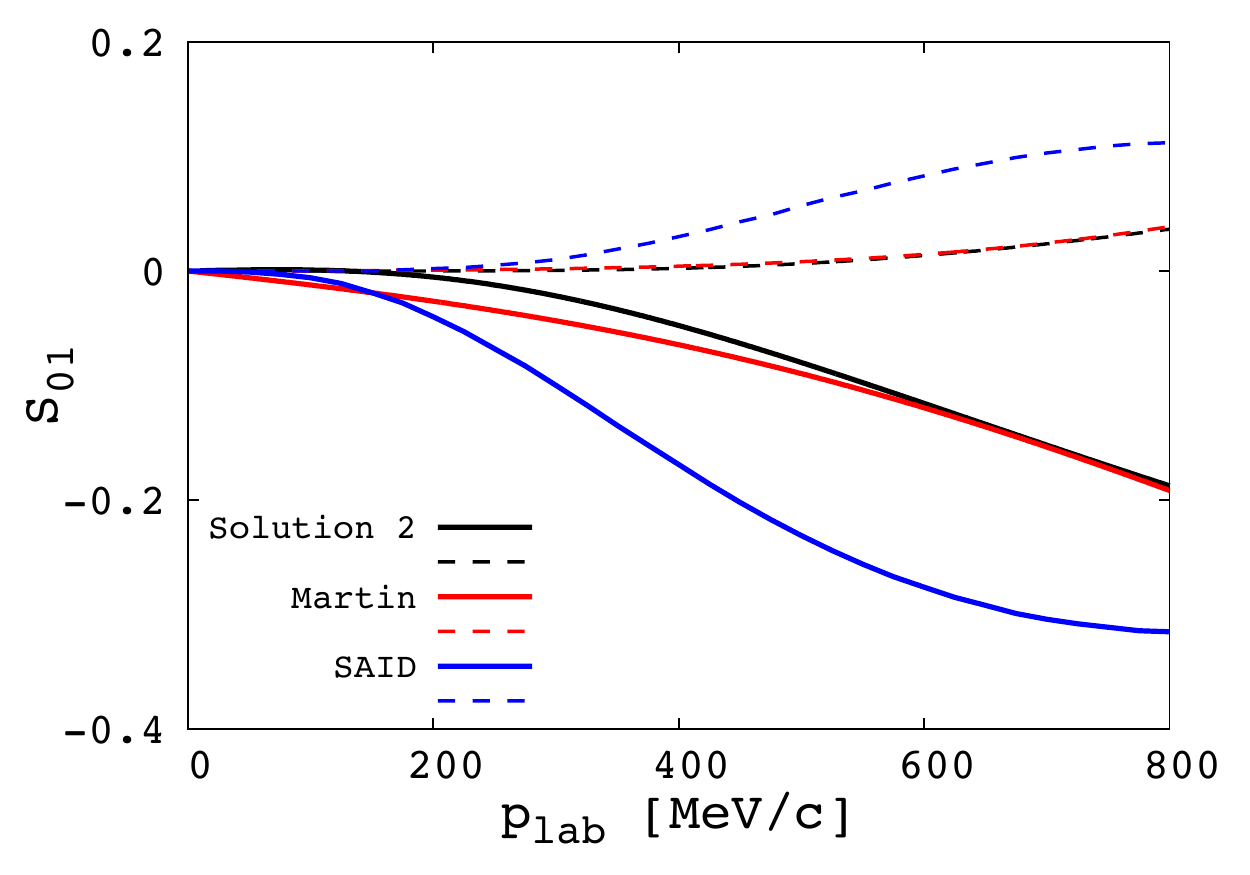}
  \end{center}
 \end{minipage}
 \begin{minipage}{0.5\hsize}
  \begin{center}
   \includegraphics[width=80mm,bb=0 0 360 252]{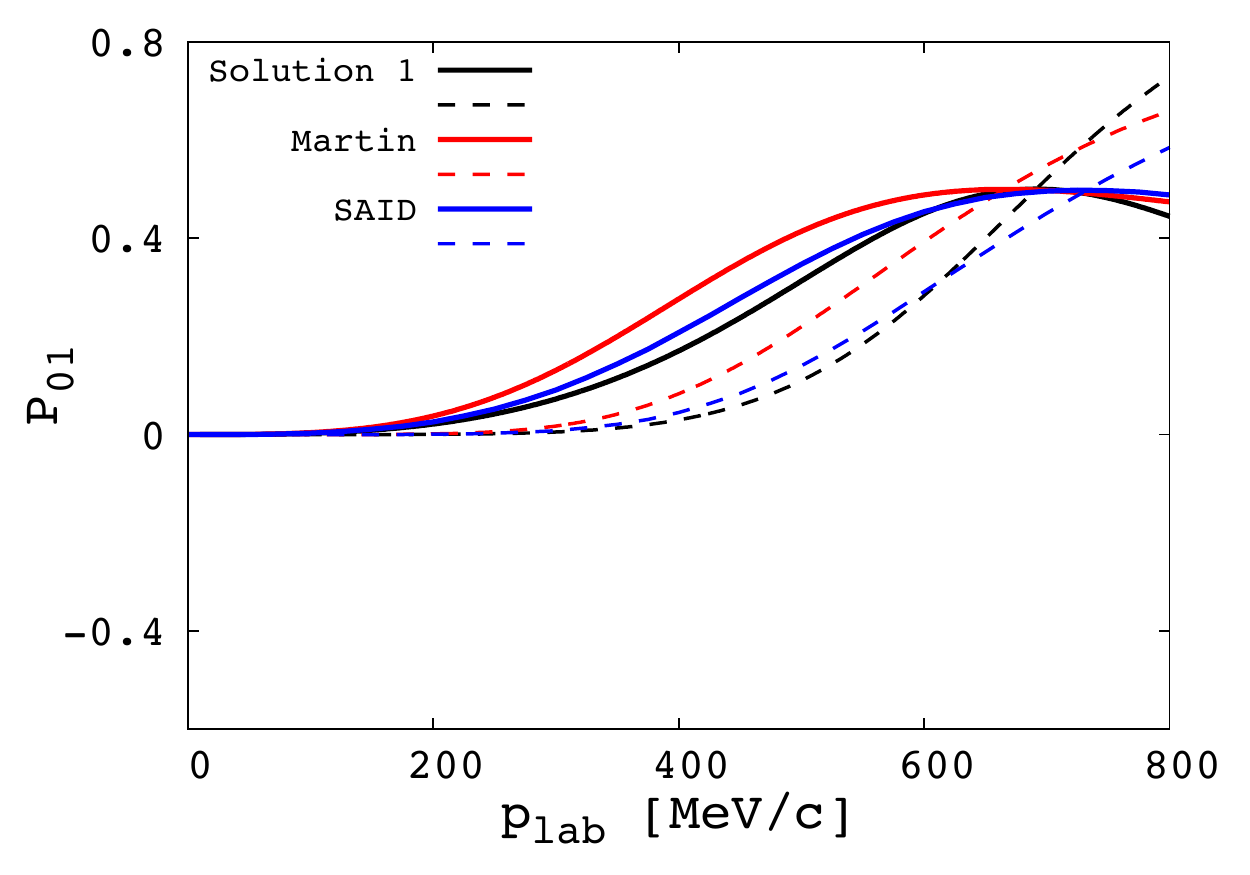}
  \end{center}
 \end{minipage}
 \begin{minipage}{0.5\hsize}
  \begin{center}
   \includegraphics[width=80mm,bb=0 0 360 252]{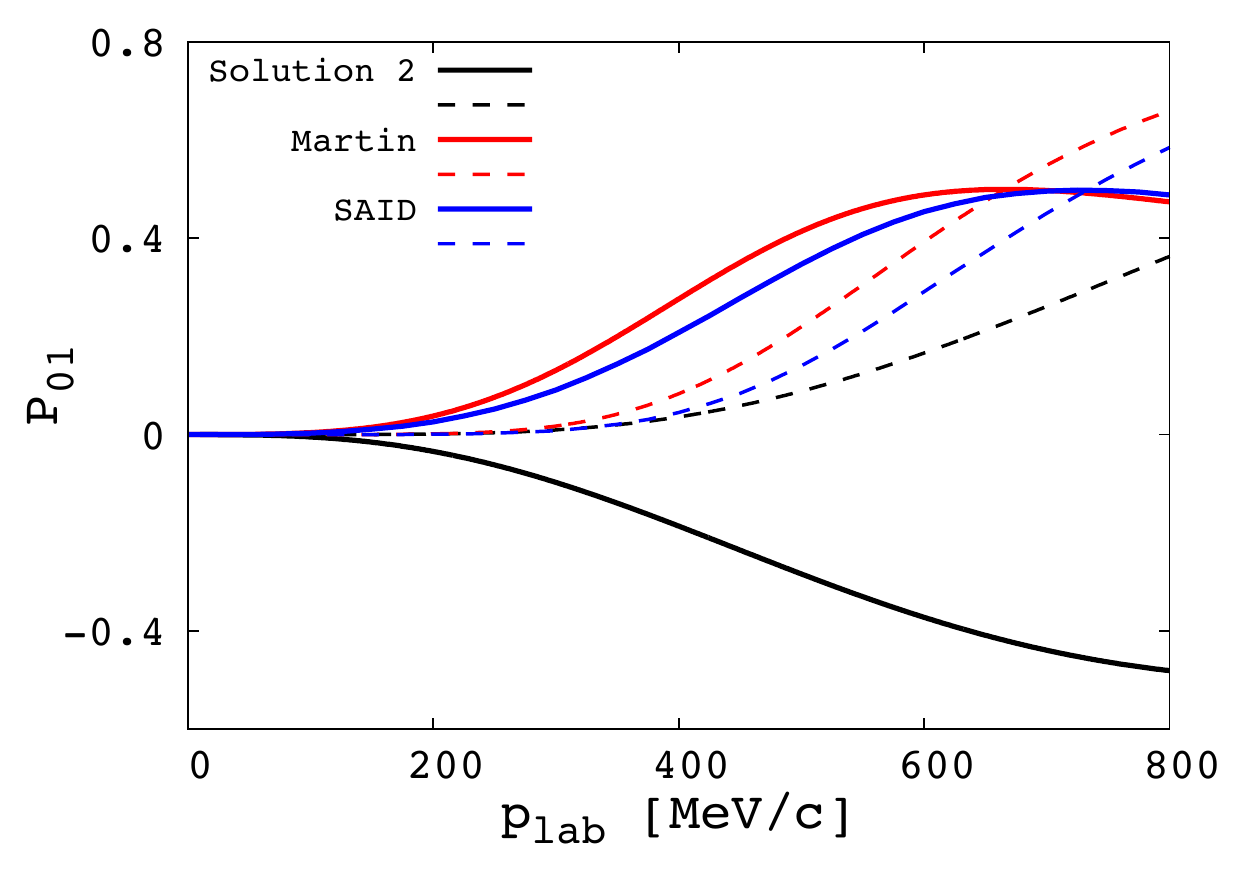}
  \end{center}
 \end{minipage}
   \begin{minipage}{0.50\hsize}
  \begin{center}
   \includegraphics[width=80mm,bb=0 0 360 252]{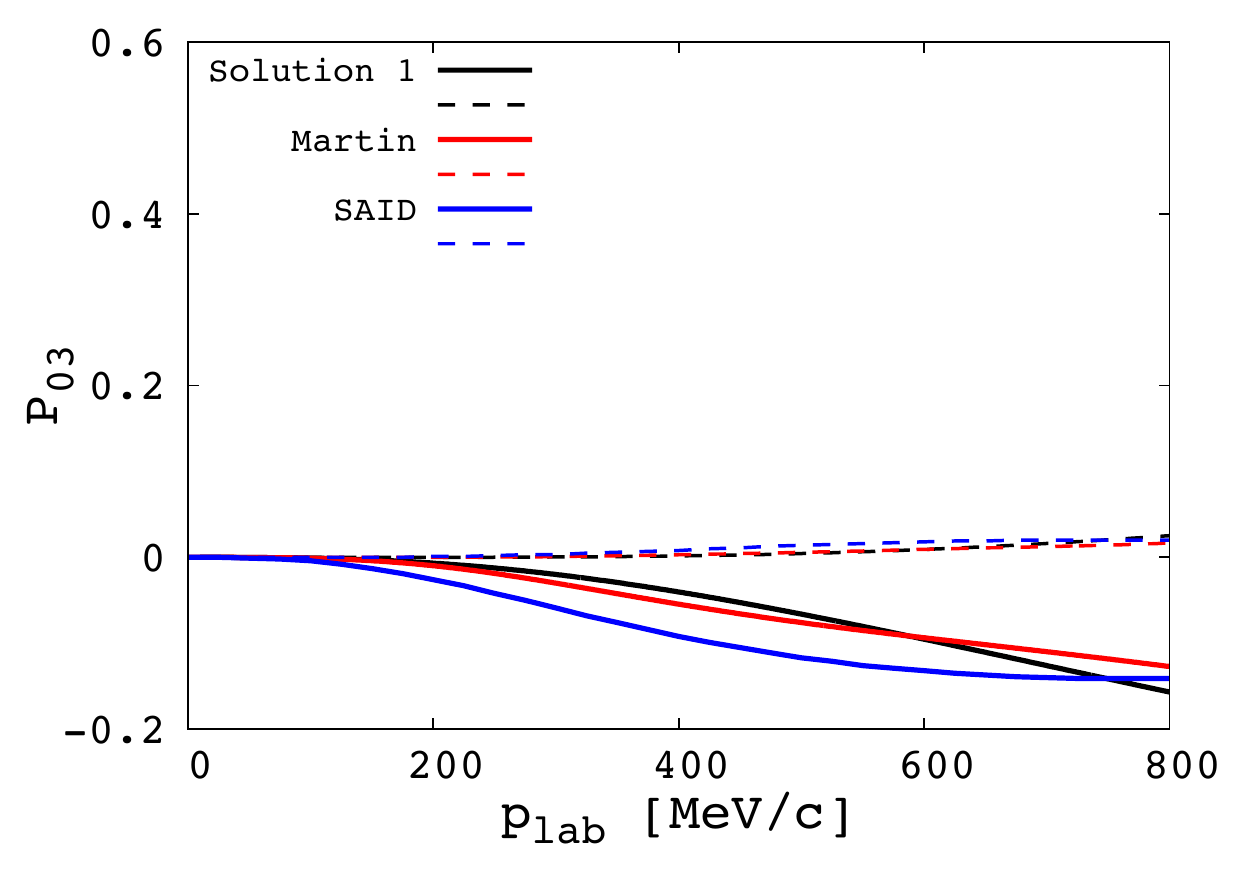}
  \end{center}
 \end{minipage}
 \begin{minipage}{0.5\hsize}
  \begin{center}
   \includegraphics[width=80mm,bb=0 0 360 252]{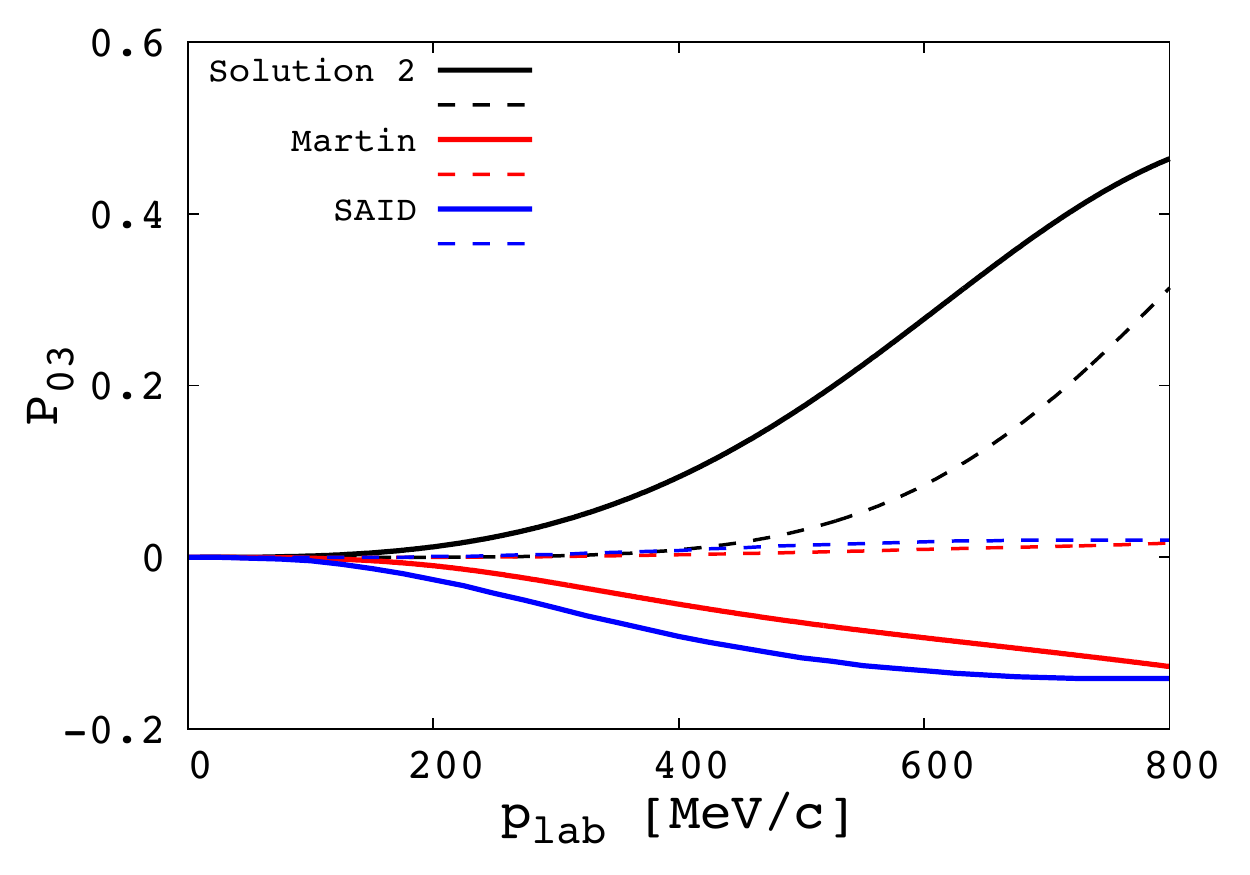}
  \end{center}
 \end{minipage}
  \vspace{-0.5cm}
 \caption{
 The dimensionless partial-wave amplitudes $T^{\prime}_{l \pm} = -kT_{l \pm}/(8\pi\sqrt{s})$
 of Solutions 1 and 2 in the $I=0$ channel up to the momentum $p_{{\rm lab}}=$800~MeV/c
 in comparison with Martin's~\cite{martin1975} and SAID~\cite{said} amplitudes.
 The solid and dashed lines stand for the real and imaginary parts of the amplitudes, respectively.}
\label{fig:energy_amp}
 \end{figure}

In Fig. \ref{fig:arg_iso0},
we show the Argand diagrams of the $I=0$ scattering amplitudes for the $S$ and $P$-wave
up to $p_{{\rm lab}}=$ 800 MeV/c using Solutions 1 and 2.
Then we compare with Martin's amplitude~\cite{martin1975} and amplitude of
SAID program~\cite{said}.
We find that the partial wave amplitudes for Solution 1 are very similar to the Martin's amplitude 
and amplitude of SAID program.
Thus, one could find a pole for a board resonance also in the Martin's amplitude
and amplitude of SAID program.
It is also interesting to point out that, for Solution 2, the $P_{03}$ channel has an attraction 
interaction and actually hold a broad resonance, while the $P_{01}$ channel is repulsive
even though some contribution of $P_{01}$ is seen in the total cross section.
In~Fig.~\ref{fig:energy_amp}, we show the momentum dependence of the $I=0$ partial-wave 
$T$-matrix~$T^{\prime}_{l \pm}$ defined by $T^{\prime}_{l \pm} = -kT_{l \pm}/(8\pi\sqrt{s})$,
where  $T_{l \pm}$ is given 
in Eq.~(\ref{eq:less_amp}), for Solutions~1 and 2 in comparison with Martin's amplitude~\cite{martin1975} and SAID amplitude~\cite{said}.
The solid and dashed lines stand for the real and imaginary parts of the amplitudes, respectively.

\begin{figure}[]
 \begin{center}
  \includegraphics[width=100mm,bb=0 0 360 252]{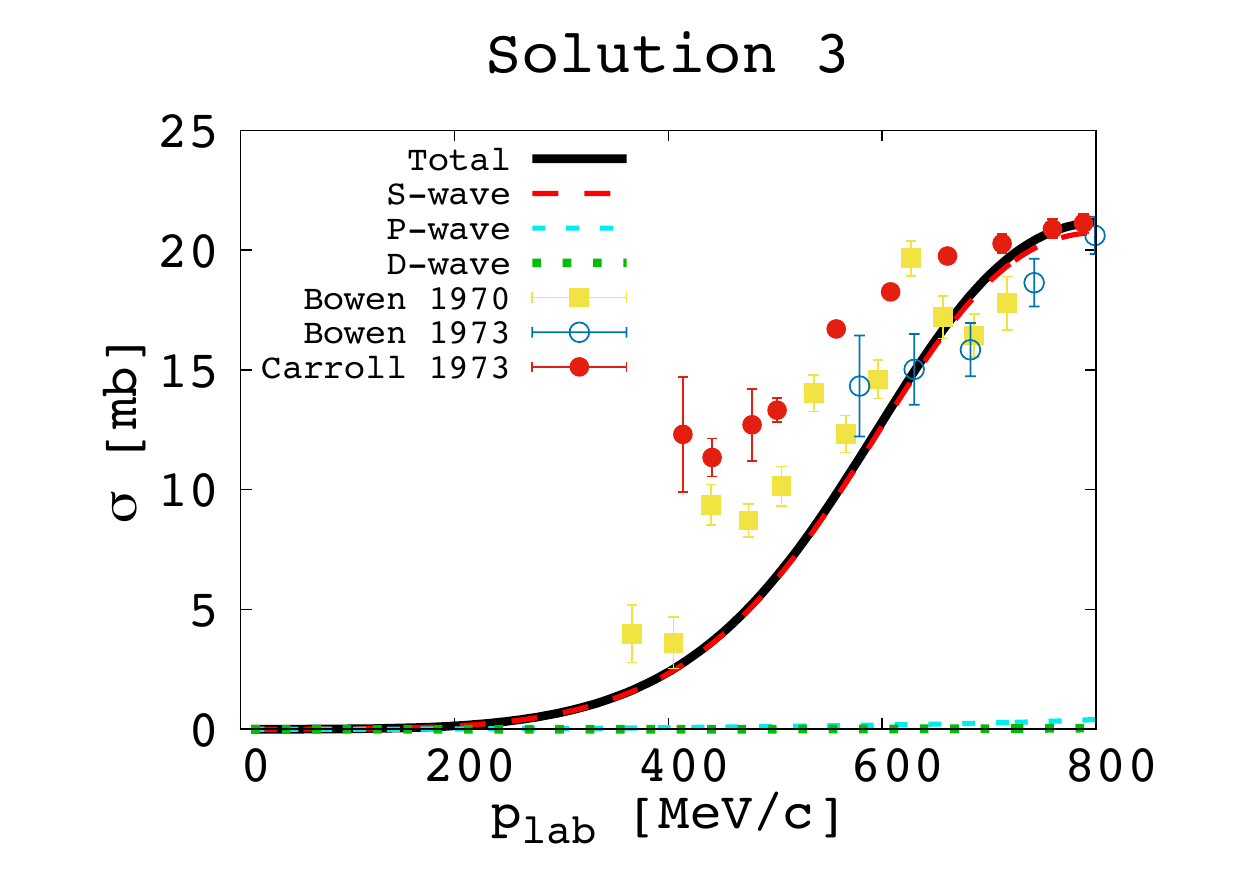}
 \end{center}
 \caption{The $I=0$ total cross section calculated using the Solution 3.}
 \label{fig:tot3}
\end{figure}

\begin{figure}[]
 \begin{center}
  \includegraphics[width=100mm,bb=0 0 360 252]{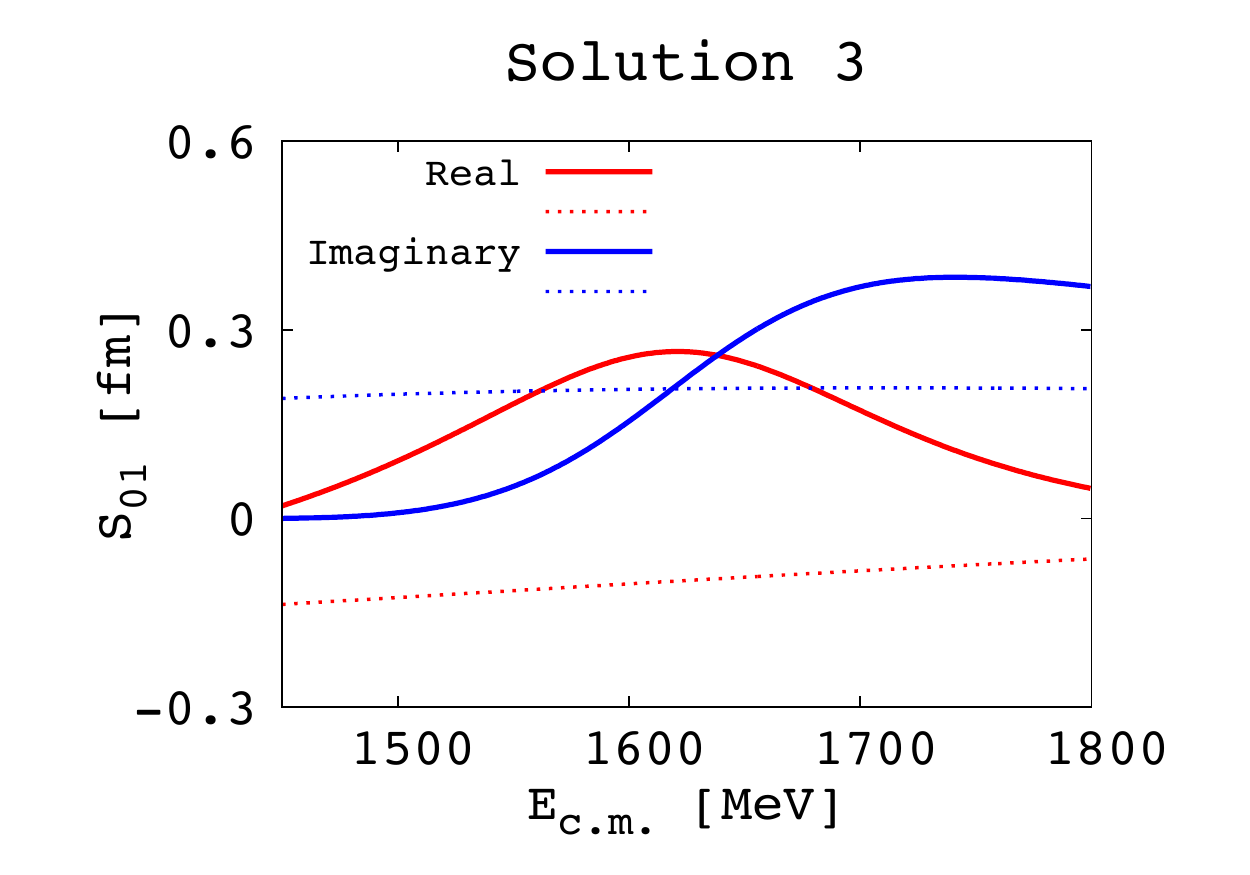}
 \end{center}
 \caption{The real and imaginary parts of the $S_{01}$ amplitude calculated using Solution 3.}
 \label{fig:amp3}
\end{figure}

It would be interesting to show a theoretical amplitude
which was the rapid increase of the $I=0$ total cross section in $S$-wave.
We find such a solution with the parameter set called Solution 3 given in Table \ref{tab:sol3}.
Figure \ref{fig:tot3} shows the $I=0$ total cross section.
It shows that the $S_{01}$ amplitude substantially contributes and
the rapid increase of the cross section stems from the $S_{01}$ amplitude.
Solution 3 cannot reproduce the angular dependence of the differential cross section
of the charge exchange, because the amplitude of Solution 3 is composed of $S$-wave contribution.
Thus, Solution 3 could be ruled out.
Nevertheless, here we discuss also Solution 3, because we want to point out
the relation between the rapid increase of the $I=0$ total cross section and existence of the
possible broad resonance.
We find a pole in the $S_{01}$ amplitude of Solution 3 at $z = 1624 - 132i$ MeV,
which corresponds to a resonance state with 1624 MeV/${\rm c^{2}}$, width 264 MeV and $J^{P}=(1/2)^{-}$.
In Fig. \ref{fig:amp3}, we show the real and imaginary part of the $S_{01}$ amplitude
for Solution 3 and find the resonance structure in the amplitude.

\begin{table}[]
\begin{center}
\caption{
The $S$-wave parameter set.
}
\begin{tabular}{c c| D{.}{.}{4}} 
\hline
           &                      & {\rm Solution 3} \\ 
 \hline  \hline
          &$b^{I=1}$         &0.30     \\
          &$d^{I=1}$         &-0.24     \\
$I=1$  &$g^{I=1}$        &0.72   \\
           &$h^{I=1}$         &1.05  \\
           & $\chi^{2}/N$   &4.06\\
\hline
          &$b^{I=0}$        &0.38     \\
          &$d^{I=0}$       &0.01  \\
$I=0$  &$g^{I=0}$      &-1.35   \\
           &$h^{I=0}$        &-0.11  \\
           & $\chi^{2}/N$   &29.7\\
\hline
\end{tabular} 
\label{tab:sol3}
\end{center}
\end{table}


\section{Conclusion}
\label{sec:sec4}
We have investigated the $KN$ elastic scattering below the energy where the 
inelastic contributions become signifiant, that is, $p_{\rm lab} < 800$ MeV/c, 
by describing the scattering amplitude in the chiral unitary approach as an analytic function. 
We utilize a next-to-leading 
chiral Lagrangian for the kernel interaction of the unitarized amplitude, and the 
low energy constants in the  amplitude are determined to reproduce
the differential cross sections of $K^{+}p \to K^{+}p$, $K^{+}n \to K^{+}n$, $K^{+}n \to K^{0} p$
and the $I=0, 1$ total cross sections.
We have obtained good scattering amplitudes which reproduce the observed
scattering cross section very well. Particularly, the $I=1$ scattering amplitude, 
namely $K^{+}p$ elastic amplitude, has been determined well thanks to less ambiguous 
experimental data with small errors, and we have found that the $I=1$ scattering 
amplitude at $p_{\rm lab}<800$ MeV/c is essentially described by 
$S$-wave contribution, which is consistent with our conventional knowledge. 
For the $I=0$ amplitude, we have proposed two possible parameter sets, 
which reproduce the $I=0$ scattering cross sections similarly and have different 
nature for the rapid increase appearing in the $I=0$ total cross section around
$p_{\rm lab} = 500$ MeV/c. In Solution 1, the rapid increase appears in $P_{01}$-wave 
contribution,  while in Solution 2 it stems from the $P_{03}$-wave contribution.
To show the example of the rapid increase of the $I=0$ total cross section
in the $S_{01}$ amplitude by Solution 3,  even though it is not a realistic solution. 

Having performed analytic continuation of the obtained $I=0$ scattering amplitudes
to the complex energy plane, we have found a pole corresponding to a broad resonance
state around 
$E_{\rm c.m.} = 1617~{\rm MeV}$ with 305 MeV width in each scattering amplitude. 
We would like to emphasize strongly that the existence of a broad resonance 
is responsible for the rapid increase of the $I=0$ total cross section around $p_{\rm lab} = 
500$ MeV/c. Thus, further investigation of the nature of the rapid increase 
of the $I=0$ total cross section reveals directly the existence of the $S=+1$ exotic resonance
state. Usually resonance states, especially narrow resonances, appear as a bump 
in the total cross section. Nevertheless, for broad resonances, because they strongly 
couple to the non-resonance background, their resonance shape seen in the cross section 
can be modified. This is known as Fano resonance. 

In order to pin down the existence of the $S=+1$ broad resonance, one needs further 
detailed investigation. First of all, the resonance found in this work has a broad width
and is located far from the real axis in the complex energy plane. The experimental 
informations are in the real axis and constrain the scattering amplitude well close to the real axis.
To make the scattering amplitude, or the position of the pole, constrained more, 
more accurate experimental data are necessary. In addition, one also needs more reliable
theoretical description. For instance, it could be necessary to introduce more terms 
into the interaction kernel. It is also important to describe $K^{+}d$ scattering with 
the deuteron wavefunction theoretically. This makes us to perform direct comparison 
of the theoretical calculation to the experimental observation.

\section*{Acknowledgment}

The authors would like to thank Dr.\ T.~Hyodo for his helpful comments. 
The work of D.J.\ was partly supported by Grants-in-Aid for Scientific Research from JSPS (17K05449). 


%

\end{document}